\documentclass[fleqn,usenatbib]{mnras}

\usepackage{newtxtext,newtxmath}

\usepackage[T1]{fontenc}
\usepackage{ae,aecompl}
\usepackage[graphicx]{realboxes}
\usepackage{longtable}

\usepackage{graphicx}
\usepackage{epstopdf}
\usepackage{amsmath}	
\usepackage{amssymb}	
\usepackage[lofdepth, lotdepth]{subfig}

\title[Sequential Star Formation in PGCC G181.84+0.31]{Sequential Star Formation in the filamentary structures of the Planck Galactic cold clump G181.84+0.31}

\author[Yuan et al.]{
Lixia Yuan,$^{1, 2, 3}$\thanks{E-mail: lxyuan@bao.ac.cn}
Ming Zhu,$^{1,3}$\thanks{E-mail: mz@nao.cas.cn}
Tie Liu,$^{4,5}$
Jinghua Yuan,$^{1}$
Yuefang Wu,$^{6}$
\newauthor
Kee-Tae Kim,$^{4}$
Ke Wang,$^{7,8}$
Chenlin Zhou, $^{1,2,3}$
Ken'ichi Tatematsu, $^{9}$
Nario Kuno$^{10,11}$
\\
$^{1}$National Astronomical Observatories, Chinese Academy of Sciences,
20A Datun Road, Chaoyang District, Beijing 100101, China \\
$^{2}$University of Chinese Academy of Sciences, 100049, Beijing, China\\
$^{3}${Key Laboratory of FAST, NAOC, Chinese Academy of Sciences, Beijing 100012, China} \\
$^{4}$Korea Astronomy and Space Science Institute, 776
Daedeokdae-ro, Yuseong-gu, Daejeon 34055, Republic of Korea \\
$^{5}$East Asian Observatory, 660 N. A'ohoku Place, Hilo, HI 96720, USA\\
$^{6}$Department of Astronomy, Peking University, 100871, Beijing China \\
$^{7}$Kavli Institute for Astronomy and Astrophysics, Peking University, 5 Yiheyuan Road, Haidian District, Beijing 100871, China \\
$^{8}$European Southern Observatory (ESO) Headquarters,
Karl-Schwarzschild-Str. 2, 85748 Garching bei M\"{u}nchen, Germany \\
$^{9}$National Astronomical Observatory of Japan, 2-21-1 Osawa, Mitaka, Tokyo 181-8588, Japan\\
$^{10}$Tomonaga Center for the History of the Universe, University of Tsukuba, 1-1-1 Tennodai, Tsukuba, Ibaraki 305-8571, Japan \\
$^{11}$Department of Physics, Graduate School of Pure and Applied Sciences, University of Tsukuba, 1-1-1 Ten-nodai, tsukuba, \\ 
Ibaraki 305-8577, Japan 
}

\date{Accepted 2019 May 03. Received 2019 May 03; in original form 2018 December 23}

\pubyear{2018}

 \hypersetup{draft}
\begin{document}
\label{firstpage}
\pagerange{\pageref{firstpage}--\pageref{lastpage}}
\maketitle

\begin{abstract}
	We present a multi-wavelength study of the Planck cold clump G181.84+0.31, which is located at the northern end of the extended filamentary structure S242. We have extracted 9 compact dense cores from the SCUBA-2 850 $\micron$ map, and we have identified 18 young stellar objects (YSOs, 4 Class I and 14 Class II) based on their Spitzer, Wide-field Infrared Survey Explorer (WISE) and Two-Micron All-Sky Survey (2MASS) near- and mid-infrared colours. The dense cores and YSOs are mainly distributed along the filamentary structures of G181.84 and are well traced by HCO$^{+}$(1-0) and N$_{2}$H$^{+}$(1-0) spectral-line emission. 
We find signatures of sequential star formation activities in G181.84: dense cores and YSOs located in the northern and southern sub-structures are younger than those in the central region.  
We also detect global velocity gradients of about 0.8$\pm$0.05 km s$^{-1}$pc$^{-1}$ and 1.0$\pm$0.05 km s$^{-1}$pc$^{-1}$ along the northern and southern sub-structures, respectively, and local velocity gradients of 1.2$\pm$0.1 km s$^{-1}$pc$^{-1}$ in the central substructure. These results may be due to the fact that the global collapse of the extended filamentary structure S242 is driven by an edge effect, for which the filament edges collapse first and then further trigger star formation activities inward. We identify three substructures in G181.84 and estimate their critical masses per unit length, which are $\sim$ 101$\pm$15 M$_{\odot}$ pc$^{-1}$, 56$\pm$8 M$_{\odot}$ pc$^{-1}$ and 28$\pm$4 M$_{\odot}$ pc$^{-1}$, respectively. These values are all lower than the observed values ($\sim$ 200 M$_{\odot}$ pc$^{-1}$), suggesting that these sub-structures are gravitationally unstable.                                                                      
\end{abstract}




\section{Introduction}

Recent dust continuum surveys, such as ATLASGAL \citep{Schuller2009}, BGPS \citep{Nordhaus2008} and Herschel infrared Galactic Plane Survey \citep{Molinari2010} have revealed that filamentary structures are ubiquitous along the Galactic plane. \cite{Wang2015, Wang2016} have identified 56 large-scale, velocity-coherent, dense filaments throughout the Galaxy, which is the first comprehensive catalog of large filaments. \cite{Li2016} also identified $\sim$ 517 filamentary structures in the inner-Galaxy. These filamentary structures are tightly correlated with the spiral arms and have a broad range of lengths and a variety of aspect ratios, widths and masses \citep{Arzoumanian2019, Hacar2018, Li2016}. In addition, the studies based on Herschel data, and other observations (e.g., near-Infrared extinction), found that more than 70\% of gravitationally bound dense cores and protostars are embedded in filaments that are supercritical \citep{Andre2013, Andre2014, Schneider2012, Contreras2016, Li2016}. There is a general consensus that large-scale supersonic flows compress the gas, driving the formation of filaments in the cold interstellar medium, then the filaments gravitationally fragment into pre-stellar cores and ultimately form protostars \citep{Andre2013, Andre2017}.

\begin{figure*}
	\includegraphics[width=16cm]{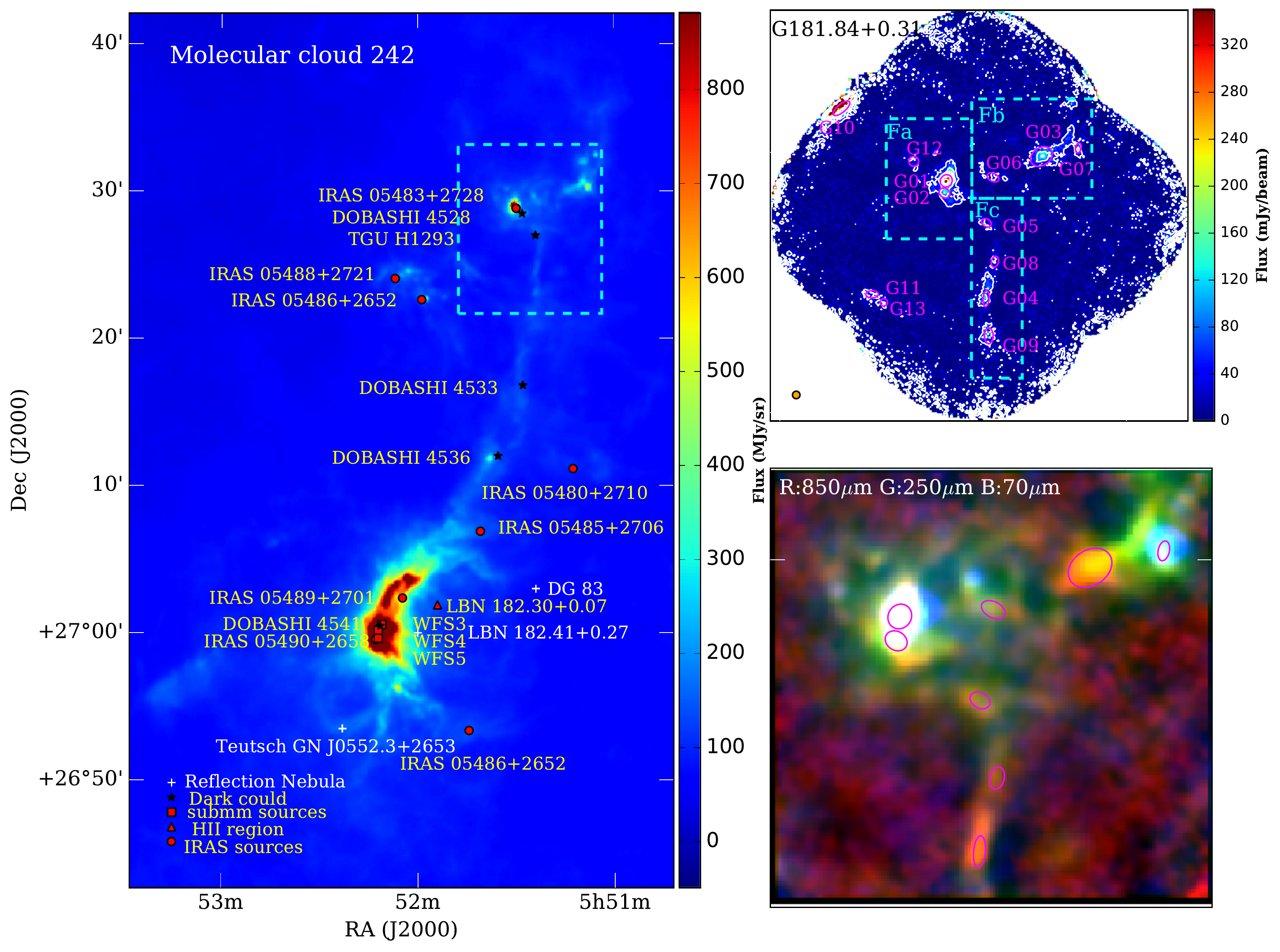}
    \caption{\textbf{Left panel:} The S242 molecular cloud is an elongated filamentary structure (length $\sim$ 25pc) \citep{Dewangan2017}. The red circle, triangle, square respectively stand for the IRAS sources, HII region, submm sources, the black stars and white crosses represent dark clouds and white reflection Nebula. All of them are retrieved from the SIMBAD Astronomical Database. The PGCC G181.84 region is outlined with dashed-cyan box. \textbf{Upper right panel:} SCUBA-2 850 $\mu$m continuum emission map for PGCC G181.84+0.31 with contour levels from 10\% to 90\% by step interval of 15\% of the peak value (350 mJy/beam). Magenta ellipses denote the compact sources extracted from 850 $\micron$ map. The orange circle in the right corner shows the SCUBA-2 effective beam size for SCUBA-2 at 850 $\micron$. The cyan-dashed boxes labeled as Fa, Fb and Fc, represent three major emission sub-structures in G181. \textbf{Lower right panel:} Three-colors map for G181, made using the  SCUBA-2 850 $\micron$ in red, SPIRE 250 $\micron$ in green, PACS 70 $\micron$ in blue.}
    \label{fig:f1}
\end{figure*}

\cite{Pon2011,Pon2012} found that global collapse in filamentary geometry (especially for those with aspect ratio A$_{o}$ $\gtrsim$ 5.0) tends to be influenced by an edge effect, such as end-dominated collapse. Their ends are preferentially accelerated, and further locally collapse in density enhancement to form stars. \cite{Johnstone2017} found a variation in evolutionary ages between the two end-regions of IC 5146. On the other hand, high-resolution observations with ALMA also revealed that edge instabilities may also affect the separation of the condensations in the edge region of filaments or pre-stellar cores \citep{Kainulainen2017,Ohashi2018}. Therefore, edge acceleration may play a role in the formation of protostars in filaments.

S242 is an elongated filamentary structure (EFS) with a length of $\sim$ 25 pc and a velocity ranging from -3 to 5 km s$^{-1}$ \citep{Dewangan2017}. The two end-regions of this filament contain most of the massive clumps and YSO clusters. Such distribution is consistent with the prediction of the end-dominated collapse scenario for star formation in filamentary clouds \citep{Dewangan2017}. The southern end of S242 is associated with an HII region, ionized by the star BD+26 980 with a spectral type of B0V \citep{Hunter1990}. The temperatures near the ionized gases and warm dusts can be up to $\sim$ 30 K. While the northern end is coincident with the Planck Galactic Cold Clump (PGCC) G181.84+0.31 (hereafter denoted as G181), having a column density of $\sim$ 10$^{22}$ cm$^{-2}$ and relatively low dust temperatures of 10 -- 12 K. G181 is located at a kinematic distance of 1.76$\pm$0.04 kpc, which was estimated using the parallax-based distance estimator for spiral arm sources \citep{Reid2016}.
 
In order to investigate the role of edge instabilities in star formation in gas filaments, we have made detailed analyses of the star formation activities and dense gas distribution in the northern end of S242 (G181), this end is not influenced by the circumambient environment. Furthermore, the PGCCs typically have significant sub-structures and are often associated with cloud filaments \citep{Rivera2016}. Follow-up observations in molecular lines and dust continuum emission have already resolved dense filaments inside some PGCCs \citep{Kim2017, Liua2018, Liub2018, Zhangchuan2018}. G181 was first detected by the \cite{Planck2011}. We divided the G181 region into three parts, as outlined by the cyan-dashed boxes of the upper left panel in Figure \ref{fig:f1}: Fa for the center sub-structure, Fb and Fc for the northern and southern sub-structures, respectively. Note that a FIR source, IRAS 05483+2728, is embedded in the Fa sub-structure.  

The paper is organized as follows: Section 2 presents the observations and data reductions. Section 3 presents the results. In Section 4, we discuss the relationship between filaments and star formation. Sections 5 presents the conclusions of this study.

\section{Observations and Data reduction}
\subsection*{2.1 Continuum emission}
\subsubsection*{2.1.1 The SCUBA-2 observation}
The 850 $\micron$ continuum observations of G181 were carried out with the JCMT\footnote{The JCMT telescope is the largest single-dish astronomical telescope in the world with a diameter of 15-m designed specifically to operate in the submillimeter wavelength. SCUBA-2 (Submillimeter Common-User Bolometer Array) is bolometer operating simultaneously at 450 and 850 $\micron$ with an innovative 10000 pixel bolometer camera \citep{Holland2013}.} in September 2015. This source was included in the ``SCUBA-2 Continuum Observations of Pre-protostellar Evolution" (SCOPE\footnote{https://www.eaobservatory.org/jcmt/science/large-programs/\%20scope/}) large project, which aims to study the roles of filaments in dense core formation and to investigate dust properties and detect rare populations of dense clumps \citep{Liua2018, Liu2016a, Eden2019}. The observations were conducted using the CV Daisy mode. The CV Daisy provided a deep 3\arcmin~(in diameter) region in the center of the map, but covered out to beyond 12$\arcmin$ \citep{Bintley2014}. We took a mean Flux Conversion Factor (FCF) of 554 Jy pW$^{-1}$ beam$^{-1}$ to convert data from pW to Jy beam$^{-1}$. The beam size at 850 $\micron$ is $\sim$ 14 arcsec and the main-beam efficiency is about 0.85. The observations were carried out under grade 3/4 weather condition. The 225 GHz opacity was between 0.1 -- 0.15. The rms noise level of the map is $\sim$ 6-10 mJy beam$^{-1}$ in the central 3\arcmin~region, and increases to 10-30 mJy beam$^{-1}$ out to 12\arcmin~area. This sensitivity is better than the value of 50-70 mJy beam$^{-1}$ in the ATLASGAL survey \citep{Contreras2013}.

\subsubsection*{2.1.2 Herschel and WISE archive data}
We also use the level 3.0 SPIRE maps and level 2.5 PACS maps available in the Herschel Science Archive. The observations for SPIRE (250-500 $\mu$m band) and PACS (70, 160 $\mu$m band) maps were carried out in parallel mode with fast scanning speed (60\arcsec/s). The FWHM beam sizes of the original Herschel maps are approximately 9.2, 12.3, 17.6, 23.9, and 35.2 arcsec at 70, 160, 250, 350, and 500 $\mu$m, respectively.

The Wide-field Infrared Survey Explorer (WISE) surveyed the entire sky in four mid-infrared bands (3.4, 4.6, 12 and 22 $\mu$m). The angular resolutions of WISE maps are respectively 6.1, 6.4, 6.5, and 12.0 arcsec at corresponding WISE bands. We retrieved the processed WISE images and catalogue data via the NASA/IPAC Infrared Science Archive (IRSA). The 5$\sigma$ point source sensitivities in the AllWISE catalog are respectively better than 0.08, 0.11, 1, and 6 mJy in unconfused regions at the corresponding WISE bands. The position precision for high signal-to-noise sources are better than 0.15 arcsec \citep{Wright2010}. 

\subsection*{2.2 Spectral-line emission}
\subsubsection*{2.2.1 Observations with the Nobeyama 45-m telescope}
We have carried out On-The-Fly (OTF) mapping observations for the G181 structures in the $J$=1-0 transition lines of HCO$^{+}$ and N$_{2}$H$^{+}$ using the Nobeyama Radio Observatory (NRO)\footnote{Nobeyama Radio  Observatory is a branch of the National Astronomical Observatory of Japan, National Institutes of Natural Sciences} 45-m telescope in 2018 January. We used the four-beam receiver (FOREST) with the Spectral SAM45 backend, which processes 16 IFs simultaneously. We used IF-A to cover the HCO$^{+}$(1-0) line with a spectral bandwidth of 125 MHz centering at the rest frequency of 89.19 GHz, and the IF-B to cover the N$_{2}$H$^{+}$(1-0) line with the same bandwidth centering at the rest frequency of 93.17 GHz. The FWHM beam sizes ($\theta_{HPBW}$) are $\sim$ 18.9$\pm$0.5 arcsec and 19.2$\pm$0.6 arcsec at 86 GHz in H and V polarizations, respectively. The main beam efficiencies ($\eta_{mb}$) are 55$\pm$5$\%$ and 58$\pm$5$\%$ at 86 GHz in H and V polarizations, respectively. The OTF mapping was performed for three areas (300\arcsec$\times$220\arcsec, 190\arcsec$\times$135\arcsec, 240\arcsec$\times$110\arcsec) covering the whole G181 region. An rms noise level of $\sim$ 0.1 K (T$_{mb}$) was achieved by co-adding all the observed maps. The telescope pointing was checked about every 1.2 hrs, and the accuracy was within 3\arcsec. The data was reduced using the software package NOSTAR\footnote{https://www.nro.nao.ac.jp/\%7enro45mrt/html/obs/otf/export-e.html} of Nobeyama Radio Observatory. By using a spheroidal function as a griding convolution function, we produced a data cube with a grid spacing of 6\arcsec~and a velocity resolution of 0.2 km s$^{-1}$. 

Furthermore, we have also performed single-point observations of the H$^{13}$CO$^{+}$(1-0) line toward the G01-G03 cores with T70 receiver in position switching mode. The beam size of the T70 receiver at this line frequency was 19\arcsec~and $\eta_{mb}$ was 55$\%$. The velocity resolution was 0.13 km s$^{-1}$ and an rms noise level of 0.03 K (T$_{mb}$) was achieved.

\begin{figure*}
	\includegraphics[width=16cm]{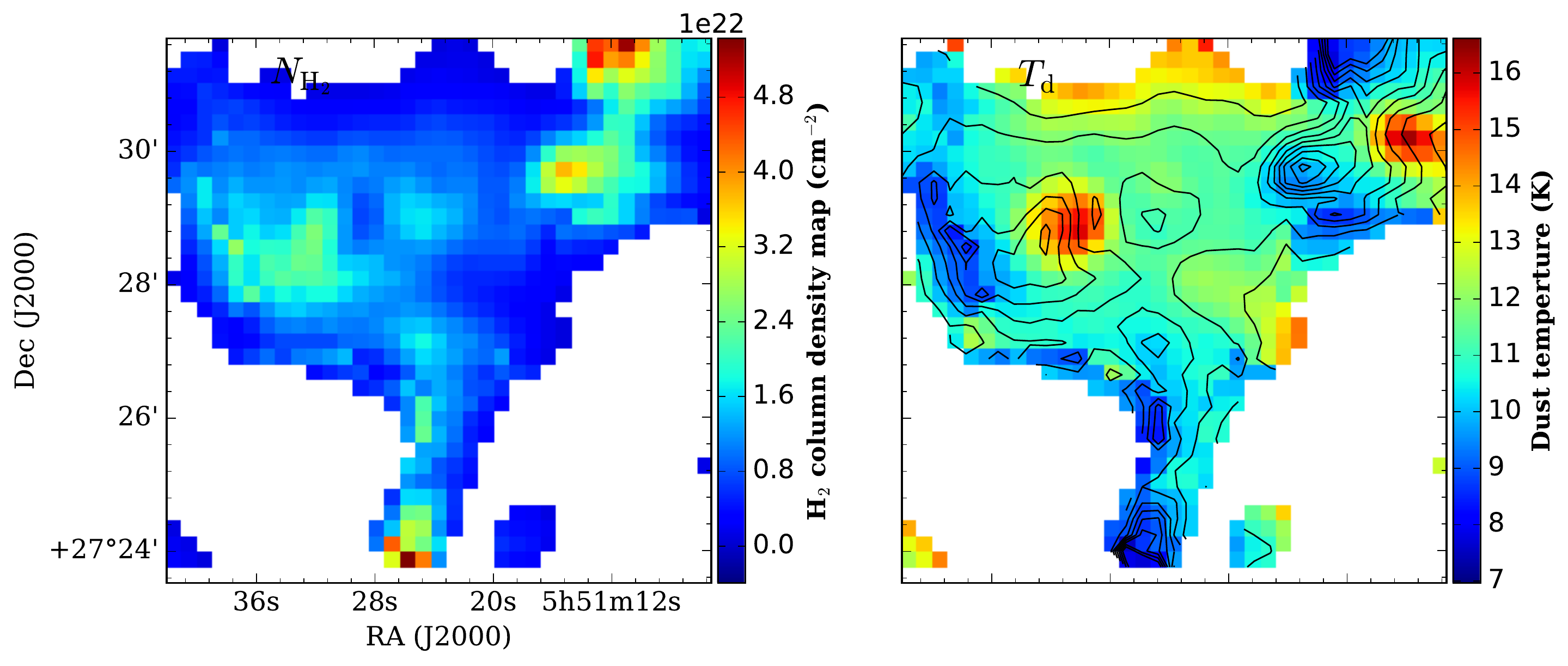}
    \caption{\textbf{Left panel}: H$_{2}$ column density map. \textbf{Right panel}: Dust temperature map. Both N$_{H_{2}}$ and T$_{d}$ are obtained from the SED fitting based on the graybody radiation mode using Herschel (160-500 $\mu$m) and SCUBA-2 850 $\mu$m photometric data. The black contours show the H$_{2}$ column density with levels stepped from 0.1 to 0.9 by 0.1 multiplied the peak value of 4.0 $\times$ 10$^{22}$ cm$^{-2}$.}
    \label{fig:f2}
\end{figure*}

\subsubsection*{2.2.2 $^{12}$CO(1-0), $^{13}$CO(1-0) spectral lines from PMO}

The mapping data of G181 in $^{12}$CO(1-0) and $^{13}$CO(1-0) were extracted from the CO line survey for the Planck Cold clumps using the 13.7-m telescope of Purple Mountain Observatory (PMO) by \cite{Wu2012}. The half-power beam width at 115 GHz is 52\arcsec~$\pm$ 3\arcsec. The main beam efficiency is about 50$\%$. The typical system temperature (T$_{sys}$) is about 110 K and varies by $\sim$ 10$\%$ for each beam. The velocity resolution is 0.16 km s$^{-1}$ for the $^{12}$CO(1-0) line and 0.17 km s$^{-1}$ for the $^{13}$CO(1-0) line. An rms noise level of 0.2 K in T$_{A}^{*}$ for the $^{12}$CO(1-0), and 0.1 K for the $^{13}$CO(1-0) line was achieved. The mapped region has a size of 22\arcmin~$\times$ 22\arcmin~centering at the PGCC G181. The scanning speed was 20\arcsec~s$^{-1}$. The noisy edges of the OTF maps were cropped, and only the central 14\arcmin $\times$ 14\arcmin~regions were kept for further analyses. The cube data were calibrated in GILDAS and gridded with a spacing of 30\arcsec~\citep{Wu2012, Liu2012, Meng2013}.

\section{Results}
\subsection*{3.1 Pixel to Pixel spectral energy distribution (SED) fitting}
\subsubsection*{3.1.1 Convolution to a Common Resolution and Foreground/Background Filtering}

We use the multi-wavelength continuum maps to estimate physical properties of G181. Firstly, the images of different wavelength bands were cropped to the same size, and transfered to the same unit for the value in every pixel. Then we re-grided maps to let them align with each other pixel-by-pixel with a common pixel size of 14\arcsec~using the imregrid algorithm of CASA\footnote{https://casa.nrao.edu/docs/TaskRef/imregrid-task.html}.

The maps were convolved with Gaussian Kernels to obtain images with an angular resolution of 35.2 arcsec, which matches the resolution of the un-convolved Herschel 500 \micron~map (whose angular resolution is the lowest). We used the convolution package of Astropy \citep{Astropy2013} with a Gaussian Kernel of 
\begin{equation}
\sqrt{35.2^{2}-\theta_{\lambda}^{2}}
\end{equation}
where $\theta_{\lambda}$ is the FWHM beam size for the corresponding Herschel or SCUBA-2 bands.

In order to reduce the atmospheric emission, the uniform astronomical signals in SCUBA-2 images beyond the spatial scales of 200$\arcsec$ have been filtered out. Following \cite{Yuan2017a}, we iteratively filtered the Herschel images using the CUPID-findback algorithm of the Starlink\footnote{http://starlink.eao.hawaii.edu/starlink/WelcomePage/} suite, which was developed for the SCUBA-2 data reduction. Further details of the algorithm are available in the online document for findback\footnote{http://starlink.eao.hawaii.edu/starlink/findback.html/}.

\subsubsection*{3.1.2 SED Fitting}
We used the smoothed and background-removed far-IR to submillimeter band maps, including Herschel (160-500 $\mu$m) data and SCUBA-2 850 $\mu$m data, to fit the intensity as a function of wavelength for each pixel through the modified blackbody model:
\begin{equation}
    I_{\nu} = B_{\nu}(T)(1-e^{-\tau_{\nu}}),
\end{equation}
where the Planck function B$_{\nu}$(T) is modified by optical depth:
\begin{equation}
\tau_{\nu} = \mu_{H_{2}}m_{H}\kappa_{\nu}N_{H_{2}}/R_{gd}.
\end{equation}
Here, $\mu_{H_{2}}$=2.8 is the mean molecular weight \citep{Kauffmann2008}, 
m$_{H}$ is the mass of a hydrogen atom and R$_{gd}$=100 is the gas to dust ratio, N$_{H_{2}}$ is the H$_{2}$ column density. 
The dust opacity $\kappa_{\nu}$ can be expressed as a power law in frequency,
\begin{equation}
\kappa_{\nu}=3.33\left(\frac{\nu}{600 GHz}\right)^{\beta} cm^{2}g^{-1},
\end{equation}
$\kappa_{\nu}$ (600 GHz) is the dust opacity, listed in Table 1 of \cite{Ossenkopf1994}, but the value has been scaled down a factor of 1.5 as suggested in \cite{Kauffmann2010}. And we took the dust emissivity index $\beta$ as 2.0, the standard value for cold dust emission \citep{Hildebrand1983}. 

The fitting was performed using the algorithm in the python package scipy.optimize\footnote{https://docs.spipy.org/doc/scipy/reference/generated/scipy.\\optimize.curve\_fit.html}. For each pixel, only the data with S/N $>$ 3 were included to perform the fitting. We considered flux uncertainties of the order of $\sim$ 15\% in all Herschel images, based on the work previously reported by \cite{Launhardt2013}.

The maps of H$_{2}$ column density (N$_{H_{2}}$) and dust temperature (T$_{d}$) are presented in Figure \ref{fig:f2}. The N$_{H_{2}}$ values across the filamentary structures are mainly around 1.0 $\times$ 10$^{22}$ -- 4.0 $\times$ 10$^{22}$ cm$^{-2}$ and the range of T$_{d}$ is from 8 to 16 K.  A single-temperature graybody model can fit the cold dust emission well, but can not account for the 70 $\micron$ emission as the 70 $\micron$ flux is mostly from the warm dust surrounding protostars. Therefore, we did not take the 70 $\micron$ data into account for the SED fitting. It should be noted that the dust temperatures could be underestimated, especially for sub-structure Fa whose 70 $\micron$ dust emission is relatively strong. Moreover, the dust opacity which is subject to a factor of two uncertainties \citep{Ossenkopf1994}, can induce uncertainties in N$_{H_{2}}$ and T$_{d}$. The dust emissivity index ($\beta$) can also significantly influence the derived results. If $\beta$ increases or decreases by 50\%, the resultant N$_{H_{2}}$ will increase by 2\% -- 35\%, or decrease by 10\% -- 35\%, while T$_{d}$ could have 5\% -- 18\% decrease or 13\% -- 28\% increase, respectively \citep{Yuan2017a}. 

Additionally, we estimated the effects of the filtering processes on the results of our SED fitting. For the sub-structures Fb and Fc, the H$_{2}$ column densities derived from unfiltered maps increase by 30 -- 50\%, when comparing with the values derived from the filtered maps. However, we found background-filter has little effect on the derived H$_{2}$ column densities in sub-structure Fa. 

From the N$_{H_{2}}$ map in Figure \ref{fig:f2}, we found that there are dense clumps and fragments unevenly distributed along the structures. Based on the overlaid contours on the temperature map in the right panel of Figure \ref{fig:f2}, we found that Fb and Fc filamentary sub-structures presented relative lower dust temperatures. The west part of Fb showing higher temperatures seems to compress the ambient matters. While, sub-structure Fa has higher dust temperatures, which may be heated by the formed protostars.

\cite{Dewangan2017} have published the H$_{2}$ column density and dust temperature maps of the molecular cloud S242. The H$_{2}$ column density in G181 is about (0.6 -- 1.5) $\times$ 10$^{22}$ cm$^{-2}$, approximately 50\% less than our fitting results (1.0 -- 4.0) $\times$ 10$^{22}$ cm$^{-2}$. The dust temperature distribution of G181 derived by \cite{Dewangan2017} is about 9 K -- 14 K, while our results is about 7 K -- 16 K. Such difference may be due to the fact that 850 $\micron$ flux is sensitive to cold dust emission and can better constrain the SED than using Herschel data alone.

Under the assumption of local thermal equilibrium (LTE), we further derived the H$_{2}$ column density maps using the $^{12}$CO(1-0) and $^{13}$CO(1-0) spectral lines from PMO through the equations (4) -- (13) in \cite{Qian2012}. As presented in Figure \ref{fig:fa1}, the CO excitation temperatures range from 8 K to 14 K. The $^{13}$CO column density is distributed in the range of 0.6 -- 1.4 $\times$ 10$^{16}$ cm$^{-2}$. Using an abundance ratio of $^{13}$CO and H$_{2}$ ([$^{13}$CO]/[H$_{2}$]) of 1.7 $\times$ 10$^{-6}$ \citep{Frerking1982}, the derived H$_{2}$ column density ranges between 1.0 -- 2.4 $\times$ 10$^{22}$ cm$^{-2}$, which is a factor of 1.5 higher than the N$_{H_{2}}$ range from \cite{Dewangan2017} (0.6 -- 1.5 $\times$ 10$^{22}$ cm$^{-2}$). This further confirms that the H$_{2}$ column density derived by SED fitting using only the Herschel data (160-500 \micron) may be underestimated.

\subsection*{3.2 Dense cores}
\subsubsection*{3.2.1 Identification of compact sources}
We have used the FellWalker\footnote{FellWalker is included the Starlink software package CUPID\citep{Berry2007}. FellWalker algorithm extracts compact sources by following the steepest gradients to reach a significant summit.} \citep{Berry2015} source-extraction algorithm to identify compact sources from the 850 $\micron$ continuum image. We follow the source-extraction processes used by the JCMT Plane Survey and the detailed descriptions are given in \cite{Moore2015} and \cite{Eden2017}.

We set a detection threshold of 3$\sigma$ and constructed a mask in the signal-to-noise map. The extracted sources were required to consist of at least 7 continuous pixels. A total of 13 compact sources have been extracted, named as G01-G13. Table \ref{tab:table1} lists the coordinates, size, peak flux density and integrated flux density at 850 $\mu$m. The identified compact sources are shown as magenta ellipses in the upper right panel of Figure \ref{fig:f1}. We manually rejected sources not associated with the filamentary structures of G181. Finally, a total of nine compact sources, G01-G09, were included in our further analysis.

\begin{table*}
	\centering
	\caption{The extracted compact sources from SCUBA-2 map in PGCC G181.84}
	\label{tab:table1}
	\begin{tabular}{lccccccccccr} 
		\hline
		Source Name & RA$_{peak}$ &  DEC$_{peak}$ & RA$_{cen}$ & DEC$_{cen}$ & $\sigma_{maj}$ & $\sigma_{min}$ & PA & S$_{int}$ & S$_{peak}$ &    \\
              & (hh:mm:ss) & (dd:mm:ss) & (hh:mm:ss) & (dd:mm:ss) & (arcsec) & (arcsec) & (degree) & ($Jy$) & (mJy/beam) & \\
		\hline
		G01 & 5:51:31.2 & 27:28:54.48 & 5:51:31.20 & 27:29:00.19 & 13.3 & 14.7 & 296.6 &1.6(0.25) & 360.9 \\
		G02 & 5:51:31.5 & 27:28:26.4 & 5:51:30.46 & 27:28:18.6 & 13.3 & 10.9 & 253.4 & 0.46(0.15) & 143.0 \\
		 G03 & 5:51:14.7 & 27:29:50.64 & 5:51:13.1 & 27:30:00 & 27.1 & 20.3 & 302.2 & 1.24(0.19) & 146.7 \\
         G04 & 5:51:24.29 & 27:24:22.68 & 5:51:24.09 & 27:24:42.68 & 6.8 & 18.8 & 347.6 & 0.47(0.12) & 126.4  \\
         G05 & 5:51:25.5 & 27:27:18.36 & 5:51:24.2 & 27:27:18.0 & 12.1 & 8.9 & 240.6 & 0.15(0.04) & 49.5  \\
         G06 & 5:51:23.09 & 27:29:02.40 & 5:51:23.7 & 27:29:09.8 & 14.7 & 9.2  &243.6& 0.21(0.04) & 52.8  \\
         G07 & 5:51:08.4 & 27:30:07.2 & 5:51:08.30 & 27:30:10.25 & 6.4 & 11.6 & 347.7& 0.16(0.05) & 113  \\
         G08 & 5:51:23.09 & 27:25:54.48 & 5:51:22.76 & +27:25:48.02 & 8.3 & 13 & 346.9 & 0.12(0.02)& 55.8 \\
         G09 & 5:51:24.3 & 27:23:06.4 & 5:51:23.83 & +27:22:57.06 & 10 & 18.3 & 180.6 & 0.37(0.04)& 102.4 \\
         G10 & 5:51:49.56 & 27:31:50.52 &5:51:49.37 & +27:31:41.75 & 23.8 & 9.3 & 309.9 & 3.2(0.25) & 618.5 \\
         G11 & 5:51:45.34 & 27:24:34.56 & 5:51:44.040 & +27:24:32.78 & 14.7 & 8.4 & 272.05 & 0.18(0.03) & 73.3 \\
         G12 & 5:51:36.62 & 27:29:18.6 & 5:51:36.85 & +27:29:39.36 & 11.6 & 11.3 & 229.4 & 0.14(0.03) & 49.4 \\
         G13 & 5:51:41.7 & 27:24:06.5 &5:51:42.00  & +27:24:14.06 & 8.7 & 8.3 & 305.5 & 0.1(0.03) & 56.2 \\
		\hline
	\end{tabular}
\end{table*}

\begin{figure*}
\centering
\subfloat[G01]{\includegraphics[width=5.5cm]{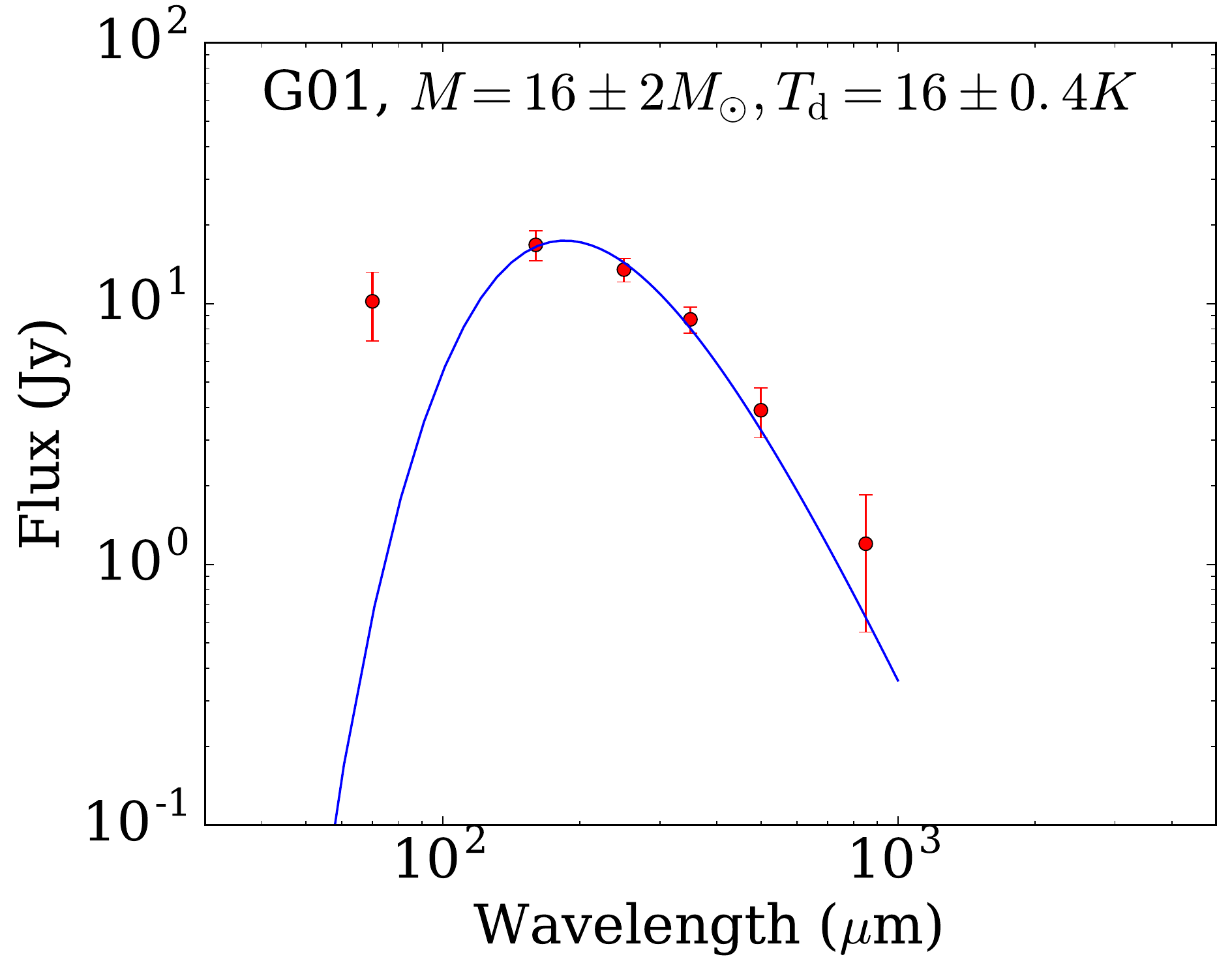}}
\subfloat[G02]{\includegraphics[width=5.5cm]{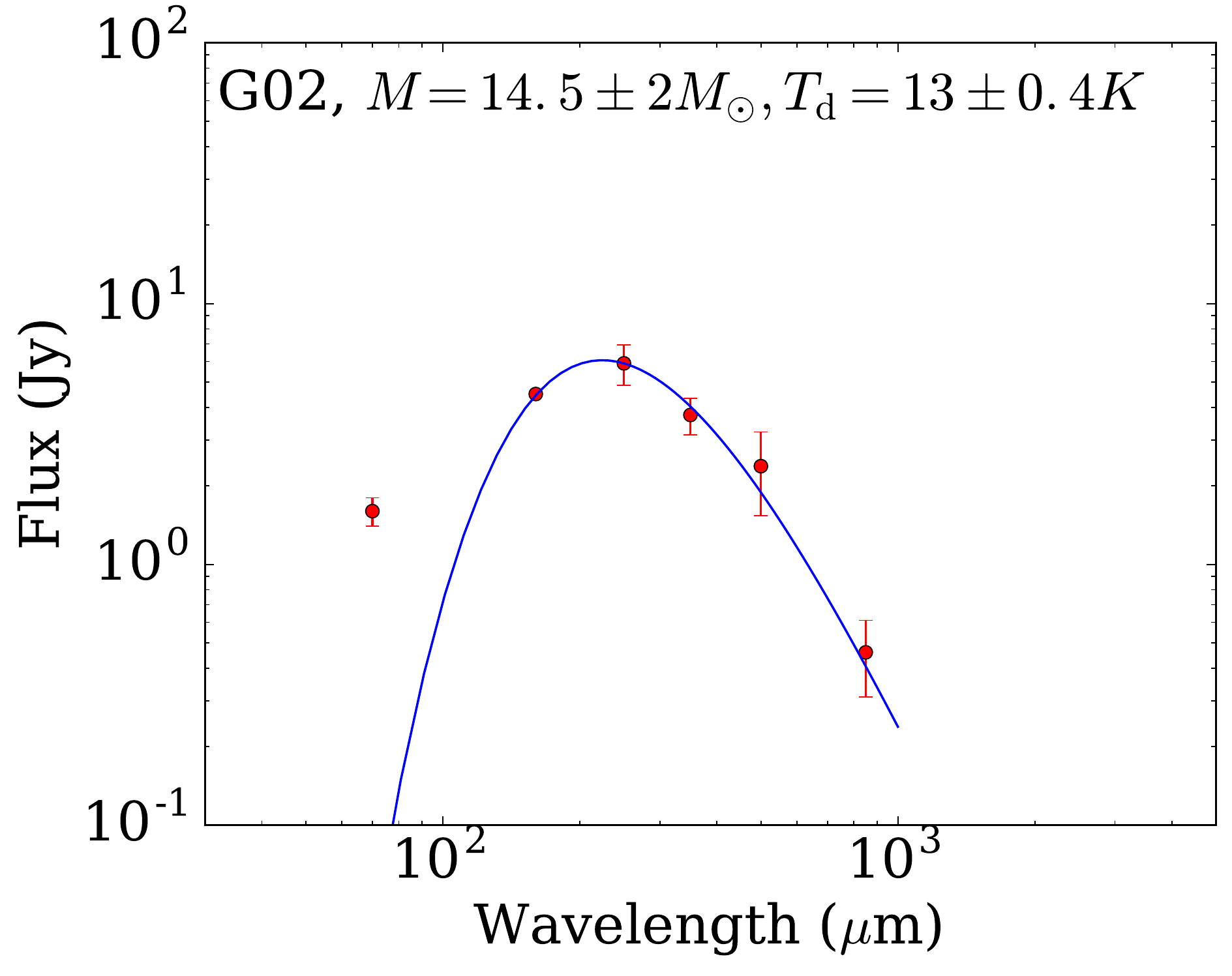}}
\subfloat[G03]{\includegraphics[width=5.5cm]{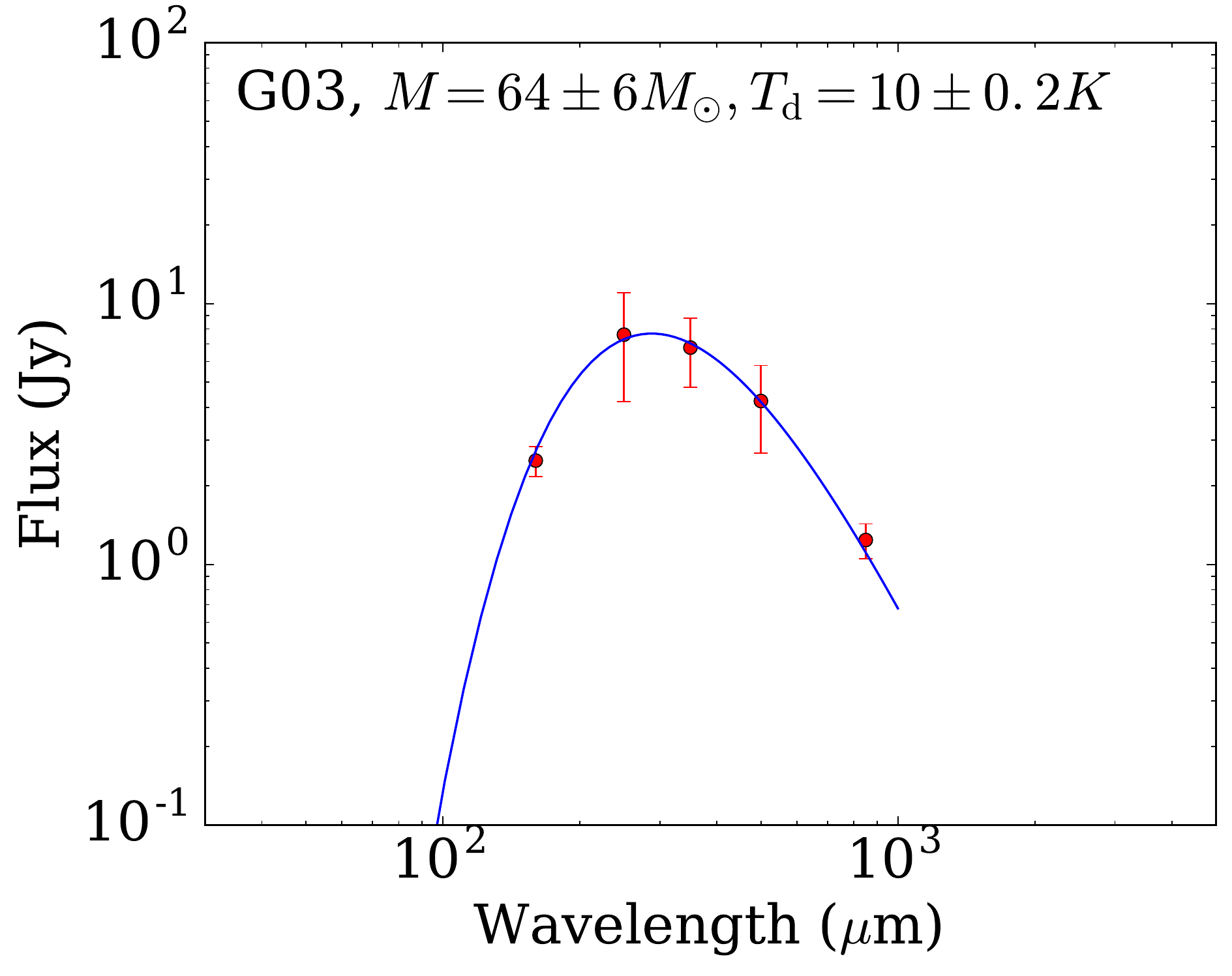}}
\qquad
\subfloat[G04]{\includegraphics[width=5.5cm]{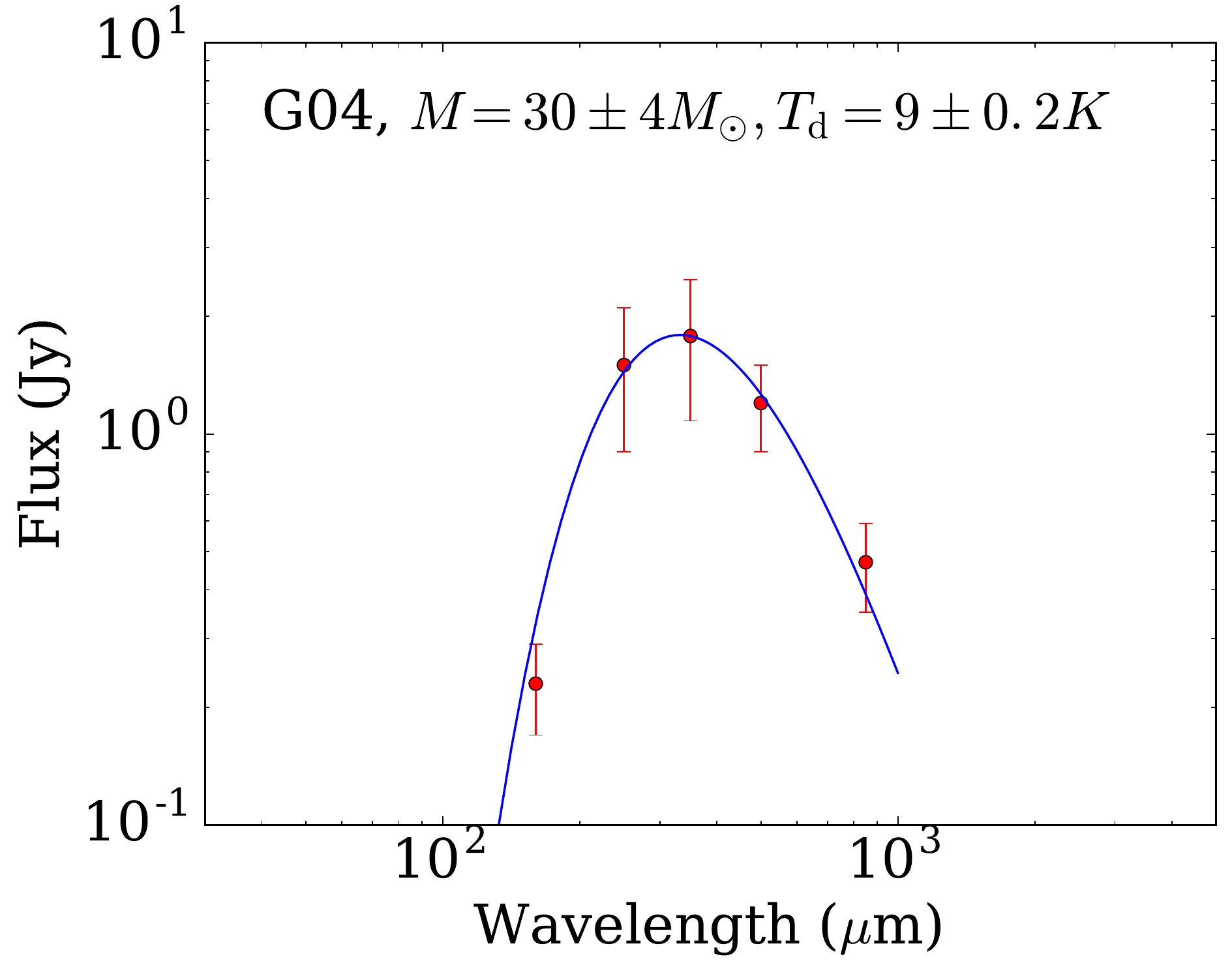}}
\subfloat[G05]{\includegraphics[width=5.5cm]{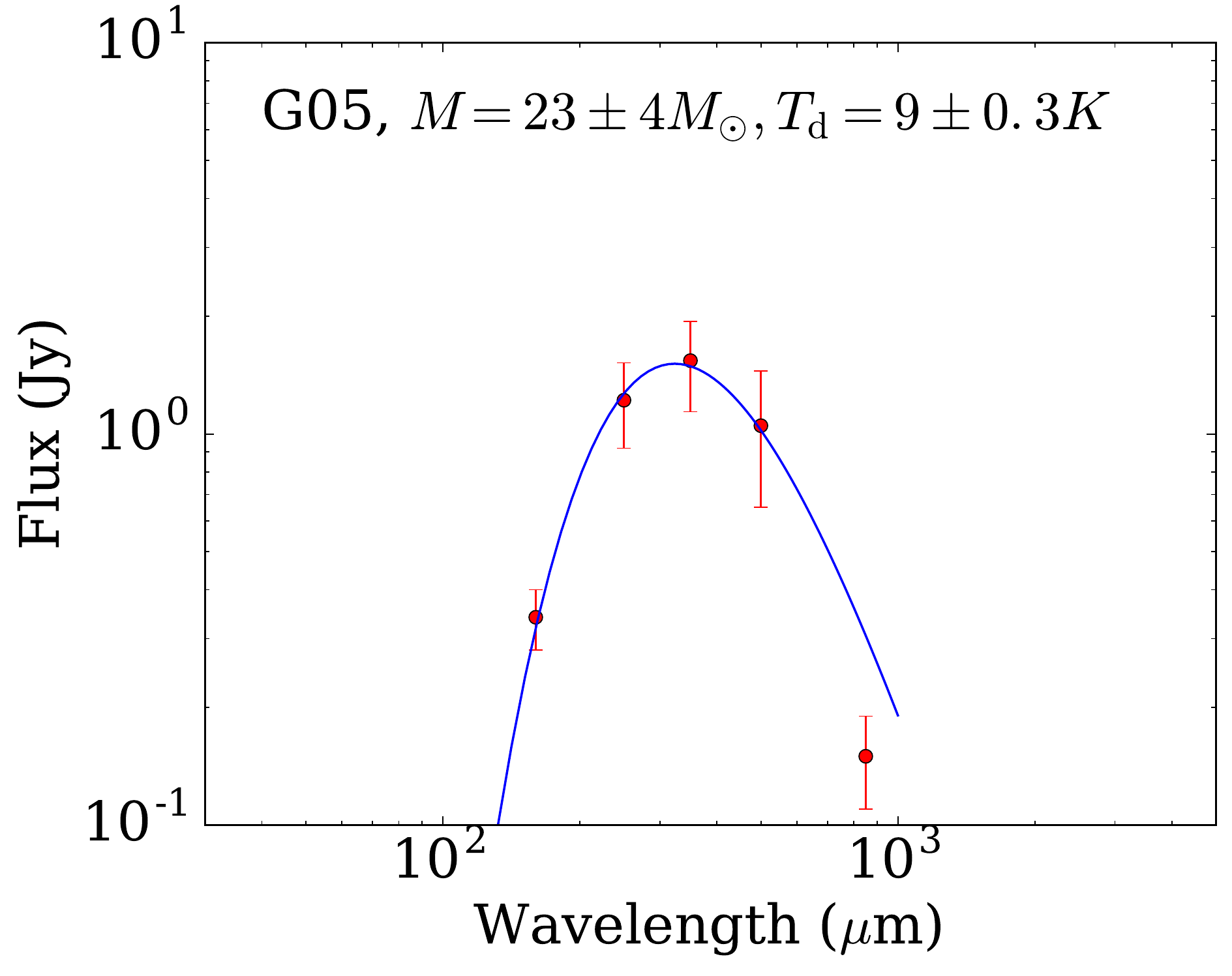}}
\subfloat[G06]{\includegraphics[width=5.5cm]{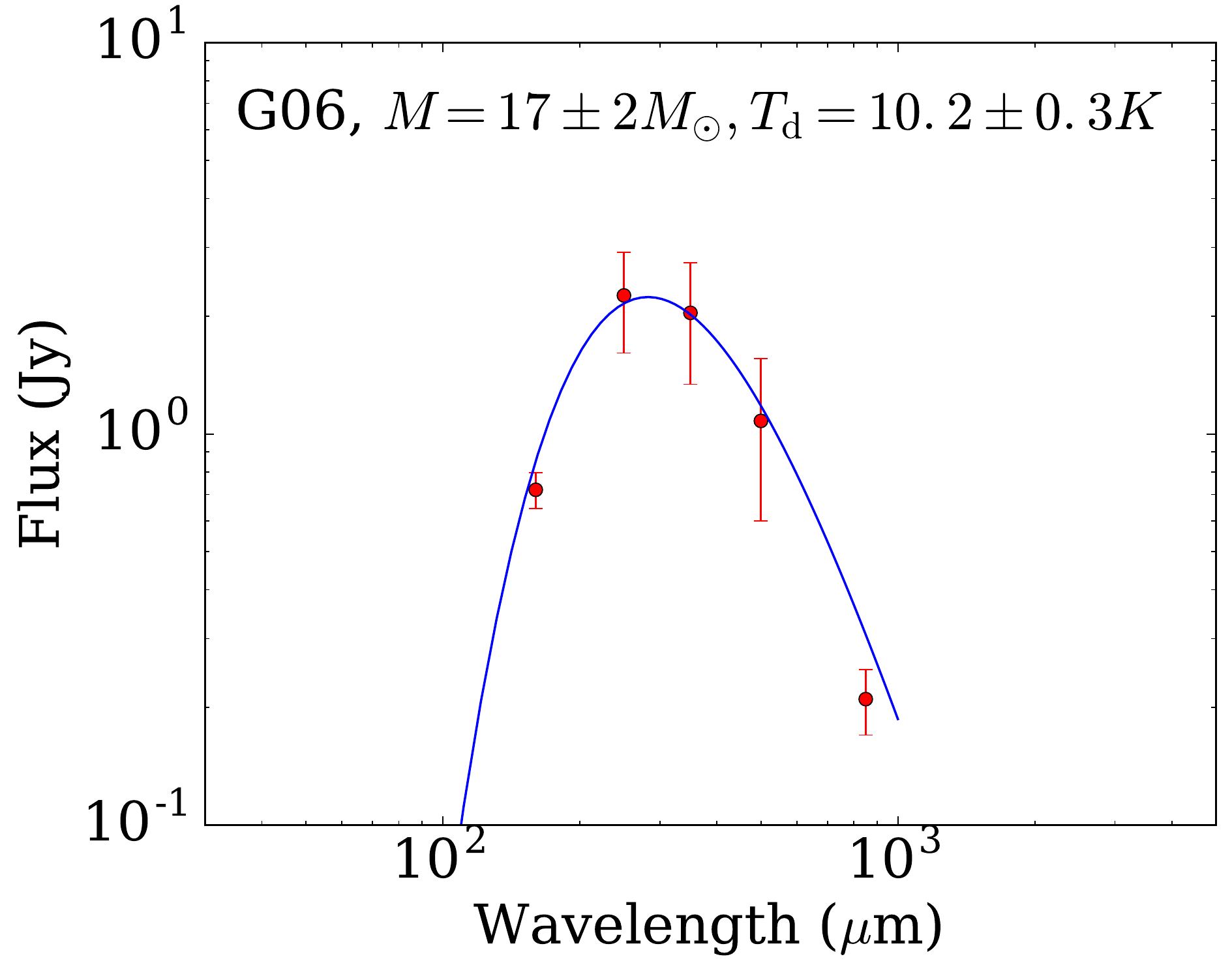}}
\qquad
\subfloat[G07]{\includegraphics[width=5.5cm]{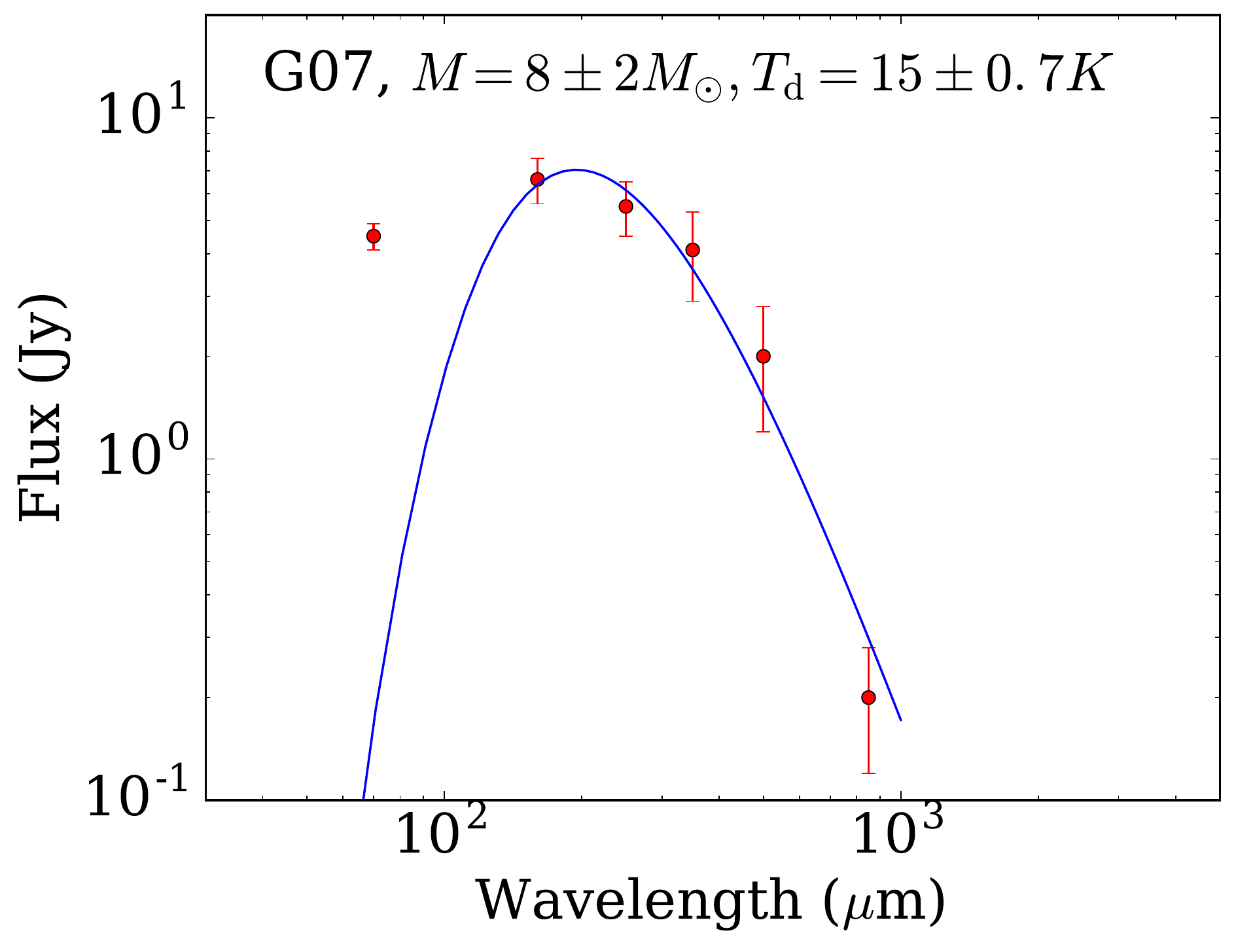}}
\subfloat[G08]{\includegraphics[width=5.5cm]{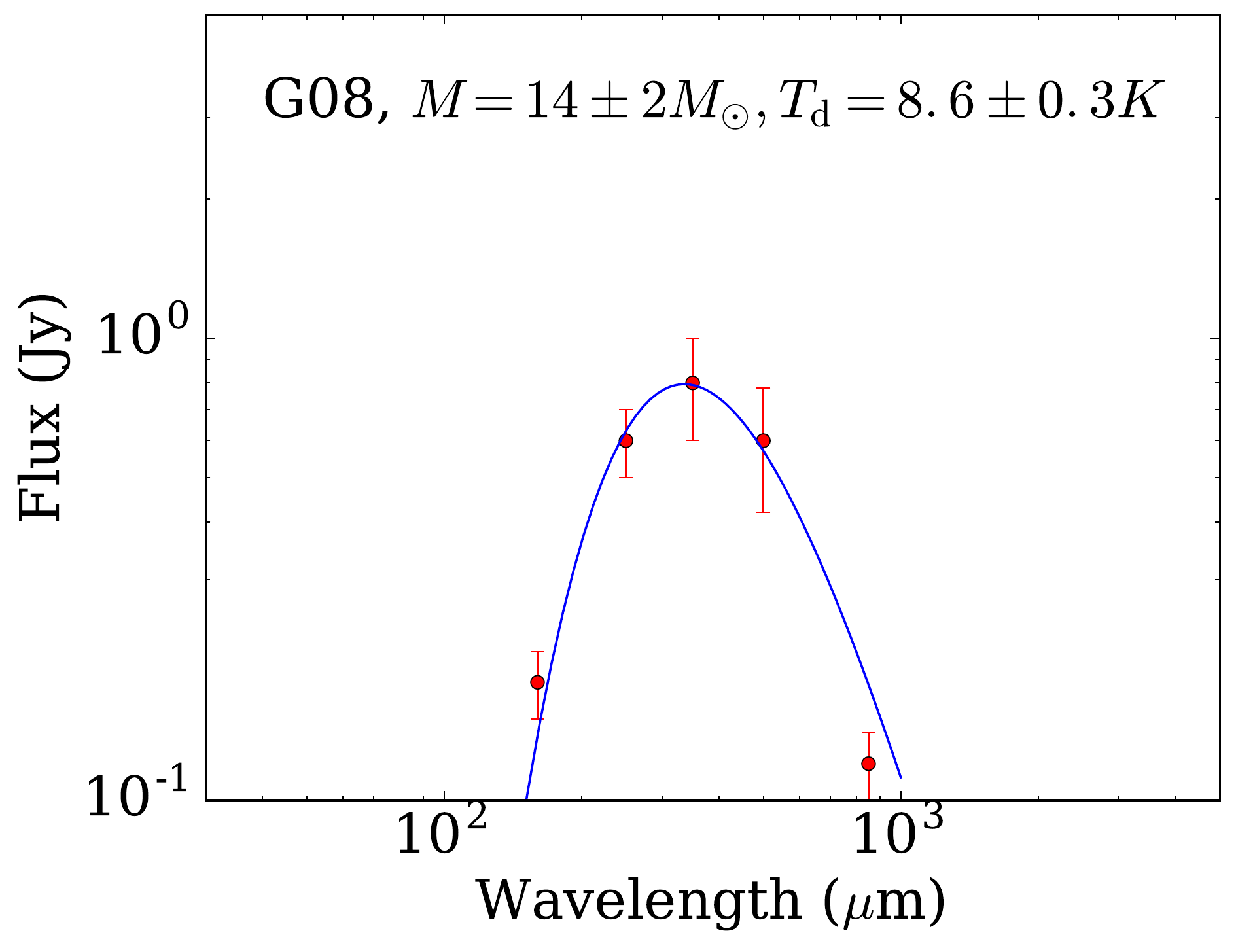}}
\subfloat[G09]{\includegraphics[width=5.5cm]{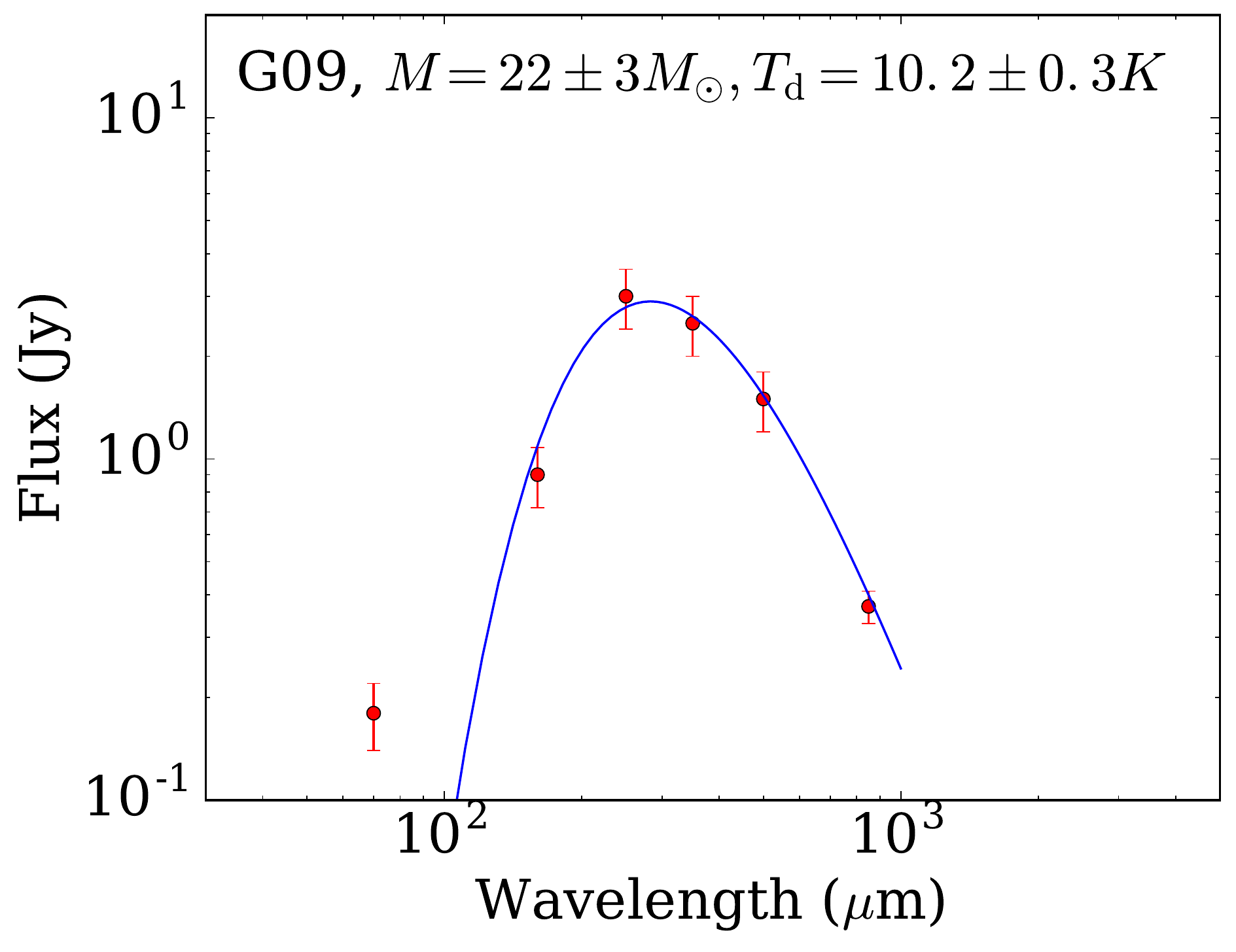}}
\caption{The SED fitting for compact sources in G181 with Herschel (160-500 $\mu$m) and SCUBA-2 850 $\mu$m photometric, based on the modified blackbody model.\label{fig:f3}}
\end{figure*}

\begin{table*}
\centering
\caption{Parameters of the compact sources}
\label{tab:table2}
\begin{tabular}{lccccccccr}
\hline
        Source Name & RA &  DEC & R$_{eff}$ & 
T$_{d}$ & N$_{H_{2}}$ & n$_{H_{2}}$ & M$_{c}$ & Luminosity &  \\
  & (hh:mm:ss) & (dd:mm:ss) & (pc) & (K) & ($10^{22}$ cm$^{-2}$) & ($10^{5}$ cm$^{-3}$) & (M$_{\sun}$)  & (L$_{\sun}$) & \\
         \hline
         G01 & 5:51:30.960 & 27:28:59.92 & 0.12 & 16(0.4) & 1.6(0.2) & 4.3 & 16(2) & 31.6\\
         G02 & 5:51:30.542 & 27:28:19.81 & 0.1 & 13(0.4) & 1.9(0.3) & 6.1 & 14.5(2) & 9\\
         G03 & 5:51:13.826 & 27:29:57.55 & 0.2 & 10(0.2) & 2.3(0.3) & 3.7 & 64(6) & 9\\
         G04 & 5:51:24.091 & 27:24:43.96 & 0.1 & 9(0.2) & 4.6(0.15) & 15.7 & 30(4) & 1.8 \\
        G05 & 5:51:24.233 & 27:27:17.80 & 0.09 & 9(0.3) & 4.1(0.6) & 15.0  & 23(3) & 1.6\\
        G06 & 5:51:23.748 & 27:29:09.53 & 0.1 & 10.2(0.3) & 2.5(0.5) & 8.0  & 17(2) & 2.7 \\
        G07 & 5:51:08.30 & 27:30:10.25 & 0.07& 15(0.7) &2.2(0.3) &9.7 &8(2) & 12  \\
        G08 & 5:51:22.76 & +27:25:48.02 & 0.09 & 8.6(0.3) &2.6(0.4) &9.5 &14(2) & 0.8  \\
        G09 & 5:51:23.83 & +27:22:57.06 & 0.1 & 10.2(0.3)& 2.4(0.5) &6.7 & 22(3)& 3.4 \\
        \hline
    \end{tabular}
\end{table*}

\begin{table*}
\centering
\caption{The parameters of compact sources from spectral lines 
\label{tab:table3}}
\begin{tabular}{lcccccccccr}
\hline
  Source Name & V$_{lsr}$ & $\tau$ & Excitation Temperature &  $\sigma_{N_{2}H^{+}}$ & $\sigma_{HCO^{+}}$  & Mass$_{vir}$ & $\alpha$ &    \\
  & (km s$^{-1}$) & & (K) & (km s$^{-1}$)  & (km s$^{-1}$) & (M$_{sun}$)  &  & \\
\hline

         G01 & 2.4(0.01) & 2.12(0.61) & 7.9(1.1) & 0.408(0.016) & 0.866(0.021) & 14(1.2) & 0.9(0.1)  \\
         G02  & 2.28(0.022) & 1(1) & 6.7(2.7) & 0.402(0.027) & 1.18(0.03) & 11.3(1.6)& 0.8(0.1)\\
         G03 & 2.9(0.008) & 3.83(0.96) & 4.33(0.26) & 0.236(0.008) & 0.482(0.011) & 7.8(0.53) & 0.12(0.01) \\
         G04 & 1.14(0.01) & 4.7(1.1) & 5.03(0.32) & 0.243(0.01) & 0.432(0.017) & 4.1(0.34) & 0.14(0.02)  \\
         G05 & 1.7(0.05) & 2.3(1.7) & 3.63(0.45) & 0.52(0.066) & 0.608(0.031) & 17.0(4.3) &  0.4(0.03)  \\
         G06 & 1.89(0.01) & 2.3(1.2) & 6.0(1.3) & 0.255(0.012) & 0.44(0.01) & 3.6(0.4) & 0.2(0.03)  \\
         G07 & 2.68(0.031) & 3.6(1.4)& 3.9(0.25) & 0.453(0.04)& 0.75(0.02)& 10(2) & 0.8(0.2) \\
\hline
\end{tabular}
\end{table*}

\subsubsection*{3.2.2 The properties of compact sources}
For each core, we calculated the flux in a convolved size at Herschel (160-500 $\mu$m) and SCUBA-2 (850 $\mu$m) bands. The uncertainties of the flux values are estimated from the background noise of different wavelength images. We derived the dust temperature (T$_{d}$), column density (N$_{H_{2}}$), volume density (n$_{H_{2}}$), core masses (M$_{c}$) based on the modified blackbody model (Equation 2, 3, 4). The effective radius is defined as R$_{eff}$=$\sqrt{ab}$, where a and b are the core semi-major/minor axes. The volume number density (n$_{H_{2}}$) and column density (N$_{H_{2}}$) are derived as $n \times ( \frac {4}{3} \pi R_{eff}^{3} \mu m_{H})=N \times ( \pi R_{eff}^{2} \mu m_{H})= M_{c}$, where $\mu$=2.8 is the mean molecular weight per H and m$_{H}$ is the atomic hydrogen mass. The fitted results are presented in Figure \ref{fig:f3}. For sources with 70 \micron~fluxes obviously deviate from the SED fitting, we exclude these data points in our fitting.

As shown in the lower right panel of Figure \ref{fig:f1}, we composed a three-color image with the 850 $\micron$, 250 $\micron$, 70 $\micron$ band images, respectively displayed as red, green, blue colors. Because the 70 $\micron$ emission is dominated by the ultraviolet (UV) heated warm dust, it can trace very well the internal luminosity of a protostar \citep{Dunham2008}. From the three colour image of Figure \ref{fig:f1} and the SED fitting plots of Figure \ref{fig:f3}, we can see that only G01, G02, G07 and G09 have 70 $\micron$ emission, a signpost of protostellar activity. There is no (or faint) 70 $\micron$ emission in the dense cores G03, G04, G05, G06 and G08, thus these cores could be prestellar candidates. Recent works have shown that infall signatures have been found in some of the massive 70 \micron~quiet clumps and prestellar cores, which provide evidence for embedded low/intermediate-mass star formation activity \citep{Traficante2017, Liub2018, Contreras2018}

\subsubsection*{3.2.3 Source characteristics from spectral-line data}\label{n2hp}
We used the average intensity of the N$_{2}$H$^{+}$ and HCO$^{+}$(1-0) lines in each core, and further fitted the emission using PySpeckit \citep{Ginsburg2011}. The dense core G09 was not covered by our line observations, and G08 was not fitted due to the low S/N ratio of the N$_{2}$H$^{+}$ emission. The fitted spectra are shown in Figure \ref{fig:fA1} and \ref{fig:fA2}. The line parameters for the N$_{2}$H$^{+}$(1-0) model are based on the analysis of \cite{Daniel2005} and \cite{Schoier2005}. The derived parameters include the optical depth, excitation temperature, centroid velocity and velocity dispersion. A Guassian model was used to fit the HCO$^{+}$(1-0) line to derive the peak intensity, centroid velocity and velocity dispersion. All these derived parameters are listed in Table \ref{tab:table3}. 

We found that the velocity dispersions derived from HCO$^{+}$(1-0) line are about 2 times larger than the values from N$_{2}$H$^{+}$(1-0). That may be due to the fact that the N$_{2}$H$^{+}$ emission, a well-known dense gas tracer for the cold dense cores, is closely associated with the densest parts of cores \citep{Caselli2002}. Besides, N$_{2}$H$^{+}$ can well trace the central region with CO depletion \citep{Bergin2002, Bergin2001}, and less affected by star-formation activities, such as molecular outflows \citep{Womack1993}. While HCO$^{+}$ emission, which is produced by the protonation of CO, is not abundant at the core center and may be more optically thick \citep{di2007}. Therefore, we used the velocity dispersion derived from the N$_{2}$H$^{+}$(1-0) line to compute the virial mass of each core. The virial mass is computed with the following Equation mentioned in \cite{MacLaren1988} and \cite{Williams1994}.
$$M_{vir} =3k\frac{r \times {\sigma _{\upsilon}}^{2}}{G}$$
where G is the gravitational constant, r is the radius for each core, and $\sigma _{\upsilon}$ represents the line-of-sight velocity dispersion. k=(5-2a)/(3-a) is a correction factor corresponding to the density profile $\rho$ $\propto$ r$^{-a}$. Assuming that the dense cores are gravitationally bound isothermal spheres, a density profile of $\rho$ $\sim$ r$^{-2}$ (a=2). The above equation could be rewritten as:
$$M_{vir}=126 \left(\frac{r}{pc}\right) \left(\frac{\Delta V_{FWHM}}{km~s^{-1}}\right)^{2} M_{\odot}$$
Here $\Delta V_{FWHM}$ is the FWHM of the N$_{2}$H$^{+}$ line along the line of sight. The resultant virial mass for each core was listed in Table \ref{tab:table3}. In addition, we derived the virial parameters ($\alpha$, M$_{vir}$/M$_{c}$) for each core. Most of our dense cores have low virial parameters ($\alpha$ $\ll$ 2), which are consistent with those observed in high-mass star forming regions \citep{Kauffmann2013}, suggesting rapid and violent collapse or magnetic field support. However, it should be noted that there could be a bias in the measurement of the kinetic energy during the initial phase of a collapsing cloud, when the core still has a lot of diffuse and relatively low-dense material \citep{Traficante2018}. The observed effective viral parameter $\alpha_{eff}$ is less than  $\alpha_{vir}$ and we should be cautious in using the parameter $\alpha$ to represent the viral state of an observed core. On the other hand,  the dense core G01, G02 and G07 with 70 \micron~detection have $\alpha$ $\sim$ 1, which may be due to an enhancement of the velocity dispersion in more evolved cores, or due to stellar feedback. Our results suggest that a gas core is strongly unstable during the initial phase of collapse, and when a protostellar object is clearly visible, the core tends to stabilize. 

\begin{figure}
	\includegraphics[width=\columnwidth]{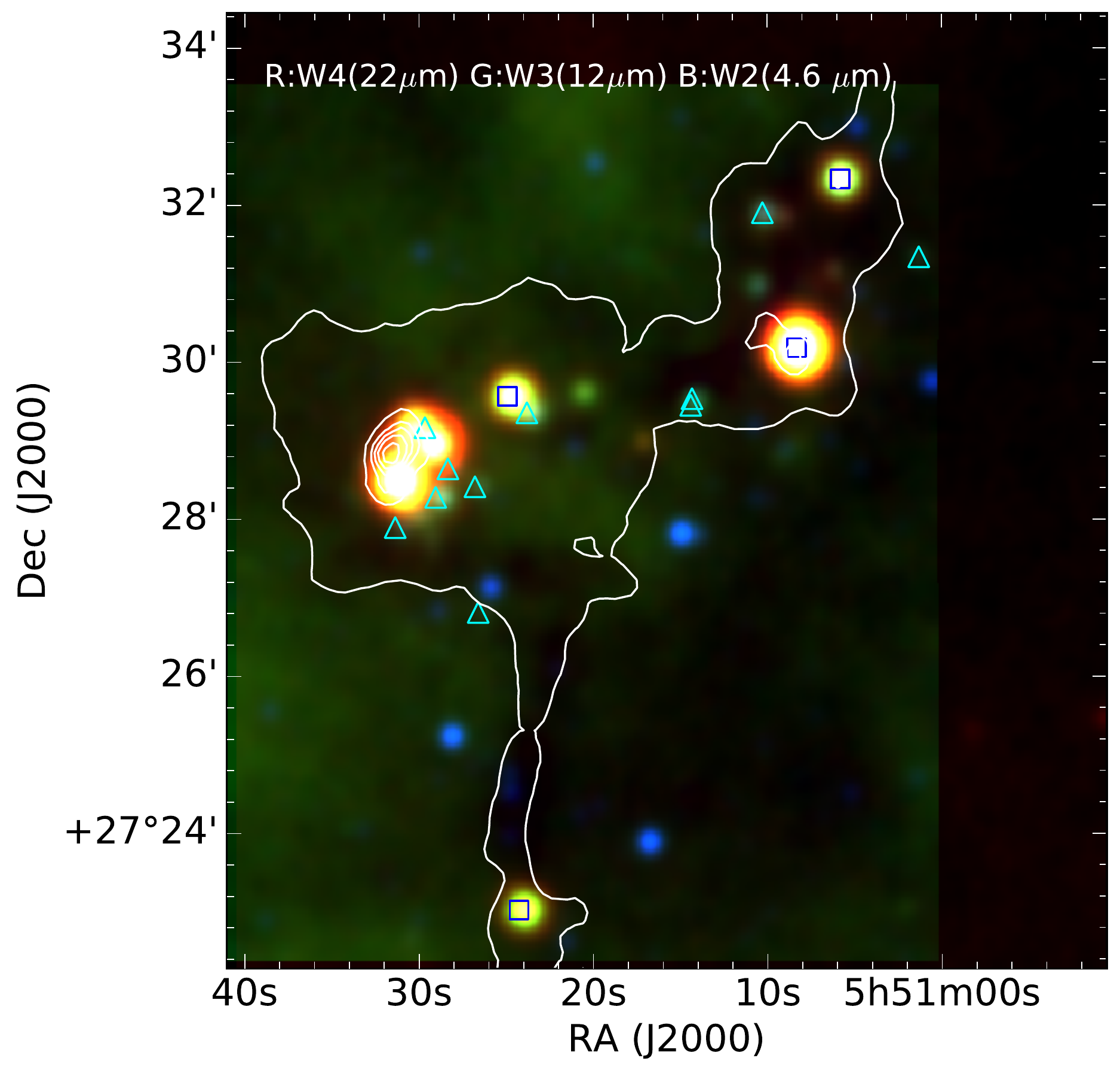}
    \caption{The three-colors map composed through the W4 (22 $\mu m$), W3 (12 $\mu m$), W2 (4.6 $\mu m$), respectively shown as red, green, blue color. The white contours display emission from 250 $\micron$ map, ranging from 10\% to 90\% stepped by 15\% of the peak value (1080 MJy $sr^{-1}$). The blue squares are for Class I YSOs and cyan triangles for the Class II YSOs.\label{fig:f4}}
\end{figure}

\begin{figure}
	\includegraphics[width=\columnwidth]{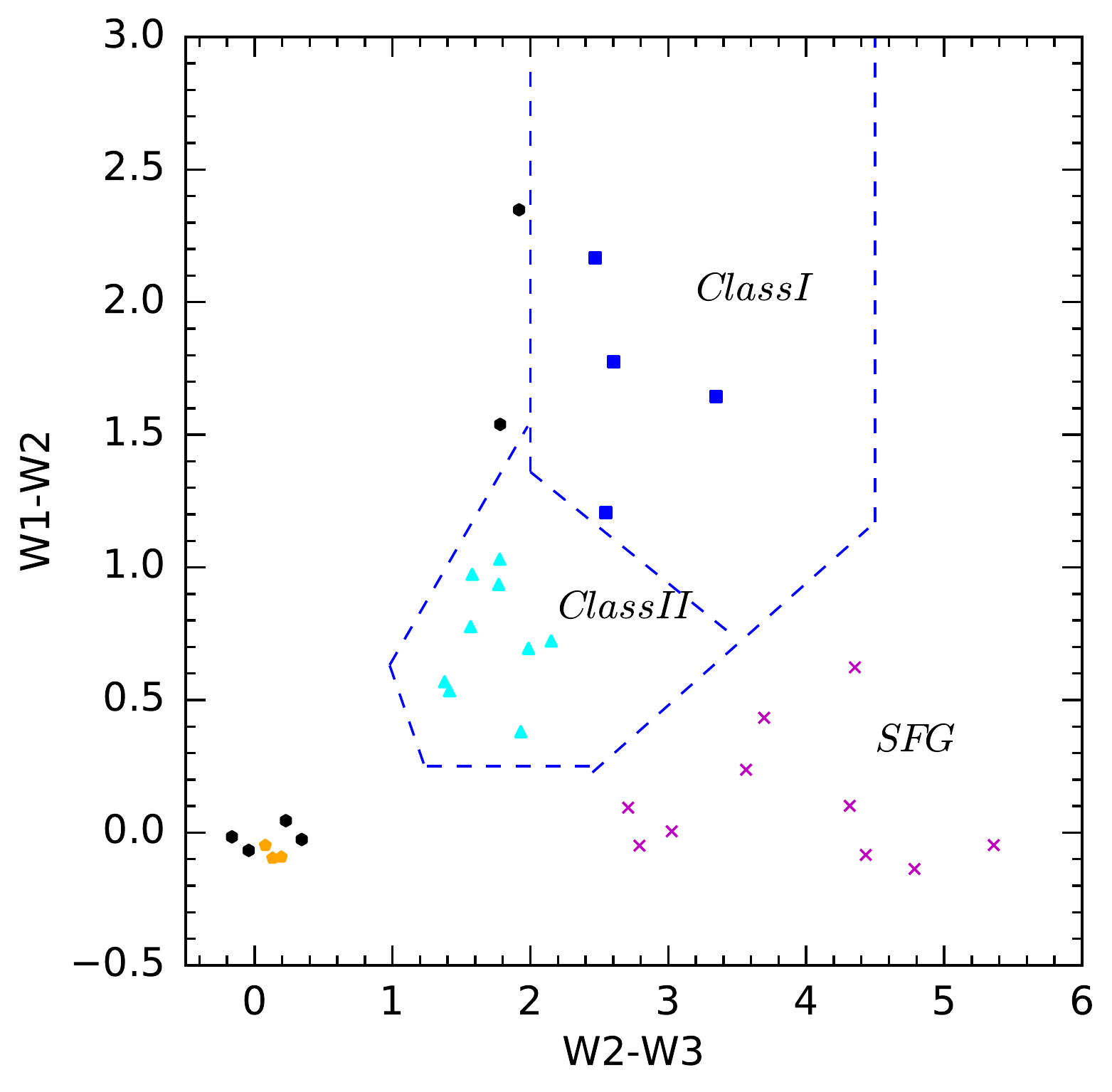}
    \caption{WISE color-color diagrams. YSOs located in regions enclosed by blue dashed lines are classified as Class I (blue squares) and Class II (cyan triangles). The black circle, orange hexagons and magenta crosses represent Field stars, AGBs and SFGs, respectively.\label{fig:f6}}
\end{figure}

\begin{figure}
\includegraphics[width=\columnwidth]{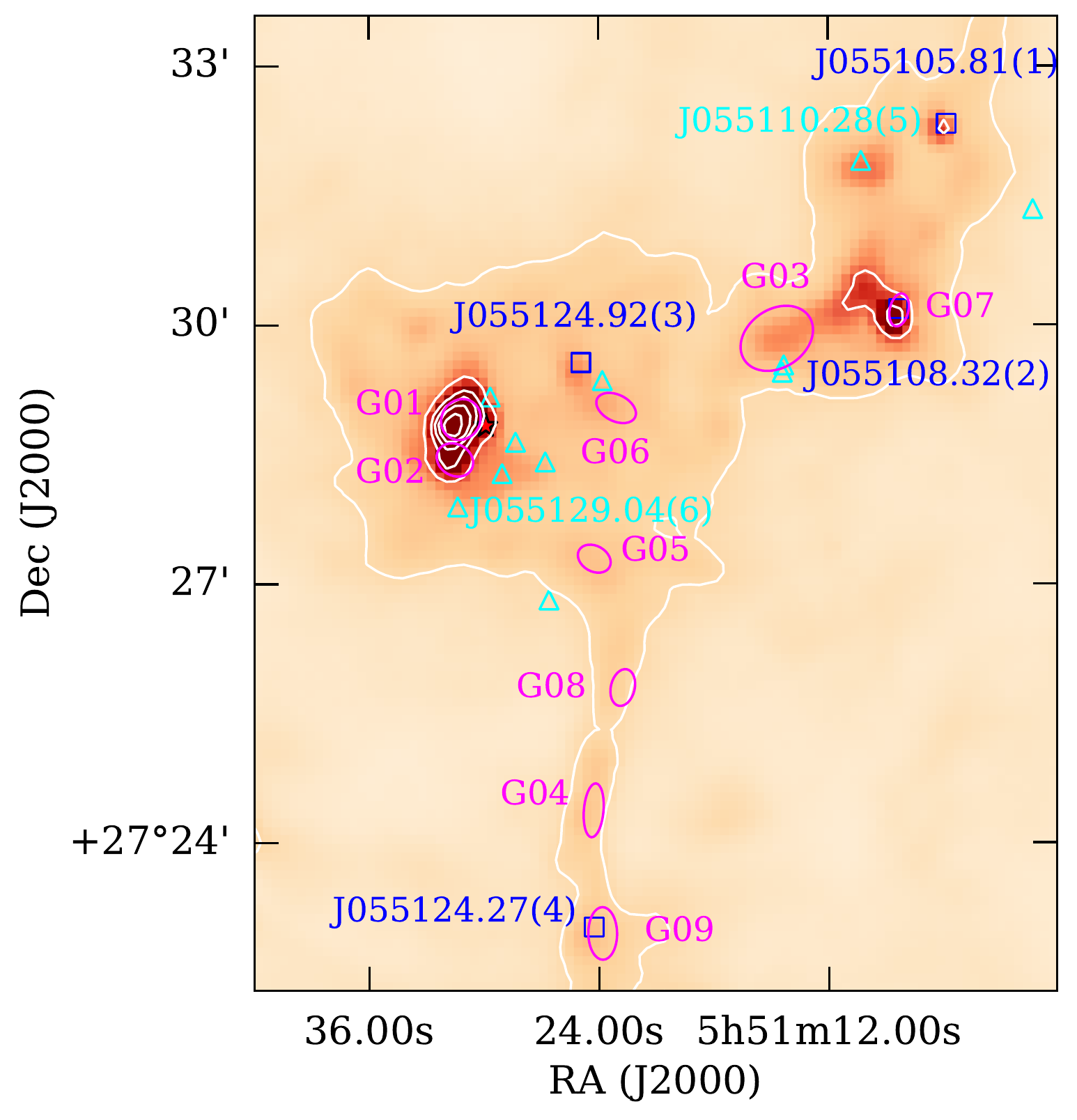}
\caption{The background is the 250 $\micron$ continuum emission map. White contours display intensities of 250 $\micron$ continuum, ranging from 10\% to 90\% stepped by the 15\% of the peak value (1080 MJy sr$^{-1}$). Blue squares, cyan triangles and magenta ellipses respectively stand for Class I YSOs, Class II YSOs and SCUBA-2 compact sources. Three red triangles represent the Class II YSOs only identified using the method in Dewangan et al. (2017). The six labeled YSOs are fitted by the R17 model (see section 3.4 for details).\label{fig:f5}}
\end{figure}

\subsection*{3.3 YSOs identification based on WISE near-IR and mid-IR data}
The near-IR and mid-IR emission in the molecular cloud is originated from warm dust heated by embedded protostars. Therefore, we could use the infrared emission to characterize star formation activities. The high-sensitivity mid-infrared images (Figure \ref{fig:f4}) from the WISE \citep{Wright2010}, including 3.4, 4.6, 12 and 22 $\mu$m wavelengths, allow us to trace the IR emission from YSOs. Note that, the WISE emission is also contributed by other types of infrared sources besides YSOs, such as foreground stars and extragalactic objects, which can be seen in projection towards the observed clouds.

To investigate the nature of the identified WISE sources, we employed the WISE color criteria listed in \cite{Koenig2014}. By using the uncertainty/signal-to-noise/chi-square criteria in \cite{Koenig2014}, we selected the reliable point-source photometry in the ALLWISE catalog within the G181 structures. The resultant catalog, including WISE and 2MASS near- and mid-infrared colors and magnitudes, was used to identify YSOs. We also removed objects matching the criteria as likely star-forming galaxies (SFG), board-line AGNs, AGB stars and CBe stars. 

As shown in Figure \ref{fig:f6}. Class I YSOs (protostellar candidates) are the reddest objects, and are classified as such if their colors match the following critieria \citep{Koenig2014}:

W2 - W3 $>$ 2.0 

and

W1 - W2 $>$ -0.42$\times$(W2-W3) + 2.2 

and

W1 - W2 $>$ 0.46$\times$(W2-W3) - 0.9 

and

W2 - W3 $<$ 4.5.

Class II YSOs (candidates of T Tauri stars and Herbig AeBe stars) are classified from the remaining pool of objects, if their colors match all the following criteria \citep{Koenig2014}:

W1 - W2 $>$ 0.25

and

W1 - W2 $<$ 0.9 $\times$ (W2 - W3) - 0.25

and

W1 - W2 $>$ -1.5 $\times$(W2 - W3) + 2.1

and

W1 - W2 $>$ 0.46 $\times$(W2 - W3) - 0.9

and

W2 - W3 $<$ 4.5

In addition, two more Class II YSOs candidates with no photometric data in WISE band 3, are identified using the photometric data in WISE bands 1, 2 and 2MASS data in the H and Ks bands, based on the additional criteria listed in \cite{Koenig2014}. In this way, we have identified 15 (4 Class I and 11 Class II ) YSOs in PGCC G181. As shown in Figure \ref{fig:f5}, all the YSOs are distributed along the filamentary structures. Five Class II YSOs are located in the Fa sub-structure, and 4 Class I YSOs lie along the sub-structures Fb and Fc. 

The previous works of \cite{Dewangan2017} have identified
13 Class II and 2 Class I YSOs in the G181 region using the color-color criteria based on the 2MASS and GLIMPSE360 point-source data \citep{Whitney2011}. For comparision, we plot their classification diagram and the distribution of their YSOs in Figure \ref{fig:fa2}. We find that our method based on the WISE colors can classified most of these YSOs consistently, except for 3 Class II YSOs which may be due to the lower resolution and sensitivity of the WISE data comparing with the GLIMPSE360 photometric data. However, three new YSOs (J055105.81(Class I), J055108.32(Class I) and J055123.81) are found with our WISE color criteria. This could be due to the fact that these YSOs, especially the Class I objects, are deeply embedded in dusty envelops, and hence our classification criteria using
the WISE data (with longer wavelength) have a better chance to discover them. If the three extra Class II YSOs identified by \cite{Dewangan2017} are included, we obtain a sample of 18 YSOs in G181. As presented in Figure \ref{fig:f5}, we find that Class I YSOs J055108.32 and J055124.27 are associated with the dense cores G07 and G09, which further confirms that these two cores are in the evolution stage of forming protostars.

\subsection*{3.4 The parameters for YSOs}
\subsubsection*{3.4.1 Models}

\cite{Robitaille2017}, hereafter R17, presented an improved set of SED models for YSOs. In the R17 models, the coverage of parameter space is more uniform and wider than the model of \cite{Robitaille2006}, and highly model-dependent parameters have been excluded. Moreover, the improved models are more suited to modeling far-infrared and sub-mm flux intensities of the forming stars. The models are split into 18 different sets with increasing complexity. The model sets span a wide range of evolutionary stages, from the youngest deeply embedded protostars to pre-main-sequence stars with little or no disks.

\begin{figure*}
\centering
\subfloat[J055129+272816]{\includegraphics[width=5.5cm]{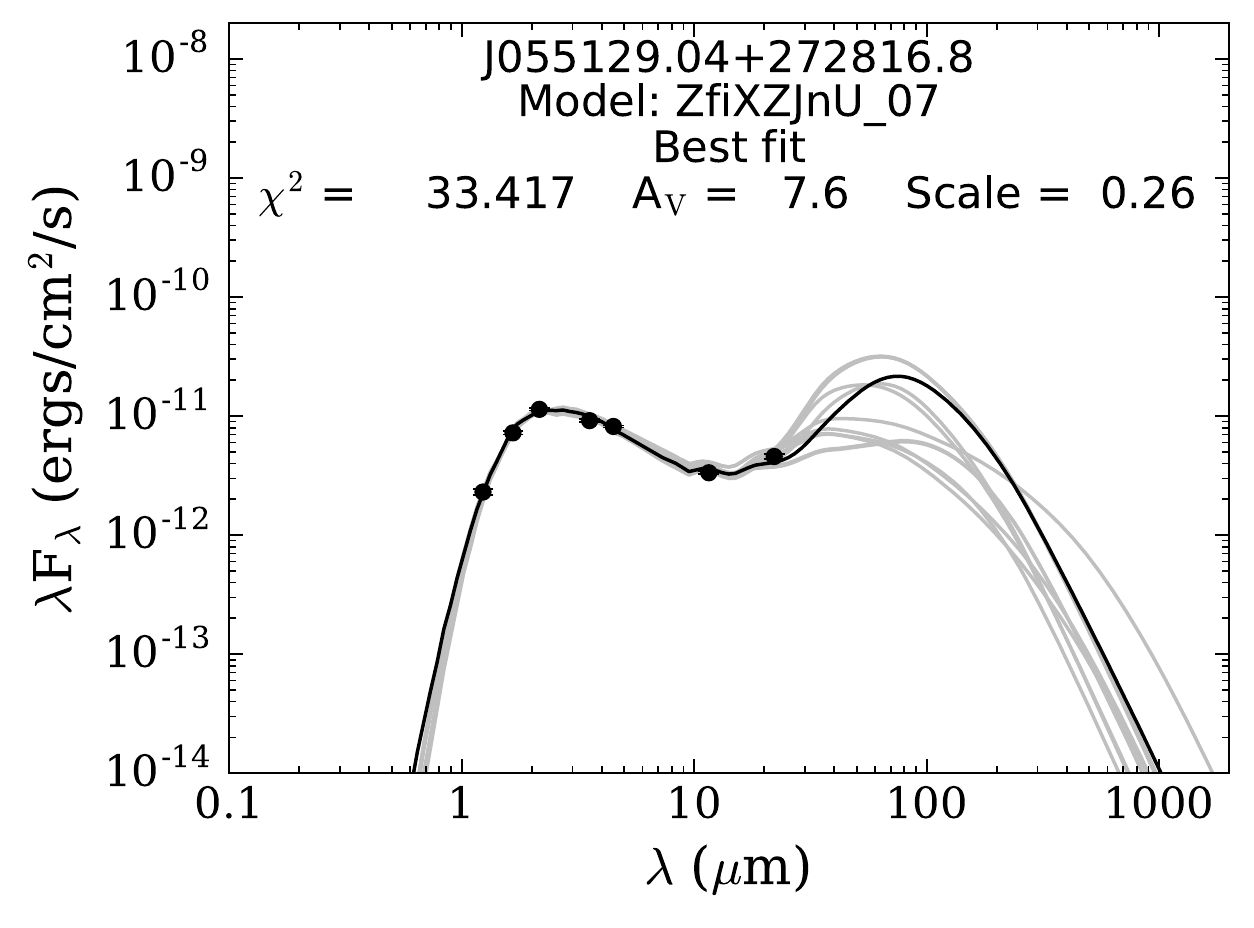}}
\subfloat[J055110+273154]{\includegraphics[width=5.5cm]{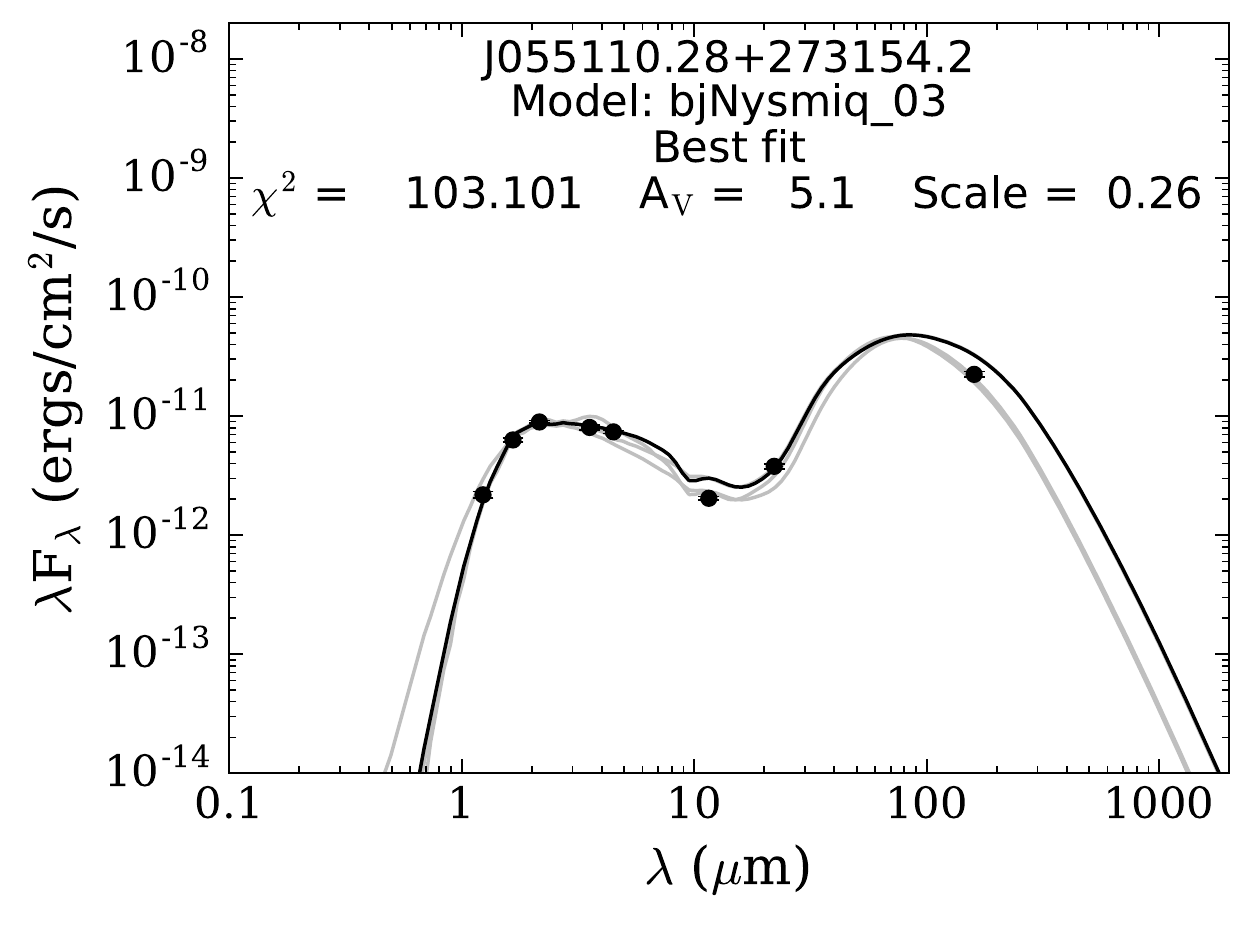}}
\subfloat[J055108+273011]{\includegraphics[width=5.5cm]{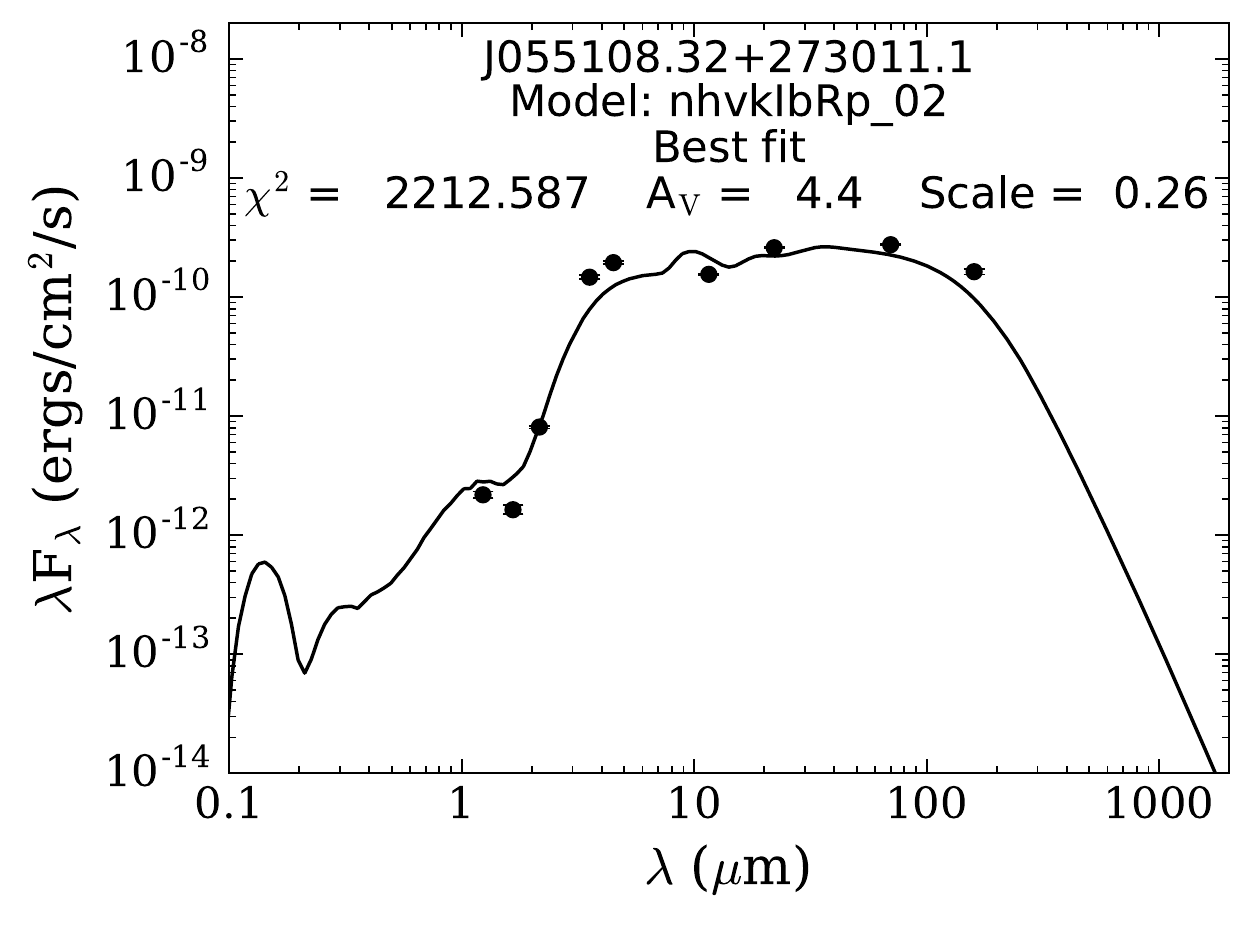}}
\qquad
\subfloat[J055105+273219]{\includegraphics[width=5.5cm]{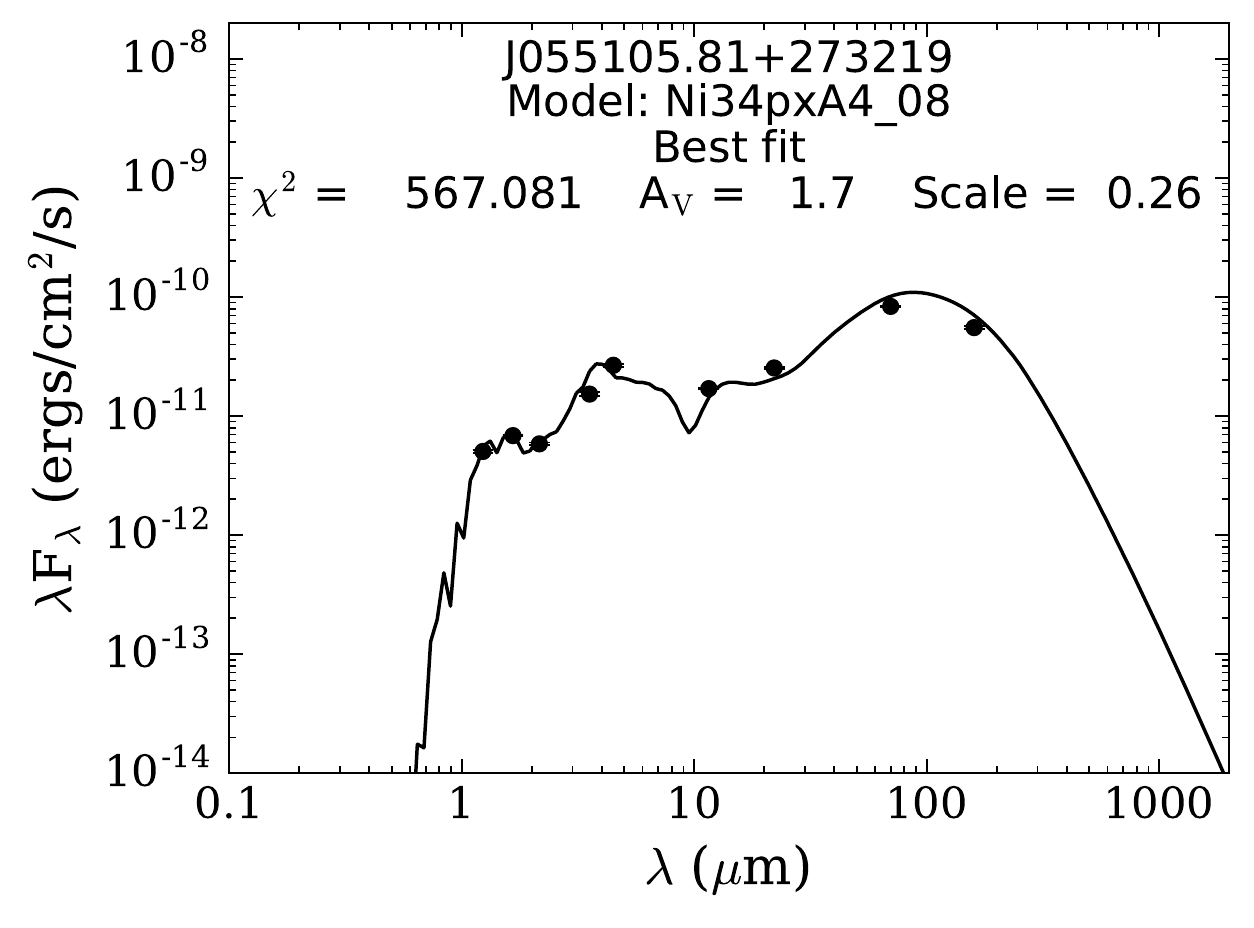}}
\subfloat[J055124+272301]{\includegraphics[width=5.5cm]{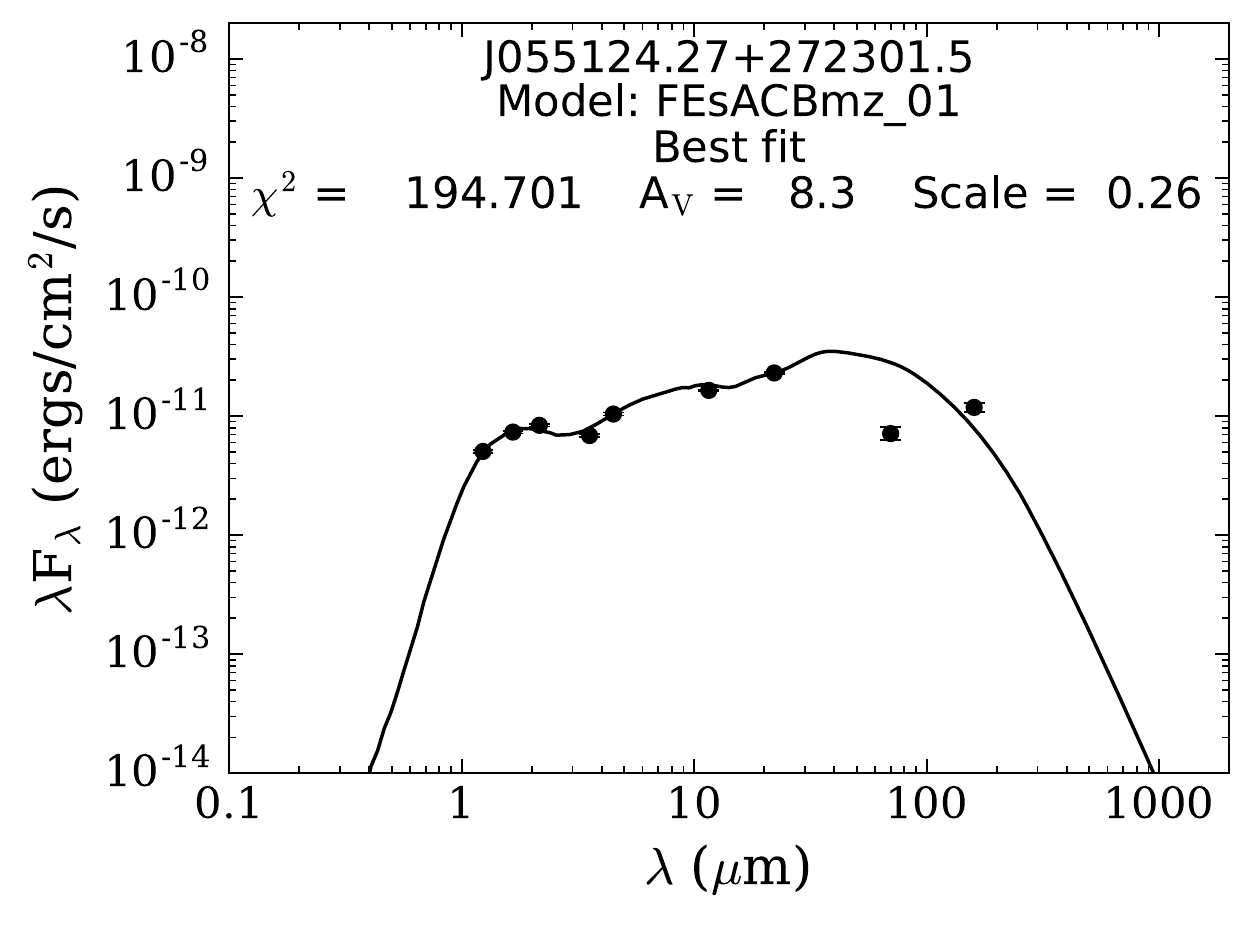}}
\subfloat[J055124+272933]{\includegraphics[width=5.5cm]{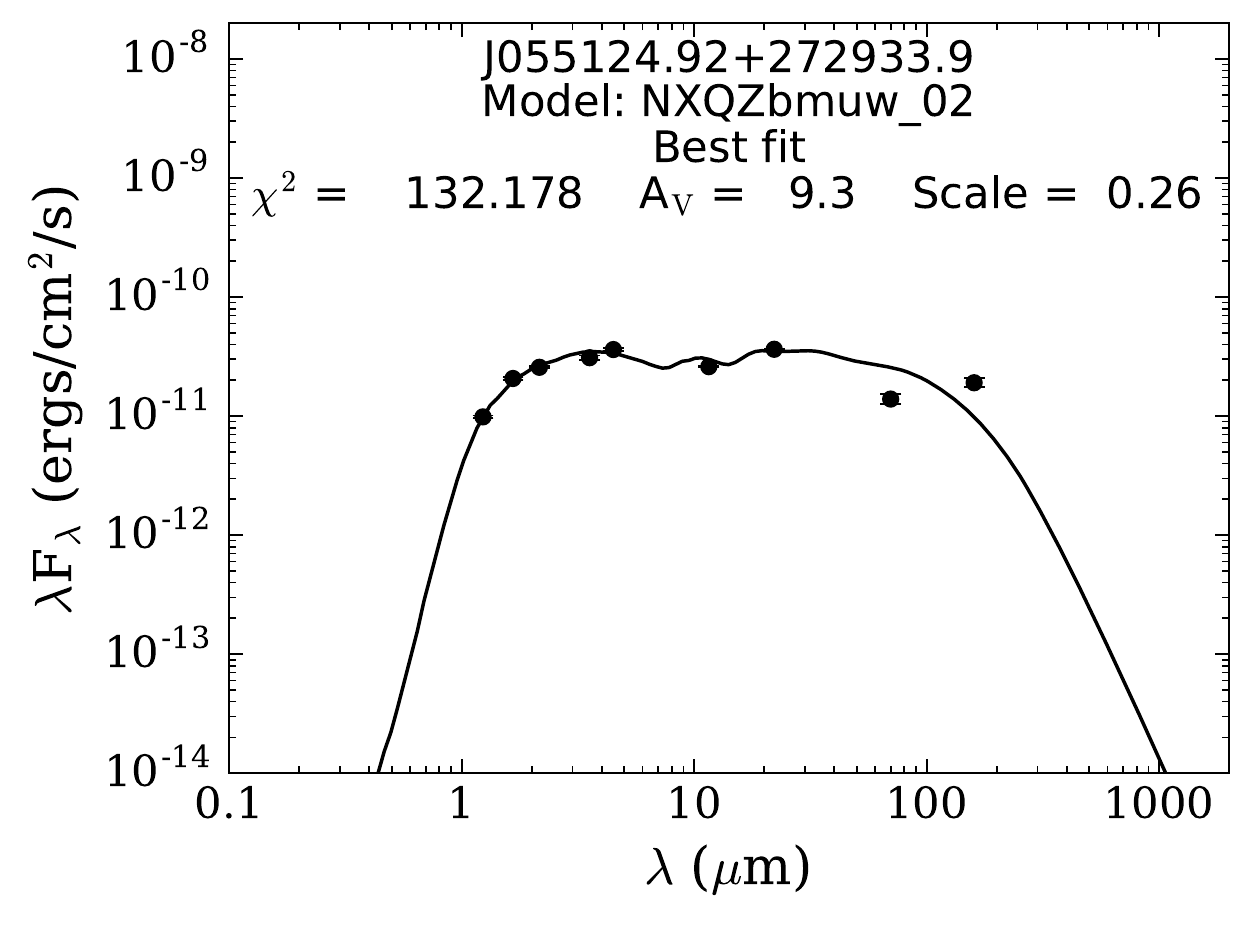}}
\caption{SED fits for 6 YSOs in G181.84 with available photometric data up to 160 $\mu$m, performed by the SED fitting tool using different YSO SED models in R17. The best fit is shown using solid black line and grey lines show all other fits that satisfy the criteria:$\chi^{2}$ - $\chi^{2}_{best}$ $<$ 3$n_{data}$. \label{fig:f7}}
\end{figure*}

\begin{figure}
	\includegraphics[width=\columnwidth]{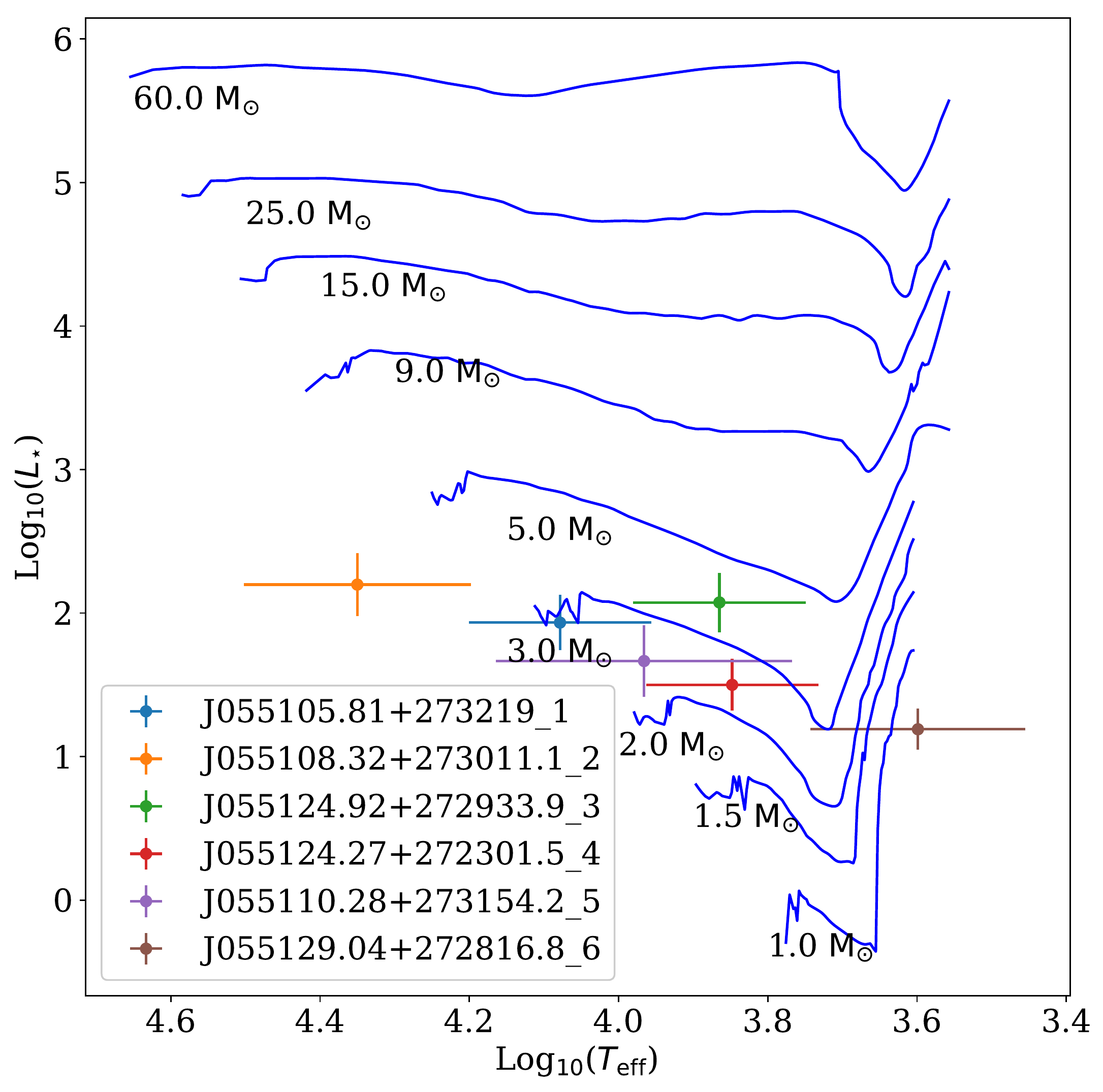}
    \caption{log10 (L$_{\star}$) vs. log10 (T$_{eff}$) diagram display the location of YSOs (cross symbols) and the pre-main sequence (PMS) stellar tracks from Bernasconi \& Maeder(1996) for different masses in blue. The errors in T$_{eff}$ and L$_{\star}$ are also shown. \label{fig:f8}}
\end{figure}

\subsubsection*{3.4.2 Input data and SED fits}
We constructed photometric SEDs for the identified YSOs using wavelengths ranging from 1.25 to 160 $\mu$m. Searching within 3\arcsec~of the YSO positions in G181, we retrieved the J, H and Ks band fluxes from the 2MASS All-Sky Point Source Catalog (PSC) \citep{Skrutskie2006}, the fluxes of IRAC bands 1 and 2 (3.6 $\mu$m and 4.5 $\mu$m) in the Spitzer GLIMPSE360 catalogue, and the mid-infrared bolometric fluxes from WISE (W3 and W4 bands) in the ALLWISE Source catalog and PACS Point Source Catalogue in Herschel. In order to yield reliable results, the SED model fitting requires to have data at $\lambda$ $>$ 12 $\micron$. For this reason, we carried out SED fitting only for sources that have data available in the Herschel PACS PSC. Such sub-sample contains six sources out of the 18 identified YSOs in G181. The selected sources have been noted by source names in Figure \ref{fig:f5}, which are distributed around the entire structures.

The distance was fixed to be 1.76$\pm$0.04 kpc, and Av range was from 0 to 40 mag. The extinction law was the same as that used in \cite{Forbrich2010}. The errors of the photometry data were adopted from the original source catalog. We select all SEDs fitting that satisfy $\chi^{2}$ - $\chi^{2}_{best}$ $<$ 3$n_{data}$ as the reasonable results, where $\chi^{2}_{best}$ is the $\chi^{2}$ of the best model for each model set, n$_{data}$ is the number of data points. For each source, the resultant model set is corresponding to the fitted SED having the lowest $\chi^{2}_{best}$ value. Figure \ref{fig:f7} shows the resultant model sets for the 6 fitted YSOs in G181.

\subsubsection*{3.4.3 Fit parameters}
The fitted parameters for YSOs using the best-fitted SED model sets are listed in Table \ref{tab:tableb1}. We estimated the mean values of A$_{V}$, T$_{eff}$, and stellar radius R$_{\star}$ from all the fits satisfying the $\chi^{2}$ cut in the chosen model sets. Then we used the Stefan-Boltzmann law, assuming a solar T$_{eff}$ of 5772 K, to estimate the stellar luminosity L$_{\star}$ using the model fitted values of T$_{eff}$ and R$_{\star}$. Based on both the stellar L$_{\star}$ and T$_{eff}$, we further estimated the stellar mass for each YSO using the pre-main sequence (PMS) tracks. The PMS tracks are for stars with metallicity of Z=0.02, its mass range is from 1.0 M$_{\odot}$ to 60 M$_{\odot}$ \citep{Bernasconi1996}. 

We roughly derived the mass for the YSOs by the least separation of each source away from the stellar track of the corresponding mass in the log$_{10}$(L$_{\star}$) - log$_{10}$(T$_{eff}$) space. 
Figure \ref{fig:f8} shows the stellar tracks and locations of 6 YSOs in the log$_{10}$(L$_{\star}$) vs. log$_{10}$(T$_{eff}$) diagram. The masses of the 6 YSOs are in the range of 1 -- 5 M$_{\odot}$. Combining with Figure \ref{fig:f5}, we find that the young YSOs (Class I) are mainly located in the sub-structures Fb and Fc, and most of the evolved YSOs (Class II) are around Fa sub-structure. Furthermore, J055105.81+273219.0(1), J055108.32+273011.1(2) and J055110.28+273154.2(5) in the Fb sub-structures, and J055124.27+272301.5(4) in the Fc sub-structure are located in evolutionary stages in the PMS tracks earlier than the YSOs J055124.92+272933.9(3) and J055129.04+272816.8(6) around the Fa sub-structure. This is a good evidence for the sequential star formation in the filamentary structure of G181. 

\subsection*{3.5 Dynamical structures}
\subsubsection*{3.5.1 The integrated intensity of HCO$^{+}$ and N$_{2}$H$^{+}$ spectral lines}
The low-$J$ HCO$^{+}$ lines are sensitive to the gas with densities of the order of 10$^{5}$-10$^{6}$ cm$^{-3}$, which could be excellent tracers of the velocity field of molecular clouds \citep{Qi2003}. In addition, N$_{2}$H$^{+}$(1-0) line, having a critical density of n$_{crit}$= 1.4$\times$ 10$^{5}$ cm$^{-3}$ at 10 K, can trace the cold dense cores well \citep{Caselli2002}. However, taking the radiative trapping into account, the effective critical density of N$_{2}$H$^{+}$(1-0) is pushed down to n$_{crit}$ $\sim$ 1.0$\times$ 10$^{4}$ cm$^{-3}$ at 10 K \citep{Shirley2015}. The FWHM beam size of the Nobeyama 45-m telescope is about 18$\arcsec$ at 90 GHz. Such resolution is high enough to resolve the filaments with a typical width of 0.1 pc \citep{Arzoumanian2011} at a distance of $\sim$ 1.76 kpc.  

\begin{figure*}

\includegraphics[width=18cm]{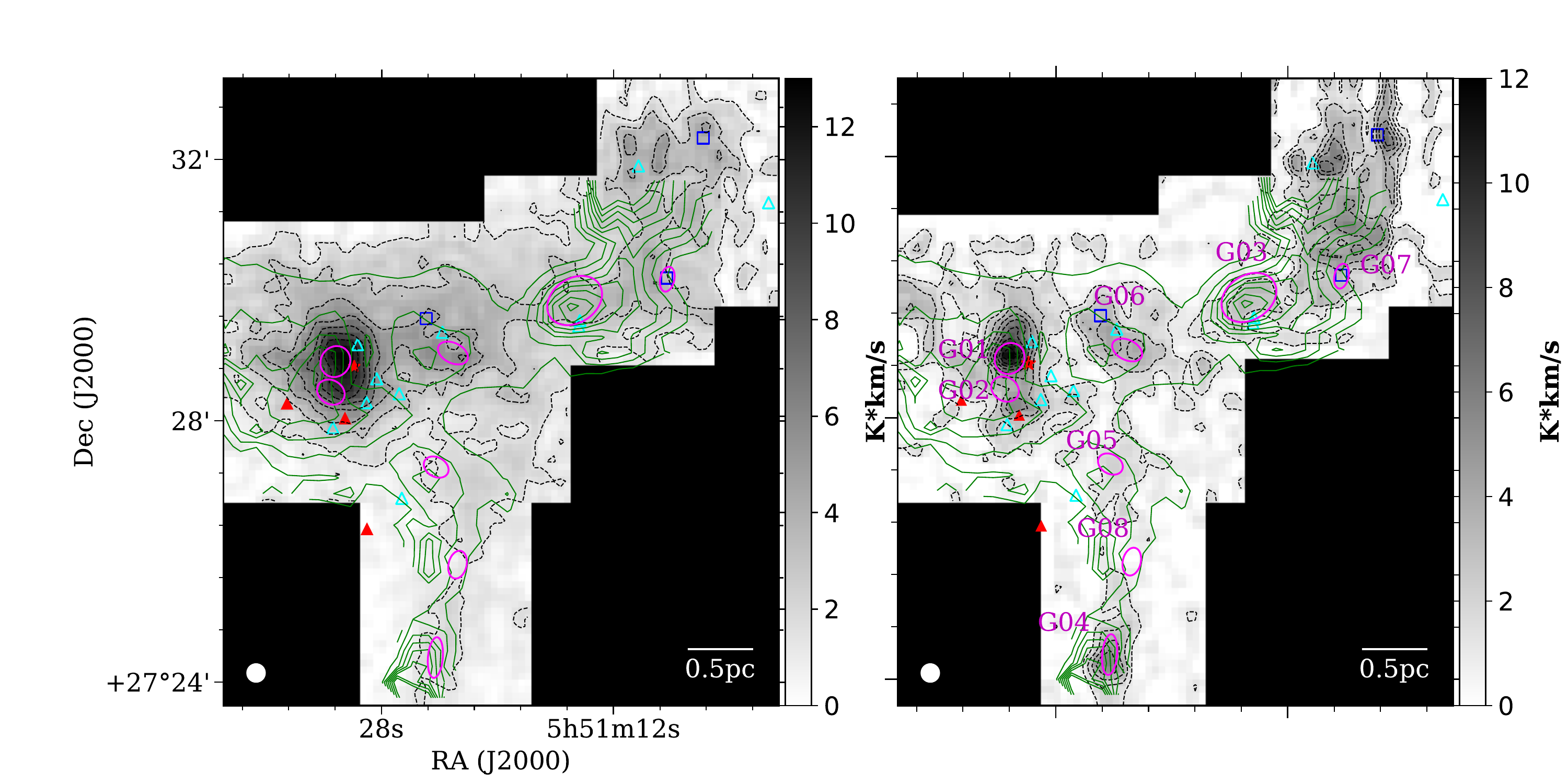}
    \caption{The gray-scale image represents the distribution of the integrated intensity between the velocity range (-15 8) km/s for HCO$^{+}$(1-0) (\textbf{Left}) and N$_{2}$H$^{+}$(1-0) (\text{Right}) line. Black contours range from 10$\%$ to 90 $\%$ stepped by the 10 $\%$ of the peak integrated intensity (13 K km $s^{-1}$ for HCO$^{+}$(1-0) and 12 K km $s^{-1}$ for N$_{2}$H$^{+}$(1-0)). The green contours means the H$_{2}$ column density map obtained from the SED fitting based on the graybody radiation mode, ranging from 10\% to 90\% stepped by the 10\% of the peak value (4 $\times$ $10^{22}$ $cm^{-2}$). The red star represents the source IRAS 05483+2728. The blue squares, cyan and red (Spitzer) triangles and magenta ellipses represent Class I YSOs, Class II YSOs and SCUBA-2 compact sources, respectively. The white circle in the left corner shows the  beam size for Nobeyama 45-m telescope at 90 GHz. \label{fig:f9}}
\end{figure*}

From Figure \ref{fig:f9}, it is clear that the distributions of HCO$^{+}$(1-0) and N$_{2}$H$^{+}$(1-0) integrated intensities are anastomotic to that of the dust very well. Dense cores and YSOs are mainly located on the region with higher column density and stronger HCO$^{+}$ (1-0) and N$_{2}$H$^{+}$ (1-0) emission. We found that HCO$^{+}$(1-0) line reveals more extended structures, while the N$_{2}$H$^{+}$(1-0) line highlights the dense cores.

\begin{figure*}
	\includegraphics[width=18cm]{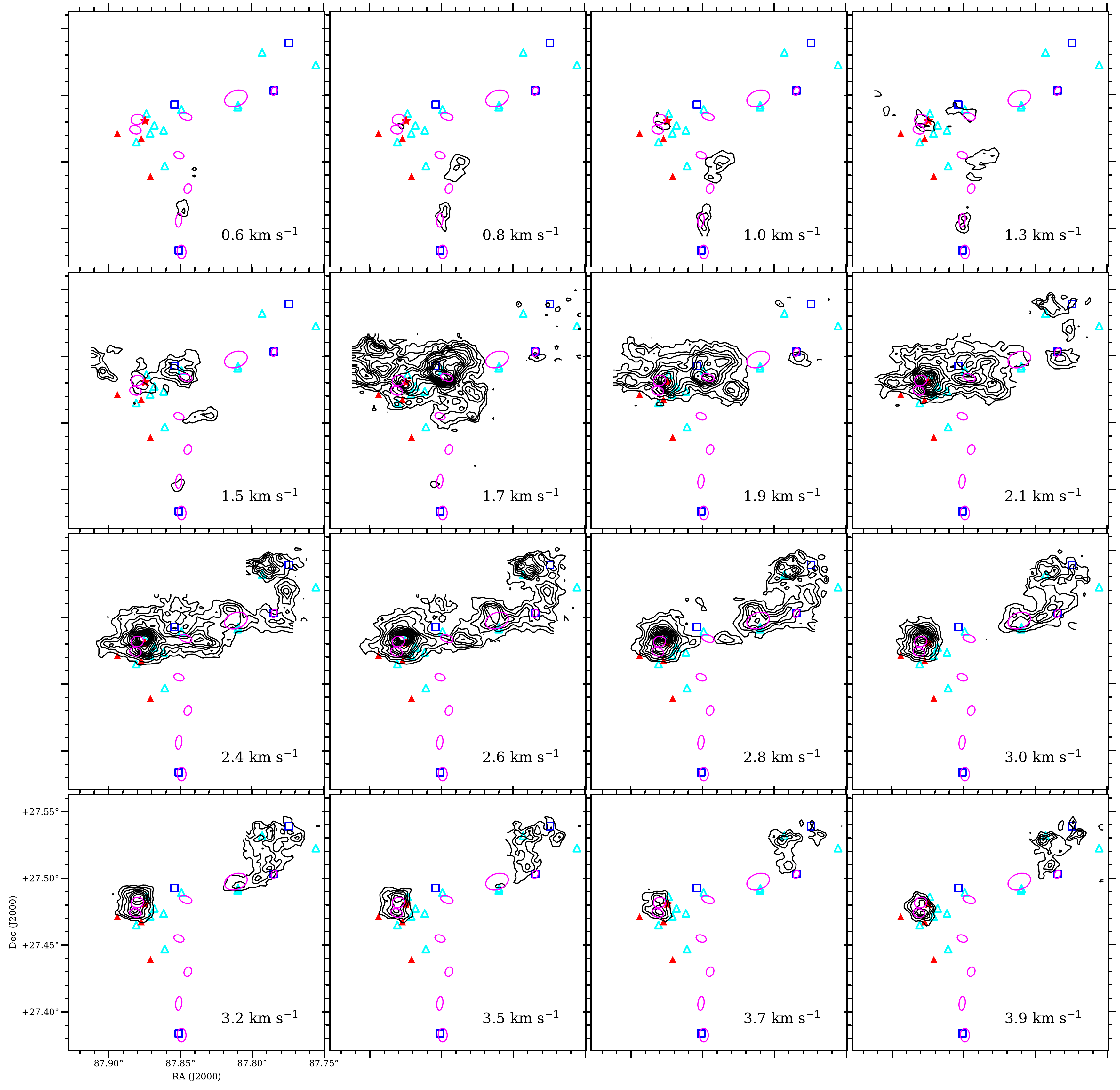}
    \caption{Channel maps of the HCO$^{+}$(1-0) line, the contour levels are from the 3$\sigma$ ( $\sim$ 0.2 K$\times$km s$^{-1}$) to the peak value by the interval of 1$\sigma$. The blue squares represent Class I YSOs, cyan and red (Spitzer) triangles stand for the Class II YSOs and magenta ellipses for the SCUBA-2 compact sources. \label{fig:f11}}
\end{figure*}

\begin{figure*}
	\includegraphics[width=18cm]{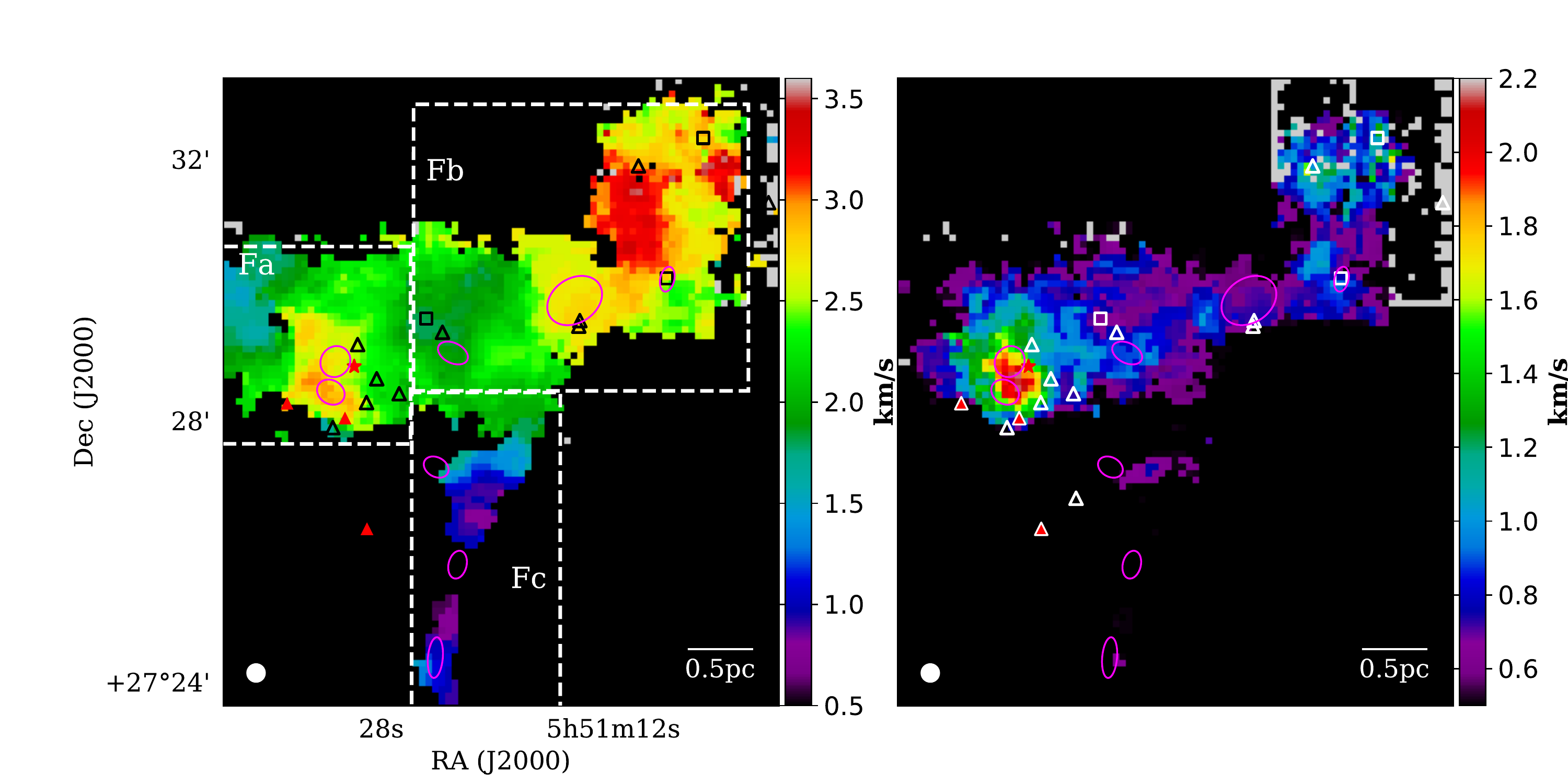}
    \caption{\textbf{Left}: The velocity field of G181 structures. \textbf{Right}: The velocity width distribution map for G181. Both are obtained from Guassian fitting of the HCO$^{+}$(1-0) line. The squares represent Class I YSOs, triangles stand for the Class II YSOs and magenta ellipses for the SCUBA-2 compact sources. The white circle in left corner shows the beam size for Nobeyama 45-m telescope at 90 GHz.\label{fig:f10}}
\end{figure*}

\subsubsection*{3.5.2 Channel Maps of $HCO^{+}$(1-0) molecular lines}
In Figure \ref{fig:f11}, we present the channel maps of the HCO$^{+}$(1-0) emission. The channel increment is 0.2 km s$^{-1}$ in the range of 0.6 -- 3.9 km s$^{-1}$. The HCO$^{+}$(1-0) emission reveals relative extended structure, in particular for the velocity range from 1.7 to 2.4 km s$^{-1}$, which harbours the crowded YSOs and cores G01, G02, G06. In the blue-shifted velocity range (0.6 1.5) km s$^{-1}$, the emission mainly distributed along the southern sub-filaments Fc. The dense cores G04, G05 and G08 are associated with this component. As velocity increases from 2.4 km s$^{-1}$ to 3.9 km s$^{-1}$, the locations of strong emission regions have an overall trend of moving from Fa sub-structure to Fb sub-structure. Fb sub-structure is mainly in the red end of the velocity range and contains G03 and 5 YSOs. The Fa sub-structure shows HCO$^{+}$(1-0) emission spanning from 1.5 to 3.9 km s$^{-1}$. It is obvious that the molecular gas is distributed velocity-coherently from the south-east to the north-west.

\subsubsection*{3.5.3 The centroid velocity and velocity width of $HCO^{+}$ line}
Figure \ref{fig:f10} shows the distribution of the line of sight centroid velocity and FWHM line-width of the HCO$^{+}$(1-0) line. We limited our view to regions with T$_{mb}$ larger than $\sim$ 6$\sigma$ (1.2 K). The value range of the V$_{LSR}$ map is between 0.5 and 3.6 km s$^{-1}$. It's clear that the dense gas velocity in the Fb sub-structure is mainly red-shifted, and vary smoothly along the sub-filaments, while the Fc sub-structure becomes blue-shifted. In the right panel, the FWHM velocity-width distributes between 0.5 and 2.2 km s$^{-1}$. We also found an notable increasing velocity width (FWHM) towards the Fa sub-structure, peaking at $\sim$ 2.15 km $s^{-1}$. The mean velocity width for Fa, Fb and Fc are 0.98$\pm$0.4, 0.69$\pm$0.2 and 0.38$\pm$0.17 km s$^{-1}$, respectively. This indicates that there are increasing star formation activities in Fc, Fb and Fa sub-structures, which inject more energy to perturb the neighbouring gas. 

The non-thermal velocity dispersion is derived from the observed line-width using the following equation \citep{Myers1983}
\begin{equation}
    \left(\sigma_{NT}\right)^{2} = \left(\sigma_{obs}\right)^{2}-\left(\sigma_{T} \right)^{2}
\end{equation}
\begin{equation}
    \sigma_{NT}=\sqrt{\frac{\Delta v_{obs}^{2}}{8ln(2)}-\frac{k_{B}T_{kin}}{m_{obs}}}
\end{equation}
where $\sigma_{NT}$, $\sigma_{obs}$, and $\sigma_{T}$  are the non-thermal, observed, and thermal velocity dispersion, $\Delta v_{obs}$ represents the observed line-width (the FWHM velocity-width of HCO$^{+}$ fitted by the Guassian model), k$_{B}$ is the Boltzmann constant and T$_{kin}$ is the kinetic temperature of the gas. The m$_{obs}$ means the mass of the observed molecule (29 a.m.u for HCO$^{+}$). According to the dust temperature map in Figure \ref{fig:f2}, we assume a gas kinematic temperature of 13$\pm$2K for Fa and 10$\pm$1 K for Fb and Fc. The thermal dispersion of the gas is corresponding to 0.04$\pm$0.003 km s$^{-1}$ (Fa) and 0.035$\pm$0.002 (Fb and Fc). The derived non-thermal velocity dispersion is $\sim$ 0.414 km s$^{-1}$ (Fa), 0.29 km s$^{-1}$ (Fb) and 0.157 km s$^{-1}$ (Fc), respectively. The corresponding mean $\sigma_{NT}$/$\sigma_{T}$ value is 10$\pm$0.3 (Fa), 8.3$\pm$0.5 (Fb) and 4.5$\pm$0.3 (Fc). The Mach number, defined as the ratio between the non-thermal velocity dispersion and the sound speed, is about 1.9 for Fa, 1.5 for Fb and 0.8 for Fc, respectively. This indicates that the sub-structures Fa and Fb are mildly supersonic, while Fc is subsonic. \cite{Konstandin2016} had studied the turbulent flows with Mach numbers ranging from subsonic (M $\approx$ 0.5) to highly supersonic (M $\approx$ 16). They confirmed the relation between the density variance and the Mach number in molecular clouds. Such relation implies that molecular clouds with higher Mach numbers present more extreme density fluctuations \citep{Konstandin2016}. This is consistent with the density enhancements in Fa, Fb and Fc sub-structures. However, the large Mach numbers, especially in Fa sub-structures, may also due to feedback from star formation activities.

It should be noted that the HCO$^{+}$ line is usually not optically thin, and consequently, its line widths can be broadened due to opacity \citep{Sanhueza2012}. In addition, HCO$^{+}$ line has usually been used to evaluate the infall properties of a given star-forming region with blue-skewed line profiles \citep{Sanhueza2012, Rygl2013, Yoo2018}. As mentioned in section \ref{n2hp}, N$_{2}$H$^{+}$ is thought to be a better high-density tracer, and less affected by star-formation activities. We have derived the velocity dispersion and centroid velocity of the N$_{2}$H$^{+}$ and HCO$^{+}$ lines for our sample of dense cores, as presented in Figure \ref{fig:fA1} and \ref{fig:fA2}. The method used to fit the hyperfine structure of N$_{2}$H$^{+}$ has taken the opacity effects into account. We revealed that there is a decreasing trend of velocity dispersion traced by N$_{2}$H$^{+}$ from dense core G01 and G02 in sub-structure Fa, G03 and G07 in sub-structure Fb, to G04 and G06 in sub-structure Fc. That further confirms the validation of velocity dispersion gradient from HCO$^{+}$ line. Due to the low S/N of N$_{2}$H$^{+}$ emission in dense core G05, its fitted velocity dispersion was not taken into account.

In addition, we found that the centroid velocities in dense cores derived from the N$_{2}$H$^{+}$(1-0) line are generally consistent with those derived from the HCO$^{+}$(1-0) line, except for G01, G02 and G03 with velocity offsets of 0.2 -- 0.6 km s$^{-1}$. This may be due to that HCO$^{+}$ traces motions of less dense gas compared with N$_{2}$H$^{+}$, because N$_{2}$H$^{+}$ tends to be more centrally distributed. We also compared the spectra of the HCO$^{+}$(1-0) with the H$^{13}$CO$^{+}$(1-0) lines for cores G01, G02 and G03. The centroid velocity traced by the two lines are consistent in G01, while H$^{13}$CO$^{+}$ $J$=1-0 profile in G02 shows double peaks and the centroid velocity of N$_{2}$H$^{+}$ matches that of the blue-shifted peak. Since the critical densities of N$_{2}$H$^{+}$ and H$^{13}$CO$^{+}$ are similar, as listed in the Table 1 of \cite{Shirley2015}, the difference between N$_{2}$H$^{+}$ and H$^{13}$CO$^{+}$ peak velocities in the G01, G02 and G03 cores may reflect the different spatial distributions of these two molecules.

\begin{figure}
	\includegraphics[width=\columnwidth]{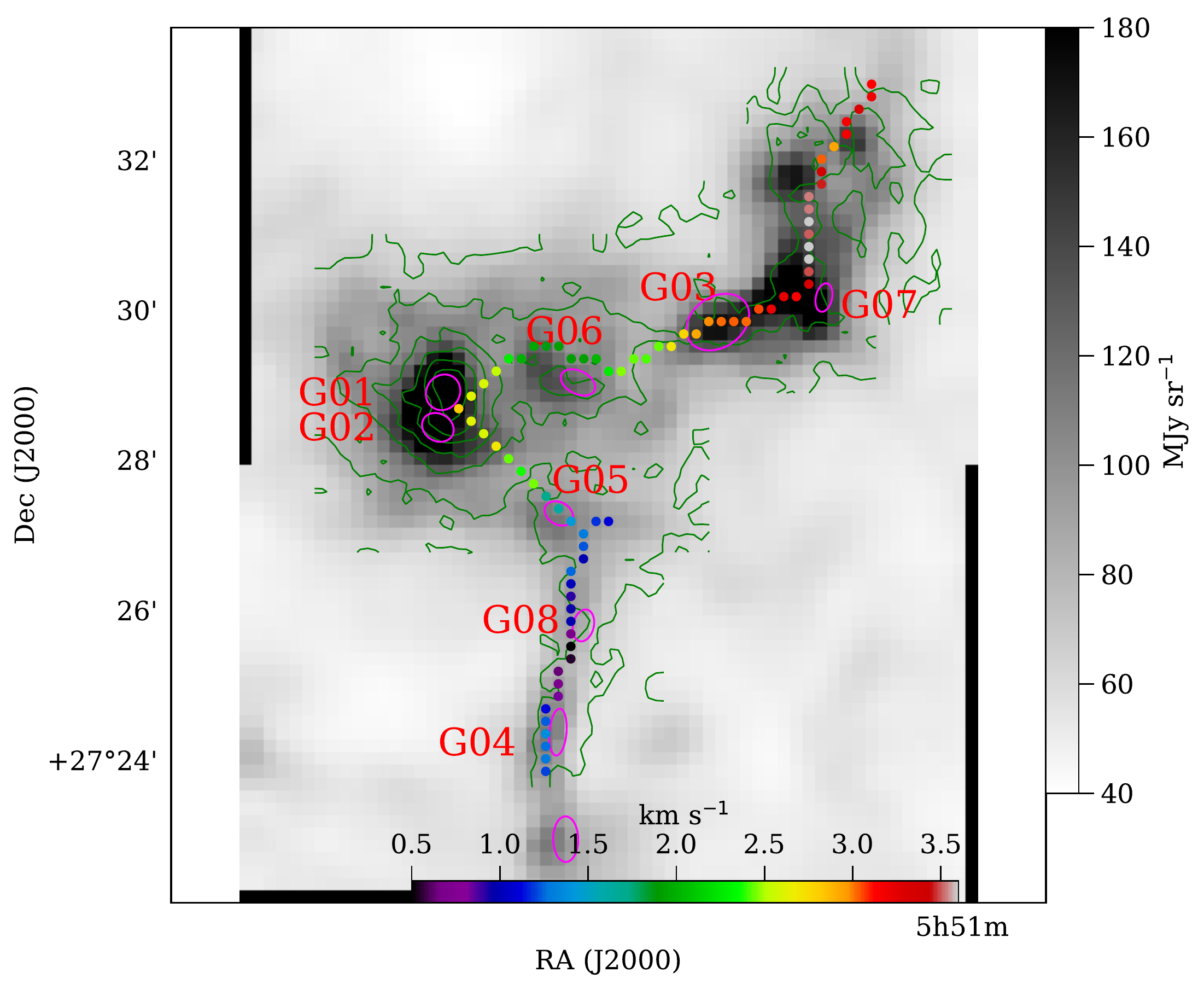}
    \caption{The skeletons of the filament network extracted with the FilFinder, color-coded with the centroid velocities of the HCO$^{+}$(1-0) spectral line. Magenta ellipses represent the SCUBA-2 compact sources. Green contours show the distribution of the HCO$^{+}$ integrated intensity, levels are from 10\% to 90\% by the interval of 15\% of the peak value (13 K km s$^{-1}$).\label{fig:f12}}
\end{figure}

\begin{figure}
	\includegraphics[width=\columnwidth]{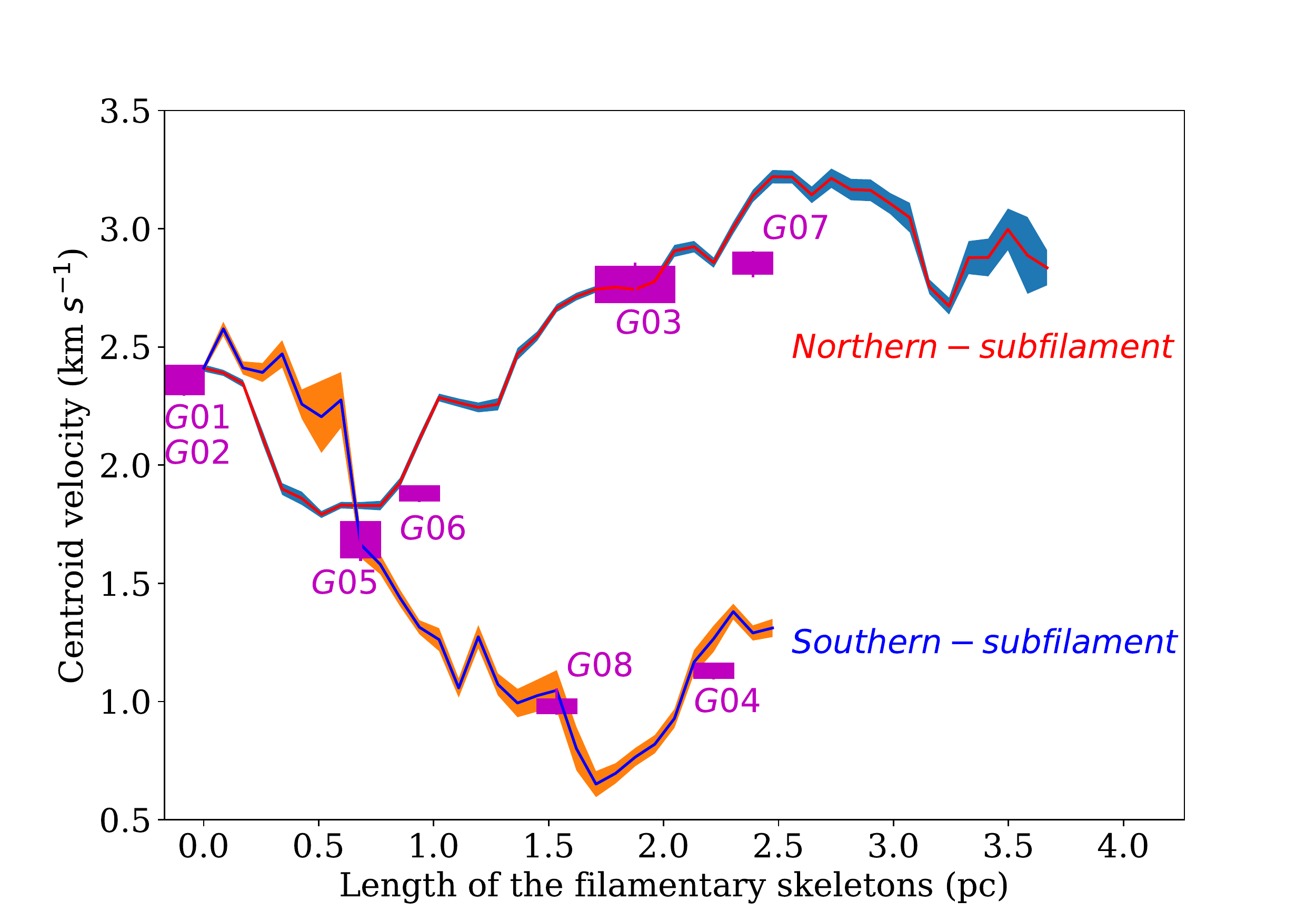}
    \caption{The centroid velocity of HCO$^{+}$(1-0) emission distributed along the skeletons of the filament network extracted by FilFinder. The zero point at x-axis is the merging point for the northern sub-structure (Fb) and southern sub-structure (Fc). The x-axis increasing direction is from the merging point at Fa sub-structure respectively along sub-skeletons to the northern and southern end. The shaded area along lines represent the error limits for the fitted centroid velocity. The magenta boxes respectively show the dense cores, and the widths of the boxes represent the spatial distribution of dense cores along the filaments, the box heights represent the range of average centroid velocity for dense cores fitted by HCO$^{+}$(1-0) and N$_{2}$H$^{+}$(1-0) lines, as shown in the Figure \ref{fig:fA1} and \ref{fig:fA2}.\label{fig:f13}}
\end{figure}

\section*{4 Discussion}
\subsection*{4.1 Hierarchical Network of Filaments in 350 $\micron$ Continuum  Emission}
We used the FilFinder\footnote{https://github.com/e-koch/FilFinder} algorithm based on the techniques of mathematical morphology \citep{Koch2015} to detect filamentary structures in the 350 $\micron$ map. Comparing to other algorithms, FilFinder can not only identify the bright filaments, but also reliably extract a population of fainter striations \citep{Koch2015}. The main process of the FilFinder algorithm includes the following steps. (1), the image was fitted a log-normal distribution and flattened using an arctangent transform, (2), the flattened data are smoothed using a Guassian filter (0.05 pc for FWHM, half of the typical width of a filament \citep{Arzoumanian2011}), (3), a mask of the filamentary structure was created in the smoothed image using an adaptive threshold, (4), the filamentary structures were detected in the local brightness thresholds within the mask. The parameters of the thresholds are listed in the Appendix. In Figure \ref{fig:f12}, we plot the skeletons of the sub-filaments, color-coded with the centroid velocities of the HCO$^{+}$(1-0) line. We found that the skeletons of sub-filaments for the Fb and Fc merged at the Fa sub-structure. In addition, combining the left panel in Figure \ref{fig:f10} with Figure \ref{fig:f12}, it is clear that the southern sub-structure (Fc) is blue-shifted (0.5 - 2.5 km s$^{-1}$), and the northern sub-structure (Fb) is red-shifted (2.5 - 3.5 km s$^{-1}$). The Fb and Fc sub-structures present clear velocity gradients. The SCUBA-2 dense cores are mainly distributed along the skeletons. 

\subsection*{4.2 The velocity gradients along sub-filaments}
In Figure \ref{fig:f13}, we present the centroid-velocity profiles of the sub-filamentary skeletons Fb and Fc. The merging point of Fb and Fc skeletons in the Fa sub-structure is defined as the zero point at x-axis. The x-axis value (Figure \ref{fig:f13}) of a pixel on the skeleton represents its distance from the zero point integrated along the spine lines (Figure \ref{fig:f12}). It is evident that the sub-filaments Fb and Fc exhibit both global and local velocity gradients. For the Fb sub-filament, the global velocity gradients are about 0.8$\pm$0.05 km s$^{-1}$ pc$^{-1}$ from 0.5 pc to 2.5 pc of x-axis value, and the local velocity gradients near the Fa sub-structure are up to 1.2$\pm$0.1 km s$^{-1}$ pc$^{-1}$ from 0.0 pc to 0.5 pc in skeleton length. The Fc sub-filament shows velocity gradients of $\sim$ 1.0$\pm$0.05 km s$^{-1}$ within a range of 0.0 pc -- 1.7 pc in x-axis, and 1.3$\pm$0.1 km s$^{-1}$ pc$^{-1}$ in the range of 0.0 pc to 0.6 pc. It is evident that the local velocity gradients around sub-structure Fa, including G01, G02 and 7 YSOs, are larger than the global velocity gradients along the sub-filaments Fb and Fc. In addition, we also find that the dense cores G01, G02 appear to locate near the maxima of velocity profiles along the sub-filaments. While there is almost a quarter phase shift between the densities (i.e., dense cores G03, G05, G04, G06, G08) and velocities. This suggests that the continuum peaks have an influence on the dynamics of the surrounding gas \citep{Hacar2011,Yuan2018, Liu2016b}.

As shown in Figure \ref{fig:f1}, G181 is located on the northern end of the extended filamentary structure (EFS) S242. For the longer filaments with the aspect ratio, A$_{o}$ $\gtrsim$ 5.0, global collapse would be end-dominated, the materials are preferentially accelerated into the both ends. The end-clumps were given more momentum, and further approach to the center of filaments in an asymptotic inward speed, sweeping up gas mass as they approach \citep{Pon2011, Pon2012}. Moreover, \cite{Clarke2015} also investigated the end-dominated collapse in filaments, and suggested that due to gravitational attraction of the end-clumps, the gas ahead of end-clumps, was immediately accelerated towards end-clumps. \cite{Dewangan2017} concluded that observed results of massive clumps and the clusters of YSOs toward both ends of EFS S242, were consistent with the prediction of the end-dominated collapse model. The global velocity gradients along Fb and Fc in the northern end of EFS S242 may be caused by the gas motion in the end-dominated collapse.

\subsection*{4.3 Gravitational fragmentation of sub-filaments}
 As shown in Figure \ref{fig:f12}, G181 contains sub-filaments with 9 dense cores embedding in them. To investigate the fragmentation of sub-filaments and hence the formation of dense cores, we compare the sub-structures of G181 to the model of a collapsing isothermal cylinder, supported by either thermal or non-thermal motions. \cite{Andre2013} has suggested that thermally supercritical filaments, where the mass per unit length is greater than the critical value (M$_{line,crit}$; i.e., M$_{line}$ $>$ M$_{line, crit}$), are associated with pre-stellar clumps/cores and star formation activities. However, there are few Herschel pre-stellar clumps/cores and embedded protostars associated with thermally subcritical filaments (M$_{line}$ $<$ M$_{line, crit}$). 
 
For thermal support, the critical mass per unit length (M$_{line, crit}$ = 2c$_{s}^{2}$/G; where c$_{s}$ is the isothermal sound speed, G is the gravitationally constant) is needed for a filament to be gravitationally unstable to radial contraction and fragmentation along its length \citep{Inutsuka1997}. The thermal critical line mass M$_{line,crit}$ is $\sim$ 16 M$_{\odot}$ pc$^{-1} \times$ (T$_{gas}$/10 K) for gas filaments \citep{Andre2014}. The filamentary structures of G181 have an observed mass per unit length of $\sim$ 200 M$_{\odot}$ pc$^{-1}$  \citep{Dewangan2017}. The observed value largely exceeds the critical value of $\sim$ 16-48 M$_{\odot}$ pc$^{-1}$ (at T = 10-30 K). 
 
Taking into account additional supports from turbulent pressure, the total 1D velocity dispersion is calculated with the following equation \citep{Myers1992}
\begin{equation}
    \sigma_{TOT}=\sqrt{\frac{\Delta v_{obs}^{2}}{8ln(2)}+k_{B}T_{kin}\left(\frac{1}{\mu m_{H}}-\frac{1}{m_{obs}}\right)}
\end{equation}
where m$_{H}$ is the mass of a Hydrogen atom, and $\mu$ = 2.33 is the atomic weight of the mean molecule. According to the observed linewidth and dust temperatures mentioned in section 3.5.2, we calculated the 1D velocity dispersions of $\sigma_{TOT}$, which are $\sim$ 0.47$\pm$0.07, 0.35$\pm$0.05 and 0.25$\pm$0.07 km s$^{-1}$, respectively for Fa, Fb and Fc. The uncertainties are estimated based on a $\sim$ 15\% error in the temperature and a $\sim$ 20\% error in the observed line-width. We calculated the critical line mass supported by both thermal and non-thermal contribution using \citep{Fiege2000}

\begin{equation}
\left(M/L\right)_{crit} = \frac{2\sigma_{TOT}^{2}}{G}
\end{equation}
The derived critical line masses supported by the thermal and non-thermal pressure are 101$\pm$15 M$_\odot$ pc$^{-1}$ (Fa), 56$\pm$ 8 M$_\odot$ pc$^{-1}$ (Fb) and 28$\pm$ 4 M$_\odot$ pc$^{-1}$ (Fc), respectively. The observed masses per unit length ($\sim$ 200 M$_{\odot}$ pc$^{-1}$) are still a factor of 2-7 times larger than the critical values. Therefore, the dense cores in G181 might be formed through cylindrical fragmentation.

\subsection*{4.4 Sequential star formation}
In section 3, based on the evolutionary states of dense cores and YSOs, we find that dense cores and YSOs around the Fa structure (2 protostars and 7 Class II YSOs) are at relatively later stages of evolution than those in the Fb (2 pre-stellar candidates, 3 Class I YSOs, 5 Class II YSOs) and Fc (3 pre-stellar candidates, 1 Class I YSOs, 2 Class II YSOs) sub-structures, which provide evidence for sequential star formation in the structures of PGCC G181. We also found inhomogenous distribution of the physical properties in Fa, Fb and Fc sub-structures of G181, including the H$_{2}$ column density, dust temperature and velocity dispersion. \cite{Heitsch2008} studied numerically the formation of molecular clouds in large-scale colliding flows including self-gravity. Their model emphasized that large-scale filaments were mostly driven by global gravity, and local collapses were triggered by a combination of strong thermal and dynamical instabilities, which further caused cores to form.

\cite{Pon2011, Pon2012} pointed out that the global collapse timescale of molecular clouds were decreasing outward in the edges. That leads to material preferentially accelerating onto the edges. Such a density enhancement occurs in the periphery of a cloud, it can further trigger local collapse to form stars. In addition, edge-driven collapse mode is mainly dominant in filaments with aspect ratios larger than 5. These filaments are also easier to induce local collapse by small perturbations \citep{Pon2012}. As pointed out by \cite{Dewangan2017}, the extended filamentary structure (EFS) S242 may be a good example for the end-dominated collapse. We also find sequential star formation activities in the northern end of S242. The dense cores and YSOs in Fb and Fc sub-structures are in younger evolutionary stages than those in Fa sub-structure (2 protostars and 7 Class II YSOs), and the star formation activities in Fb, including 2 pre-stellar candidates, 3 Class I YSOs, 5 Class II YSOs  are more active than Fc sub-structures (3 pre-stellar candidates, 1 Class I YSOs, 2 Class II YSOs). Therefore, we suggest that the star formation activities in Fa, Fb and Fc are progressively taking places in a sequence, following the order of the local density enhancements being built up gradually in the Fa, Fb and Fc sub-structures. Such picture is consistent with the scenario of the end-dominated collapse model.

\section{Conclusions}

We have used the Herschel (70 -- 500 $\mu$m), SCUBA-2 (850 $\mu$m), WISE (3.4, 4.6, 12 and 22 $\mu$m) continuum maps and HCO$^{+}$(1-0), N$_{2}$H$^{+}$(1-0) line data to study the star formation in the filamentary structures of PGCC G181, which is sited at the northern end of the extended filamentary structures S242. The main results are as follows:

1. We obtained an H$_{2}$ column density map (1.0,  4.0) $\times$ 10$^{22}$ cm$^{-2}$ and a dust temperature map (8, 16) K from SED fitting of Herschel (160-500 $\mu$m) and SCUBA-2 850 $\mu$m data, based on the graybody radiation model.

2. Nine compact sources have been identified from the SCUBA-2 850 $\mu$m map, including 4 protostellar and 5 pre-stellar candidates. Their physical parameters, including equivalent radius, dust temperature, column density, volumn density, luminosity and mass are derived with SED fitting. The characteristics of the velocity dispersion, excitation temperature and optical depth are estimated by fitting the N$_{2}$H$^{+}$(1-0) spectral line. In addition, 18 YSOs, including 4 Class I YSOs and 14 Class II YSOs, are found in the G181 region based on the near- and mid-infrared colors obtained from the WISE, Spitzer and 2MASS catalogs. 

3. We report observations of the HCO$^{+}$(1-0) and N$_{2}$H$^{+}$(1-0) spectral lines toward G181. We find that the distribution of the HCO$^{+}$(1-0) integrated intensities is well anastomotic to that of the dust emission, while the N$_{2}$H$^{+}$(1-0) emission is concentrated near the dense cores. The dense cores and YSOs are mainly distributed along the filamentary structures of G181, and located in regions with higher column densities and stronger HCO$^{+}$(1-0) and N$_{2}$H$^{+}$(1-0) emission.

4. We find an notable increasing velocity dispersion towards the Fa sub-structures. The sub-structures Fa and Fb are mildly supersonic, while the Fc sub-structure is subsonic. In addition, we also detected velocity gradients of 0.8$\pm$0.05 km s$^{-1}$ pc$^{-1}$ for Fb and 1.0$\pm$0.05 km s$^{-1}$ pc$^{-1}$ for the Fc sub-filament, while the value for the Fa sub-structure is up to 1.2 km s$^{-1}$ pc$^{-1}$. The global velocity gradients along the sub-structures Fb and Fc may be caused by the gas motion in the end-dominated collapse of filaments.

5. We derive the critical mass per unit length for each sub-filaments in both thermal and non-thermal support cases. The values are 101 $\pm$ 15 M$_\odot$ pc$^{-1}$ (Fa), 56 $\pm$ 8 M$_\odot$ pc$^{-1}$ (Fb) and 28 $\pm$ 4 M$_\odot$ pc$^{-1}$ (Fc), respectively. The observed masses per unit length ($\sim$ 200 M$_\odot$ pc$^{-1}$) are a factor of 2-7 times larger than the critical value of each sub-filament, suggesting that  the dense cores in G181 might be formed through cylindrical fragmentation.

6. We find evidence for sequential star formation in the filamentary structures of G181. The dense cores and YSOs located in the sub-structures Fb and Fc are younger than those in the Fa sub-structure. In addition, the star formation activities are gradually progressive in the Fa, Fb, Fc sub-structures, which are consistent with the materials being accelerated into the ends and then further inward the center of the filament S242 in an end-dominated collapse.

\section{acknowledgments}

Lixia Yuan, Ming Zhu and Chenlin Zhou are supported by the National Key R$\&$D Program of China No.2017YFA0402600, the NSFC grant No.U1531246 and also supported by the Open Project Program of the Key Laboratory of FAST, NAOC, Chinese Academy of Sciences. Jinghua Yuan is supported by the NSFC funding No.11503035 and No.11573036. Ke Wang acknowledges the support by the National Key Research and Development Program of China (2017YFA0402702),
the National Science Foundation of China (11721303),
and the starting grant at the Kavli Institute for Astronomy and Astrophysics, Peking University (7101502016).

The spectral lines HCO$^{+}$(1-0) and N$_{2}$H$^{+}$ are based on observations at the Nobeyama Radio Observatory (NRO), Nobeyama Radio Observatory is a branch of the National Astronomical Observatory of Japan, National Institutes of Natural Sciences. We acknowledge the observational supports from Dr.Kazufumi Torii in NRO. 
The James Clerk Maxwell Telescope is operated by the East Asian Observatory on behalf of The National Astronomical Observatory of Japan; Academia Sinica Institute of Astronomy and Astrophysics; the Korea Astronomy and Space Science Institute; the Operation, Maintenance and Upgrading Fund for Astronomical Telescopes and Facility Instruments, budgeted from the Ministry of Finance (MOF) of China and administrated by the Chinese Academy of Sciences (CAS), as well as the National Key R\&D Program of China (No. 2017YFA0402700). Additional funding support is provided by the Science and Technology Facilities Council of the United Kingdom and participating universities in the United Kingdom and Canada.

This research has made use of the NASA/ IPAC Infrared Science Archive, which is operated by the Jet Propulsion Laboratory, California Institute of Technology, under contract with the National Aeronautics and Space Administration. 
This research also made use of Montage. It is funded by the National Science Foundation under Grant Number ACI-1440620, and was previously funded by the National Aeronautics and Space Administration's Earth Science Technology Office, Computation Technologies Project, under Cooperative Agreement Number NCC5-626 between NASA and the California Institute of Technology. This research made use of Astropy, a community-developed core Python package for Astronomy.




\bibliographystyle{mnras}
\bibliography{reference_G181.bib} 



\appendix

\section{The critical parameters of the thresholds in FilFinder}
verbose=True, 

smooth\_size=0.05*u.pc,

size\_thresh=2000*u.arcsec$^{2}$, 

glob\_thresh=50,

fill\_hole\_size=0.3*u.pc$^{2}$,

border\_masking=False

\begin{figure}
	\includegraphics[width=\columnwidth]{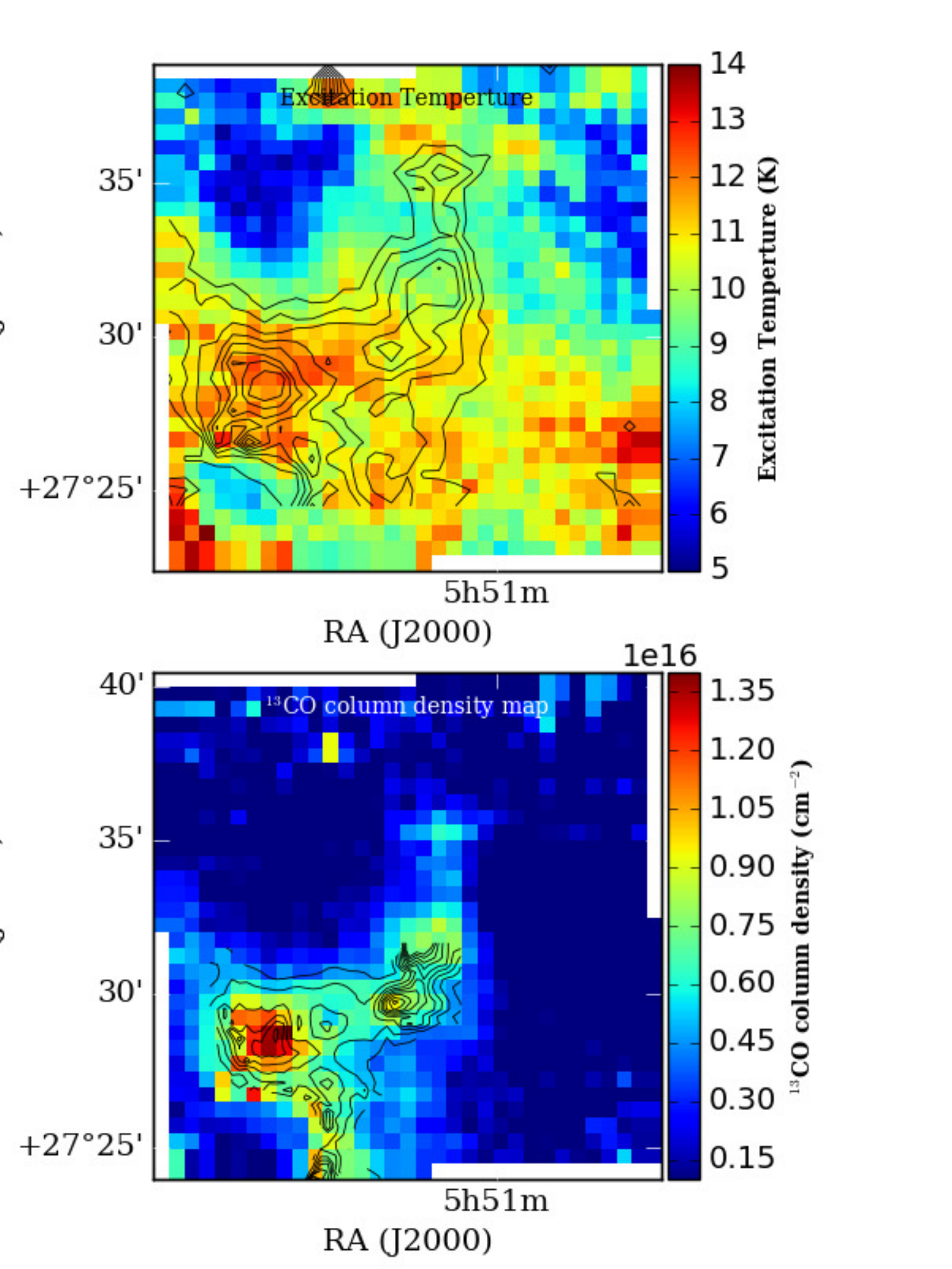}
    \caption{ \textbf{Upper} panel: Excitation temperature map, black contours represent the $^{13}$CO column density, levels are from 20\% to 90\% by the interval of 10\% of the peak value of 1.4 $\times$ 10$^{16}$ cm$^{-2}$. \textbf{Lower} panel: $^{13}$CO column density map. The overlaid black contours represent the H$_{2}$ column density derived by the SED fitting, levels are from 10\% to 90\% by the interval of 10\% of the peak value (4.0 $\times$ 10$^{22}$ cm$^{-2}$). \label{fig:fa1}}         
\end{figure}
\onecolumn
\begin{figure*}
\centering
\subfloat{\includegraphics[width=7cm]{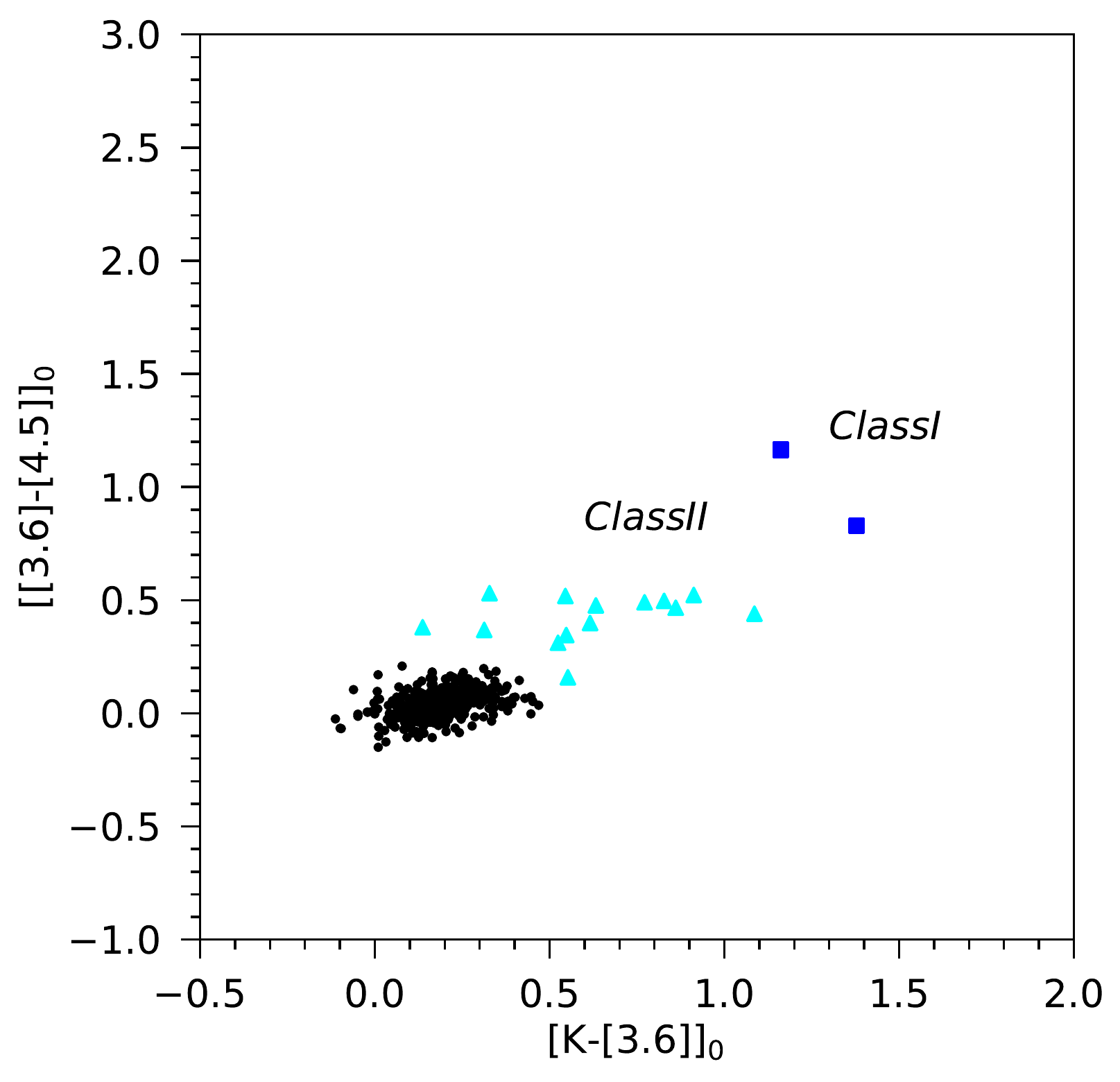}}
\subfloat{\includegraphics[width=7cm]{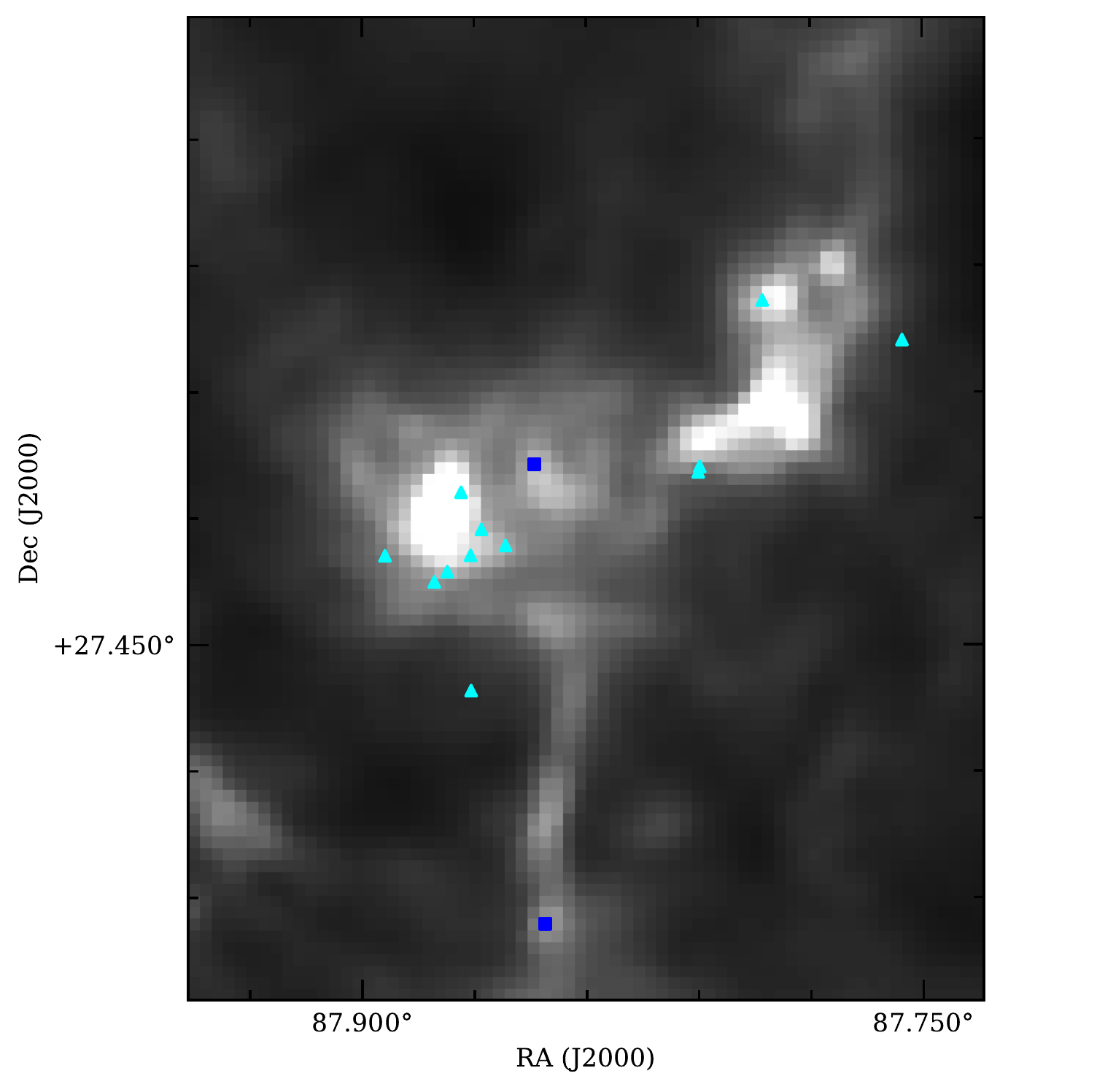}}
\caption{\textbf{Left plane}: The YSOs were identified by the 2MASS and GLIMPSE360 photometric data from 1 to 5 $\micron$ by the dereddened color-color space ([K-[3.6]]$_{0}$ and [[3.6]-[4.5]]$_{0}$). The detailed process was presented in \citep{Dewangan2017, Gutermuth2009}. \textbf{Right plane}: The distribution of the YSOs identified by the 2MASS and GLIMPSE360 photometric data in the 250 \micron~continuum map. The blue squares represent Class I YSOs, cyan triangles stand for Class II YSOs.\label{fig:fa2}}
\end{figure*}

\begin{figure*}
\centering
\subfloat[G01]{\includegraphics[width=5.5cm]{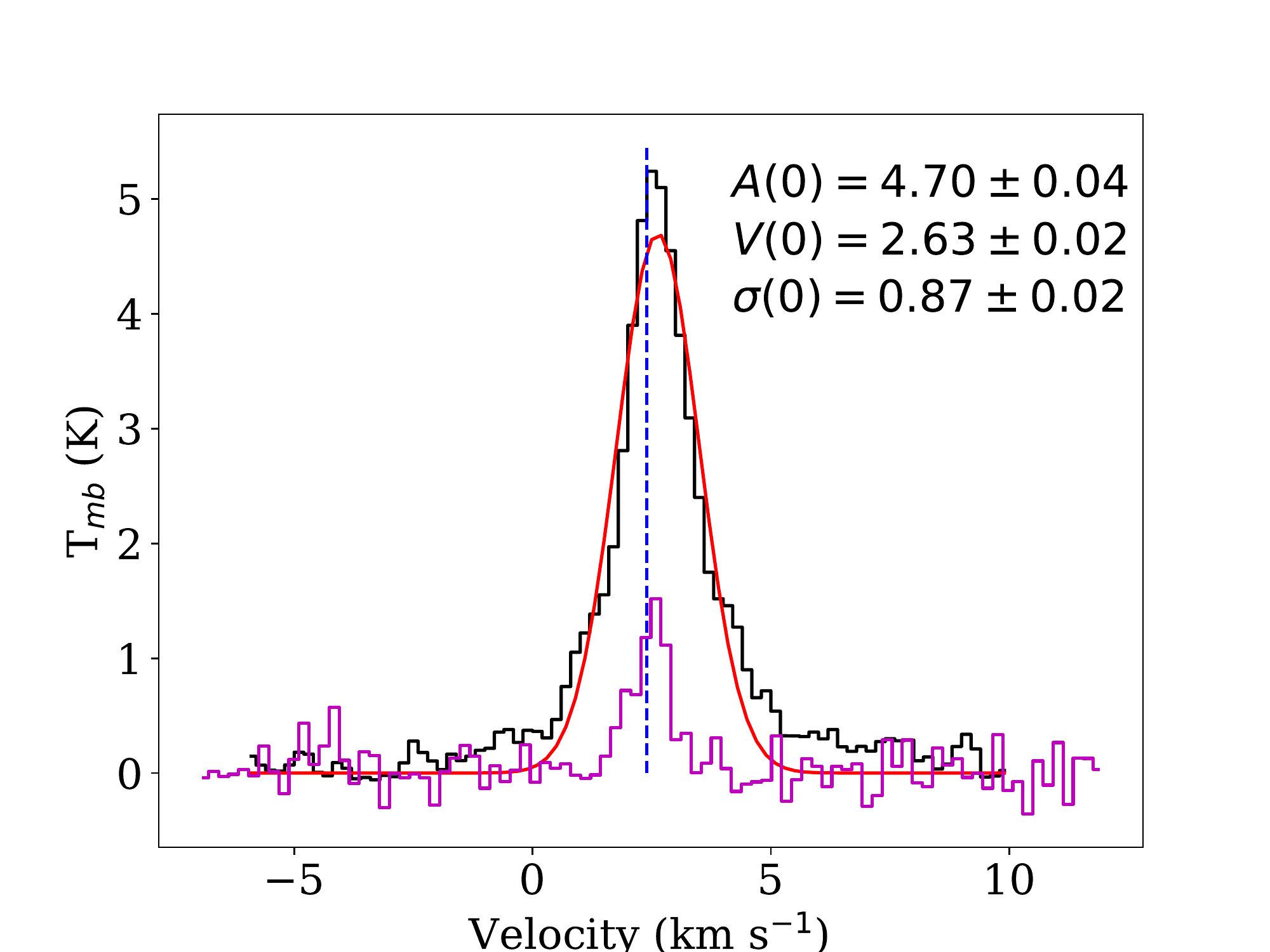}}
\subfloat[G02]{\includegraphics[width=5.5cm]{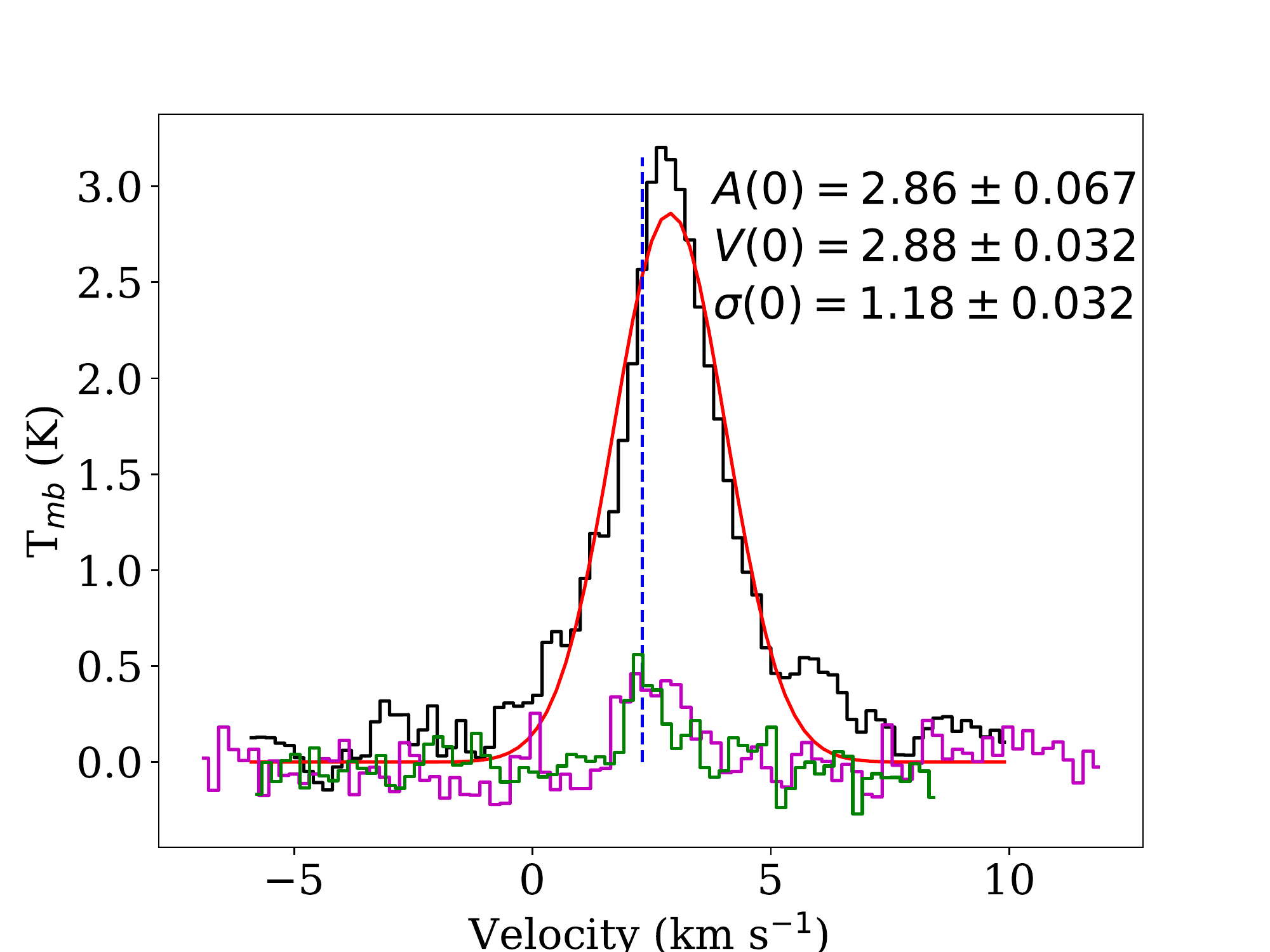}}
\subfloat[G03]{\includegraphics[width=5.5cm]{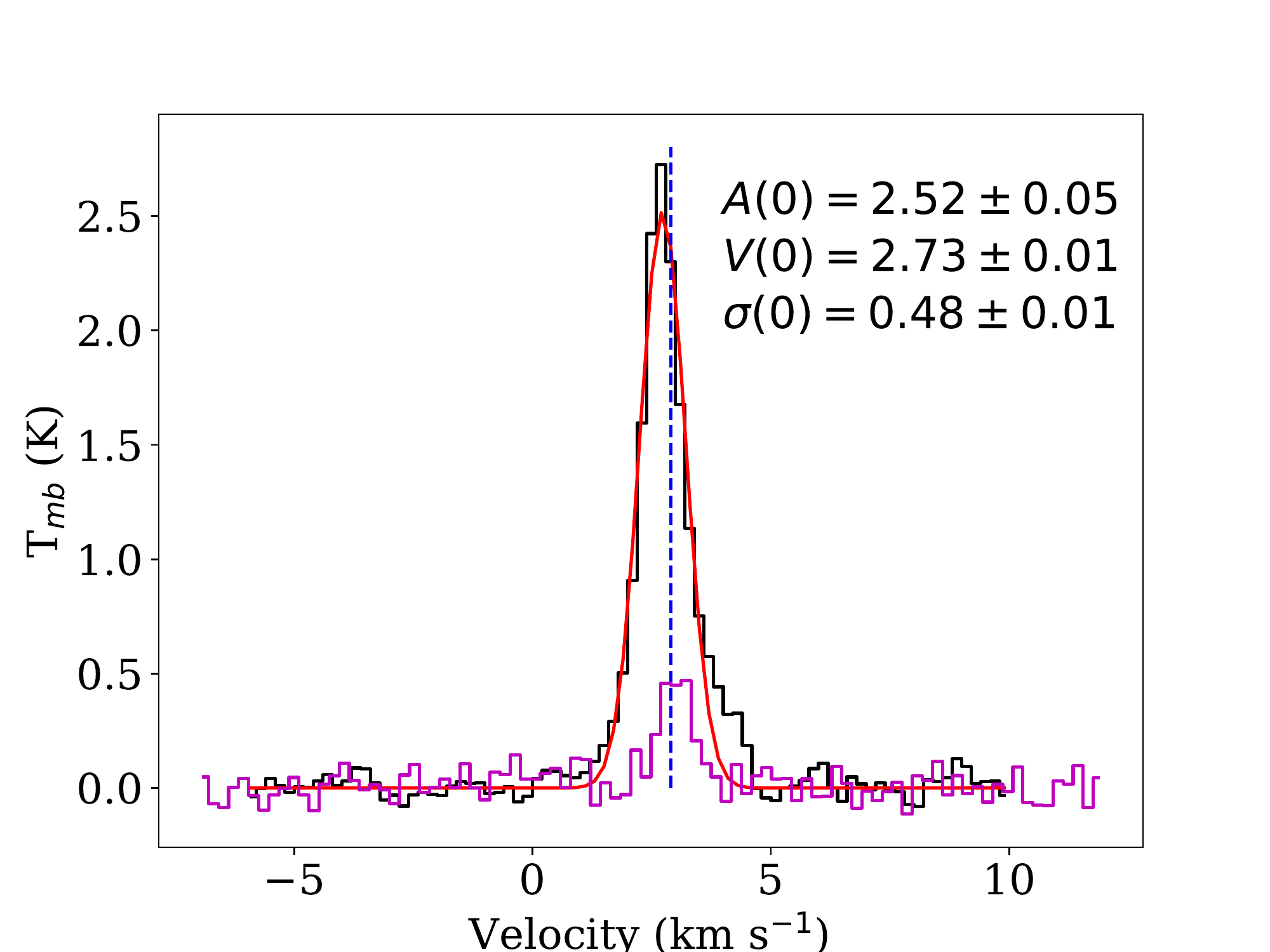}}
\qquad
\subfloat[G04]{\includegraphics[width=5.5cm]{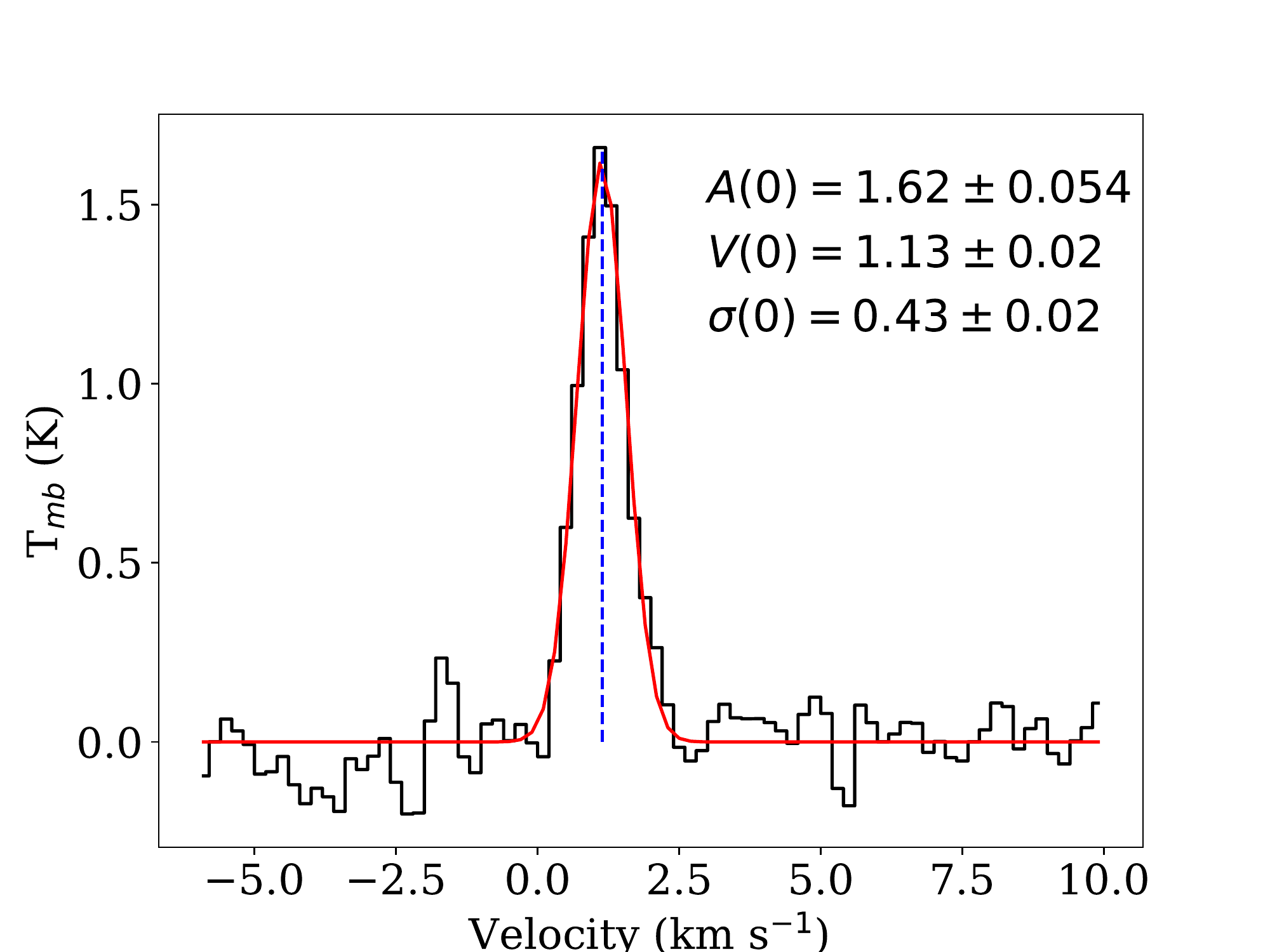}}
\subfloat[G05]{\includegraphics[width=5.5cm]{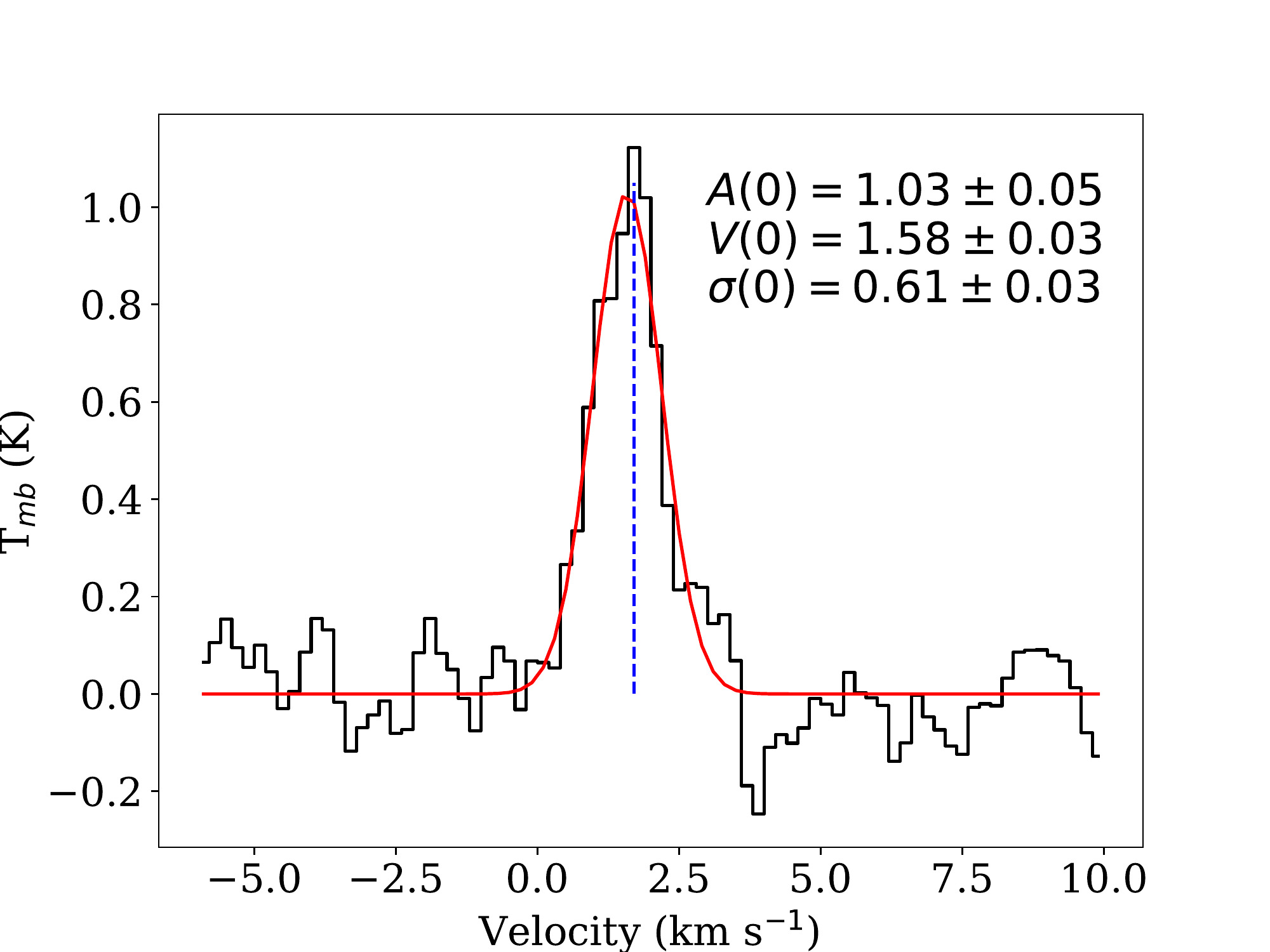}}
\subfloat[G06]{\includegraphics[width=5.5cm]{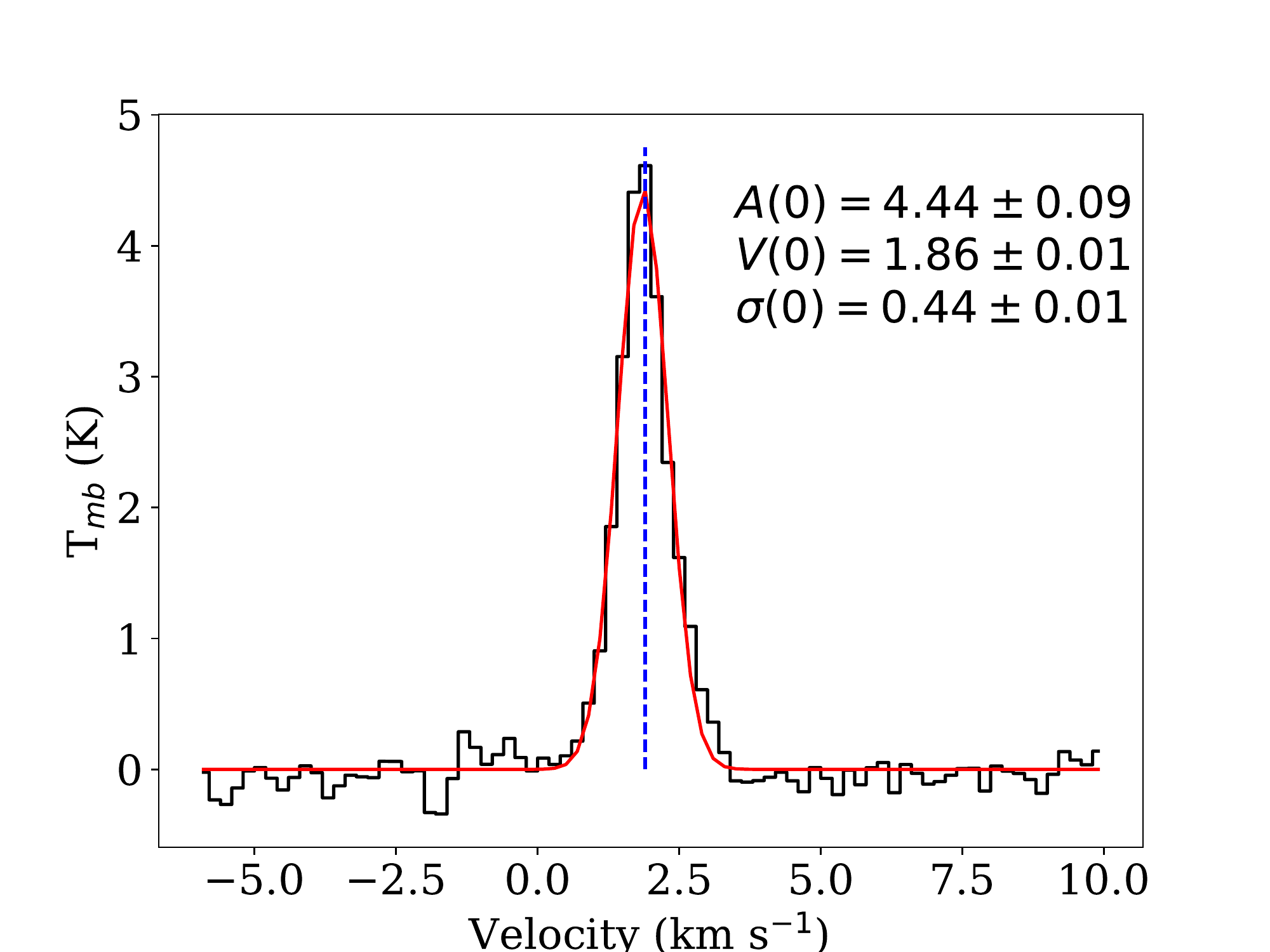}}
\qquad
\subfloat[G07]{\includegraphics[width=5.5cm]{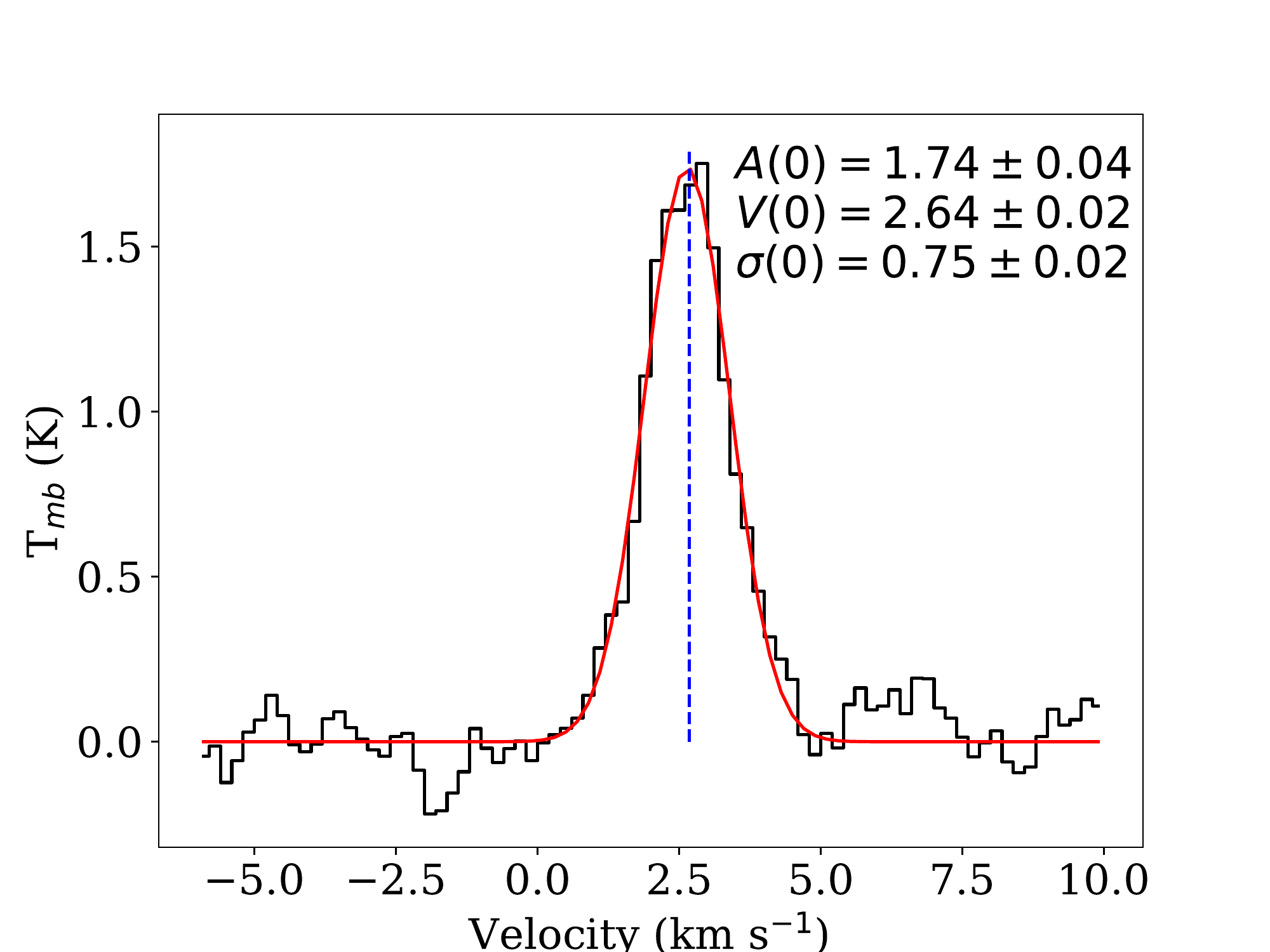}}
\subfloat[G08]{\includegraphics[width=5.5cm]{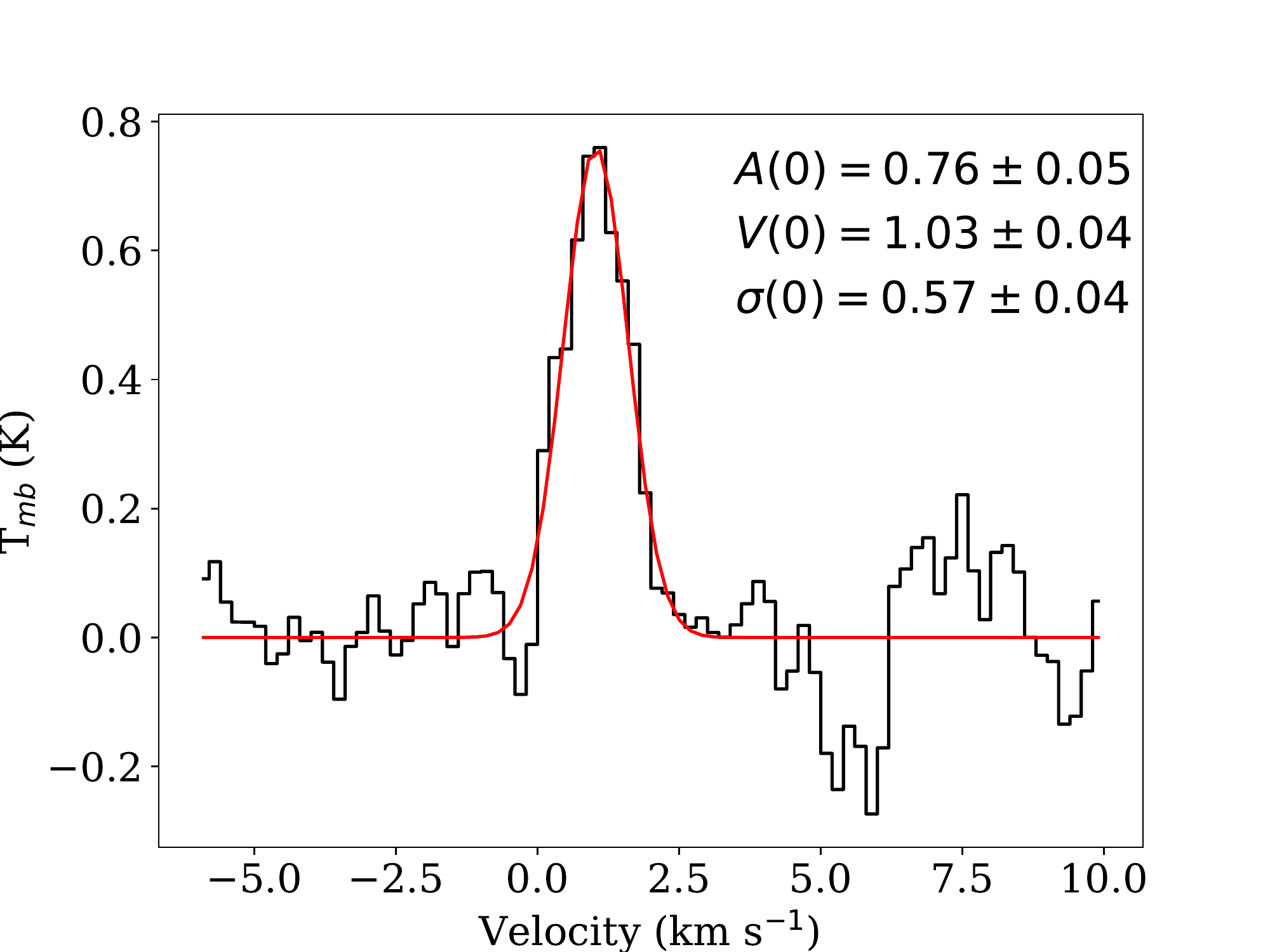}}
\caption{HCO$^{+}$(1-0) line profile for G01-G08 dense cores in G181. The blue-vertical lines represent the corresponding velocity fitted by the hyperfine structure of the N$_{2}$H$^{+}$(1-0) line. The magenta profiles in G01-G03 are the H$^{13}$CO$^{+}$(1-0) line. The green profile in G02 is the highest-frequency component of the N$_{2}$H$^{+}$(1-0) line.\label{fig:fA1}}
\end{figure*}

\begin{figure*}
\centering
\subfloat[G01]{\includegraphics[width=5.5cm]{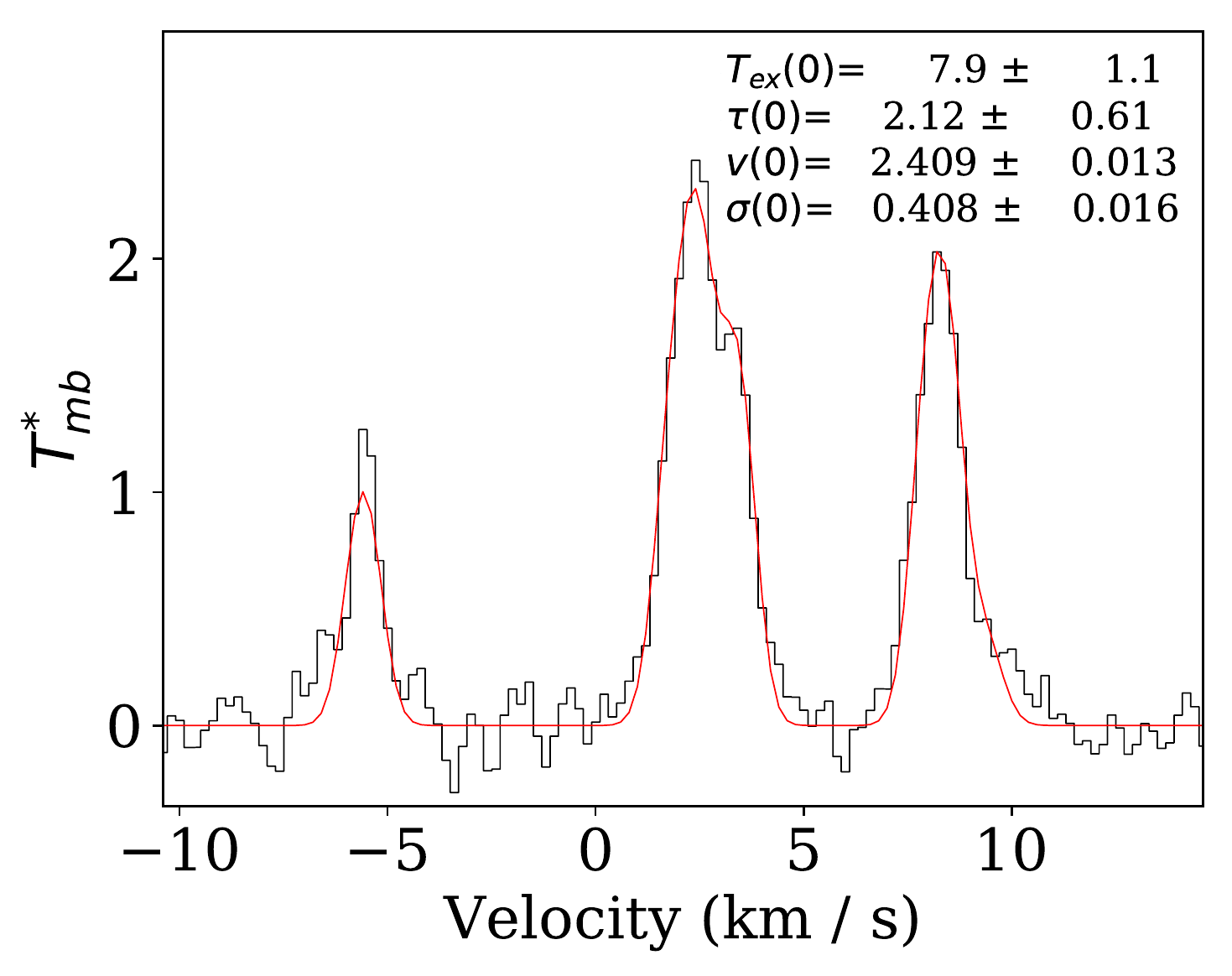}}
\subfloat[G02]{\includegraphics[width=5.5cm]{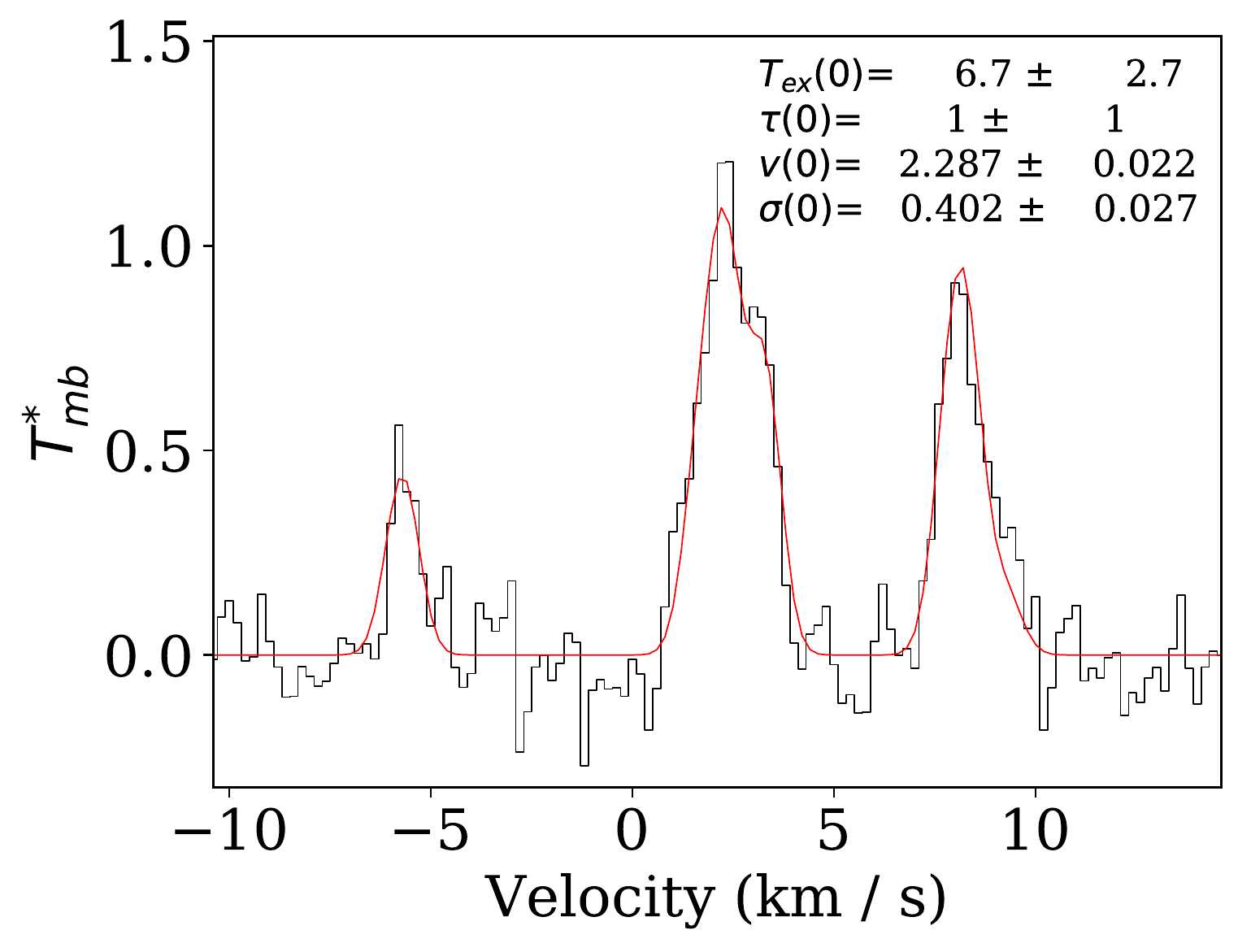}}
\subfloat[G03]{\includegraphics[width=5.5cm]{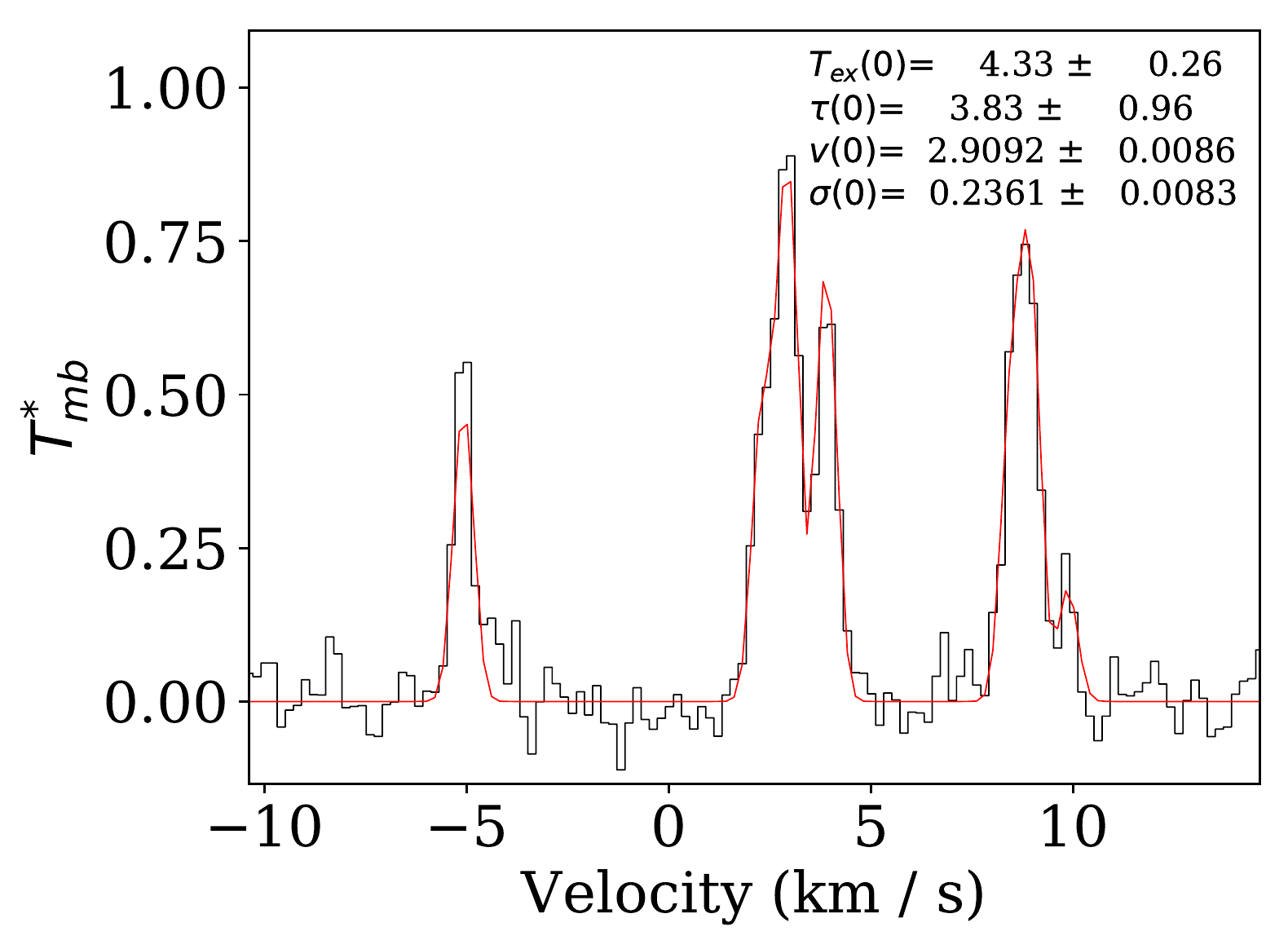}}
\qquad
\subfloat[G04]{\includegraphics[width=5.5cm]{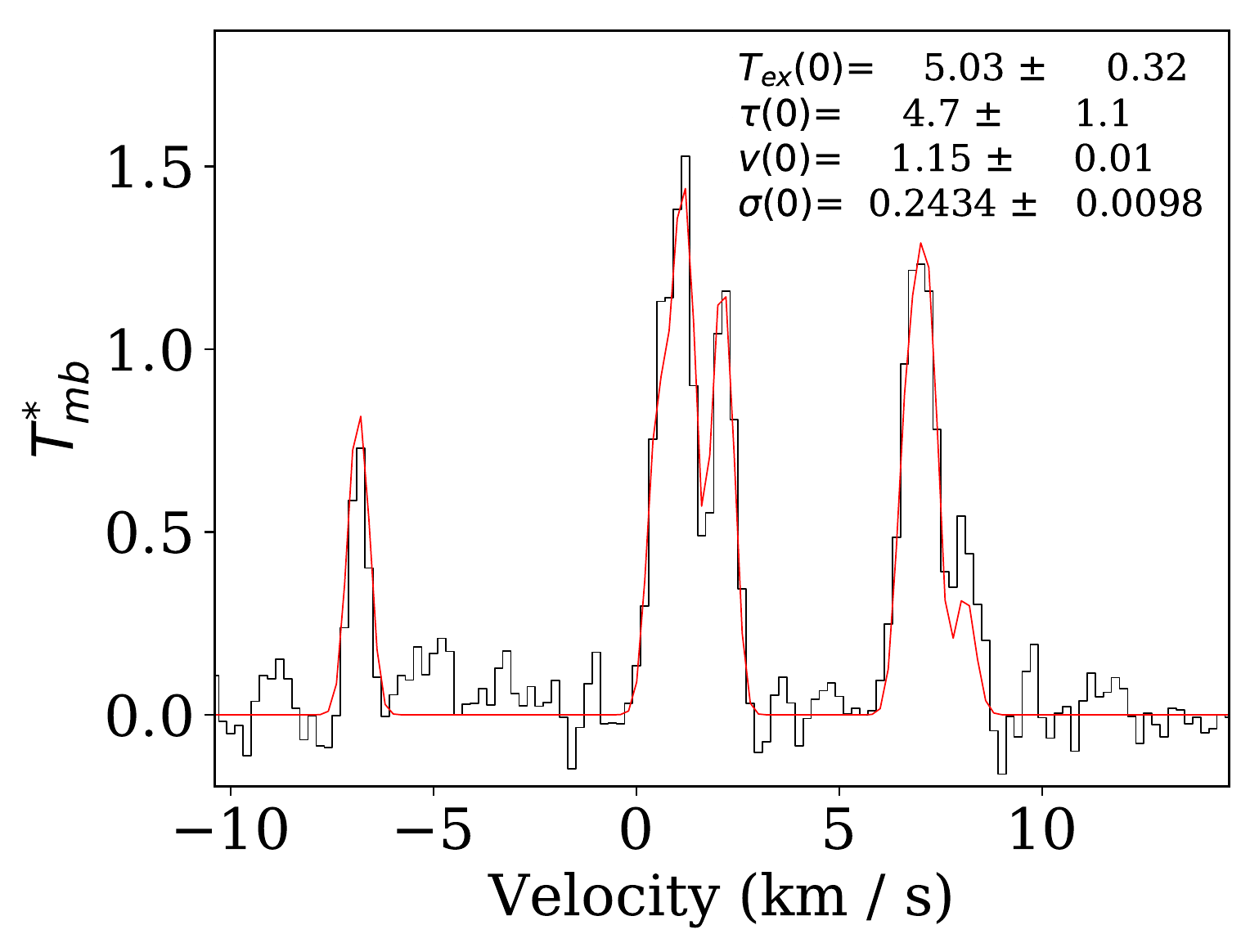}}
\subfloat[G05]{\includegraphics[width=5.5cm]{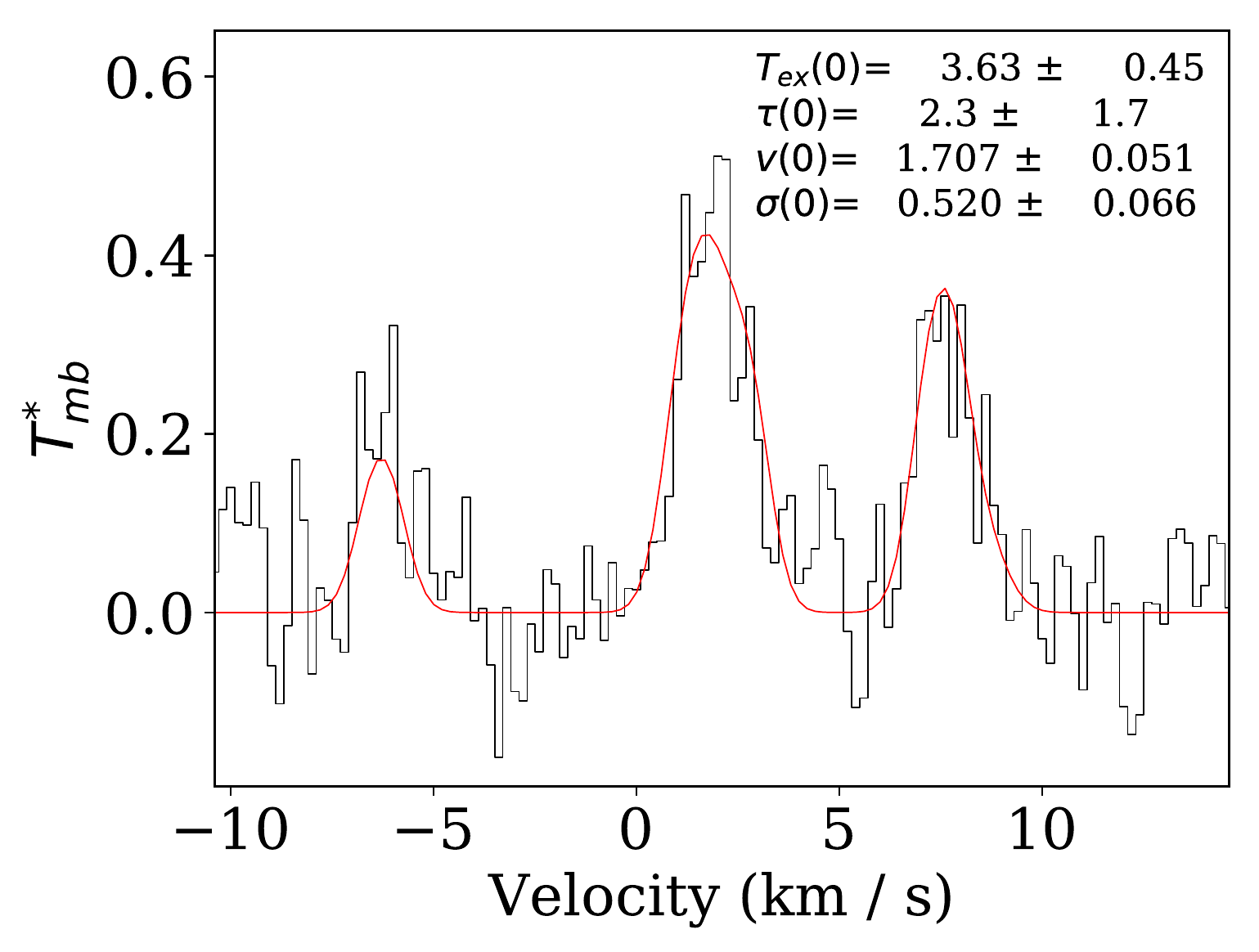}}
\subfloat[G06]{\includegraphics[width=5.5cm]{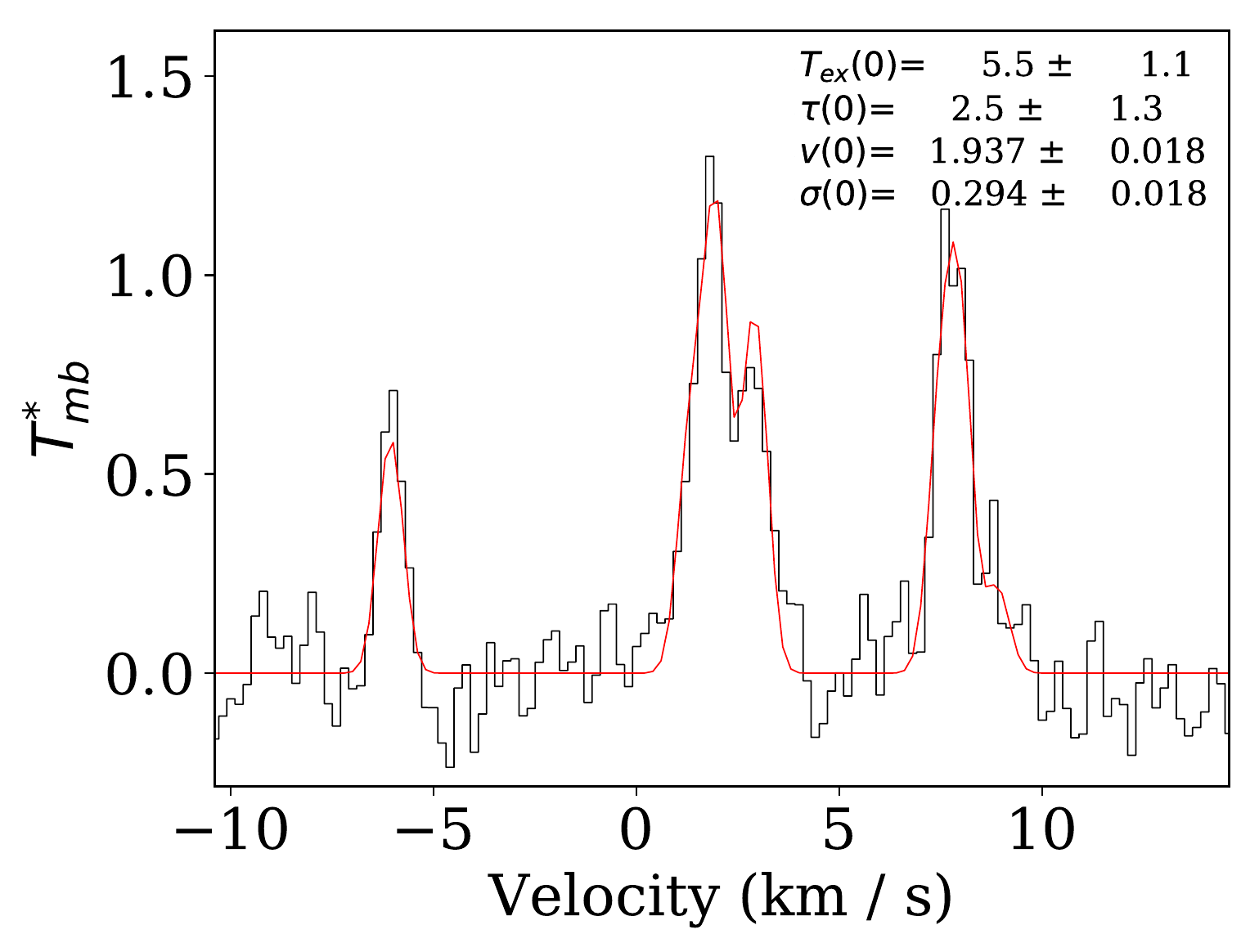}}
\qquad
\subfloat[G07]{\includegraphics[width=5.5cm]{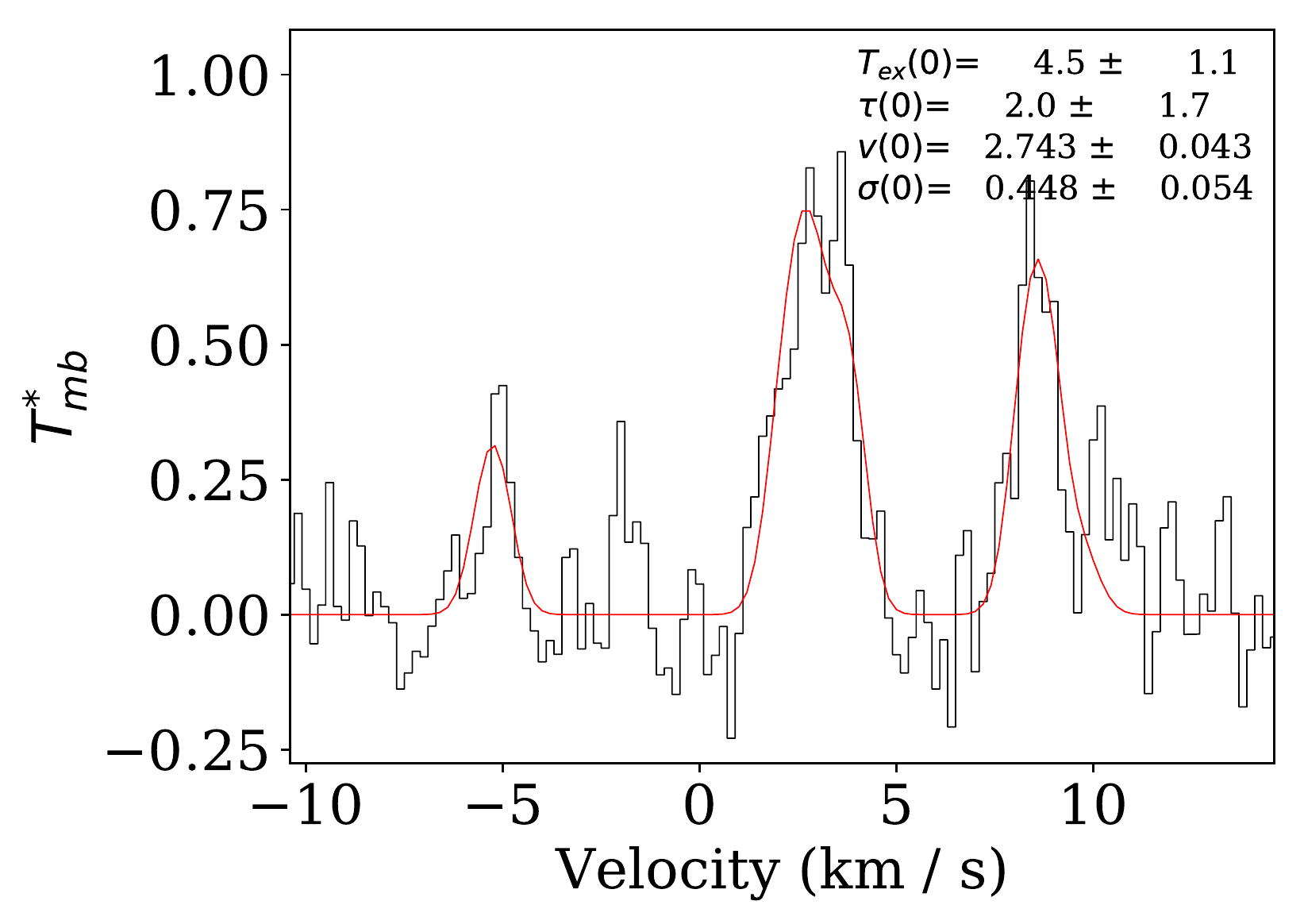}}
\caption{N$_{2}$H$^{+}$(1-0) line profiles for G01-G07 dense cores in G181.\label{fig:fA2}}
\end{figure*}

\begin{table}
\Rotatebox{90}{
\noindent\hrulefill
\normalsize
\begin{tabular}{ccccccccccc}
\\
\hline
Source name & Av & Scale & $Star_{r}$ & $Star_{T}$ &
$Disk_{mass}$ & $Disk_{rmax}$ &
$Disk_{\beta}$ & $Disk_{p}$ & $Disk_{h100}$ &
$Envelope_{\rho_{0}}$ \\

 & $Envelope_{rc}$ & $Cavity_{power}$ & $Cavity_{\theta_{0}}$ & $Cavity_{\rho_{0}}$ &  $Disk_{rmin}$ & $Envelope_{rmin}$ & $Ambient_{density}$ & $Ambient_{T}$ & Scattering & Inclination\\ 
\hline
\hline

J055105.81+273219.0 & 6.832 & 0.255 & 2.153e+00 & 1.198e+04 & 1.085e-02 & 1.044e+03 & 1.145e+00 & -5.010e-02 & 3.416e+00 & 1.185e-22  \\
& 1.044e+03 & 1.961e+00 & 7.617e+00 & 1.782e-23 & 1.805e+00 & 1.805e+00 & 1.000e-23 & 1.000e+01 & 1.000e+00 & 1.145e+00 \\
\hline
J055108.32+273011.1 & 5.607 & 0.255 & 8.348e-01 & 2.239e+04 & 3.554e-02 & 1.979e+03 & 1.151e+00 & -9.120e-01 & 8.123e+00 & 7.537e+00  \\
&  &  &  &  &  &  & 1.000e-23 & 1.000e+01 & 1.000e+00 & 2.631e+01 \\
\hline
J055124.92+272933.9 & 9.294 & 0.255 & 6.762e+00 & 7.325e+03 & 7.344e-04 & 5.733e+01 & 1.211e+00 & -1.768e+00 & 4.691e+00 & 1.373e-23  \\
& 5.733e+01 & 1.249e+00 & 6.471e+00 & 1.409e-22 & 2.322e+00 & 2.322e+00 & 1.000e-23 & 1.000e+01 & 1.000e+00 & 1.501e+01 \\
\hline
J055124.27+272301.5 & 8.317 & 0.255 & 3.777e+00 & 7.043e+03 & 1.709e-05 & 4.927e+02 & 1.220e+00 & -9.799e-01 & 9.738e+00 & 6.235e-21  \\
& 4.927e+02 & 1.777e+00 & 4.771e+01 & 1.831e-22 & 8.606e+00 & 8.606e+00 & 1.000e-23 & 1.000e+01 & 1.000e+00 & 5.674e-01 \\
\hline
J055110.28+273154.2 & 8.449 & 0.255 & 2.655e+00 & 9.242e+03 & 9.852e-02 & 2.247e+02 & 1.071e+00 & -9.608e-01 & 1.277e+00 & 3.427e-2 \\
& 2.247e+02 & 1.175e+00 & 1.755e+01 & 1.802e-22 &  &  & 1.000e-23 & 1.000e+01 & 1.000e+00 & 8.073e+01 \\
\hline
J055129.04+272816.8 & 7.613 & 0.255 & 8.326e+00 & 3.970e+03 & 1.719e-03 & 5.571e+01 & 1.166e+00 & -1.419e+00 & 6.015e+00 & 1.048e-18 \\
& 5.571e+01 & 1.493e+00 & 4.787e+01 & 3.391e-22 &  &  & 1.000e-23 & 1.000e+01 & 1.000e+00 & 6.581e+01 \\
\hline

\end{tabular}
}
\caption{The fitted parameters for YSOs using the best-fitted SED model set.}
\label{tab:tableb1}
\end{table}


\bsp	
\label{lastpage}
\end{document}